\documentclass[preprint,aps,amsmath,nofootinbib]{revtex4-1}
\usepackage{graphicx,array,dcolumn}
\usepackage{calc,tabularx, epsfig,mathrsfs}
\usepackage{hyperref}

\usepackage{amsmath,verbatim,enumerate}
\usepackage{amssymb}
\usepackage{wasysym}
\usepackage{xcolor}
\usepackage{multirow}
\usepackage{xspace}
\usepackage{slashed}
\usepackage{mathtools}
\usepackage{cancel}
\allowdisplaybreaks[1]
\newlength{\figurewidth}
\newcommand{\beq}{\begin{equation}}
\newcommand{\eeq}{\end{equation}}
\newcommand{\bea}{\begin{eqnarray}}
\newcommand{\eea}{\end{eqnarray}}
\newcommand{\ba}{\begin{array}}
\newcommand{\ea}{\end{array}}

\newcommand{\mn}{{\mu\nu}}

\newcommand{\pt}{\partial}

\newcommand{\Bl}{\biggl}
\newcommand{\Br}{\biggr}
\newcommand{\bl}{\bigl}
\newcommand{\br}{\bigr}
\newcommand{\al}{\alpha}
\newcommand{\bt}{\beta}
\newcommand{\g}{\gamma}
\newcommand{\ep}{\epsilon}
\newcommand{\ta}{\theta}
\newcommand{\lam}{\lambda}
\newcommand{\Lam}{\Lambda}
\newcommand{\G}{\Gamma}
\newcommand{\nb}{\nabla}
\newcommand{\de}{\delta}
\newcommand{\D}{\Delta}

\newcommand{\om}{\omega}
\newcommand{\sg}{\sigma}

\makeatother
\begin{document}
%
%%%%%%%%%%%%%%%%%%%%%%%%%%%%%%%%%%%%%%%%%%%%%%%
\title{
Boundary choices and one-loop complex gravitational path integral
}
\setlength{\figurewidth}{\columnwidth}
%%%%%%%%%%%%%%%%%%%%%%%%%%%%%%%%%%%%%%%%%%%%%%%
%
\author{Manishankar Ailiga}
\email{manishankara@iisc.ac.in}

\author{Shubhashis Mallik} 
\email{shubhashism@iisc.ac.in}

\author{Gaurav Narain}
\email{gnarain@iisc.ac.in}

\affiliation{
Center for High Energy Physics, Indian Institute of Science,
C V Raman Road, Bangalore 560012, India.
}

\vspace{20mm}
\begin{abstract}
The path integral of 4D Einstein-Hilbert gravity for the de Sitter-like Universe with fluctuations is investigated, and the transition amplitude from one boundary configuration to another is computed. The gravitational system is described by lapse, scale factor and metric-fluctuation field. Variational consistency demands augmenting the bulk theory with suitable boundary action. A given boundary choice on scale factor is seen to be achievable via an infinite family of covariant boundary actions, each restricting the boundary choices for the fluctuation field. General covariance intimately ties the two boundary choices, which no longer can be chosen independently. For vanishing metric fluctuations at the boundaries, the gauge-fixed gravitational path integral disintegrates into path integral over scale factor and metric-fluctuation field, connected via only lapse integration. While the former is exactly doable, the latter is computed up to one loop, leading to one-loop corrected lapse action. Ultraviolet (UV) divergences are systematically extracted and removed by the addition of suitable counterterms, leading to finite effective action for the lapse. The lapse effective action is then utilized for computing finite transition amplitude. Contributions from virtual gravitons are seen to be secularly growing with Universe size, leading to an infrared divergent transition amplitude. The presence of nonvanishing metric fluctuation at the boundaries implies that the ``no-boundary'' saddles of the theory without metric fluctuations are no longer the saddles of the one-loop corrected action. The corrected saddles have the Universe starting from a nonzero size.

\end{abstract}
\vspace{5mm}

%%%%%%%%%%%%%%%%%%%%%%%%%%%%%%%%%%%%%%%%%%%%%%%

\maketitle

\tableofcontents

%%%%%%%%%%%%%%%%%%%%%%%%%%%%%%%%%%%%%%%%%%%%%%%

%%%%%%%%%%%%%%%%%%%%%%%%%%%%%%%%%%%%%%%%%%%%%%%
\section{Introduction}
\label{intro}
%%%%%%%%%%%%%%%%%%%%%%%%%%%%%%%%%%%%%%%%%%%%%%%

Studies over the decades have highlighted the strengths and importance of using the path integral formalism to investigate quantum systems, especially in field theories. Being a Lagrangian-based formalism, it is not only manifestly covariant but also more convenient in dealing with symmetries. However, path integral is usually mathematically ill defined due to a variety of reasons: singularities, regularization, renormalizability, measure and choice of contour of integration. The path integral of gauge-invariant theories needs further treatment of gauge fixing and corresponding Faddeev-Popov ghosts. Lorentzian path integrals, being highly oscillatory, are not absolutely convergent. By going from real-time to imaginary time, one converts the original flat spacetime Lorentzian path integral to a Euclidean one, which is convergent (Wick rotation). This can be achieved in standard flat spacetime quantum field theories (QFTs). 

In cases when gravity is involved, these successes are hard to replicate, where it is not just the complications of divergences, renormalizability and gauge fixing, but also the choice of contour along which the path integral needs to be suitably performed to make it convergent. The standard Wick rotation may not always work in such situations. Furthermore, the pathological condition of the conformal factor problem in the Euclidean gravitational path integral needs a separate treatment (the path integral over the conformal factor is unbounded from below \cite{Gibbons:1978ac}). This issue can be avoided if one works directly with the Lorentzian path integral. On a manifold ${\cal M}$ with boundary $\pt{\cal M}$, the gravitational path integral with metric $g_\mn$ as the field variable is given by
\beq
\label{eq:PI_LI_g}
{\cal Z} = \int_{{\cal M} + \pt {\cal M}} {\cal D} g_\mn \, e^{iS[g_\mn]/\hbar} \, ,
\eeq
where $S[g_\mn]$ is the gravitational action. The quantity in Eq. (\ref{eq:PI_LI_g}) expresses the summation over all possible geometries with a specific boundary condition (BC), each coming with a ``weight-factor". Doing derivative expansion of the gravitational action up to second order yields 
\bea
\label{eq:grav_act}
S_{\rm grav}[g_\mn] = \frac{1}{16 \pi G} \int d^4x \sqrt{-g}\bigl[
\underbrace{-2\Lambda}_{\pt^0 g} 
+ \underbrace{R}_{\pt^2 g} 
+ \underbrace{\cdots}_{{\cal O}(\pt^4 g)} \bigr]
+ S_{\rm bd},
\eea
where $G$ is Newton's gravitational constant, $\Lam$ is the cosmological constant term, and $S_{\rm bd}$ is the boundary action added to have a consistent variational problem. 

It should be highlighted that in the process of defining convergent path integral in standard flat spacetime QFTs when one Wick rotates from real to imaginary time, one is eventually allowing for the consideration of field theories on complex spacetime geometries in the path integral. It is natural to ask what are then the allowed possible complex geometries on which QFTs can be meaningfully defined? Kontsevich and Segal (KS) \cite{Kontsevich:2021dmb} have recently argued that physically meaningful QFTs should be the limit of theories that are well defined on a class of fixed ``allowable'' complex geometries. Inspiration for this comes from the observation that QFTs complexified in this manner satisfy the usual causality axiom: commutation of spacelike separated operators in the correlation function. For the real free closed $p$-form field $F_{(p)} = {\rm d} A_{(p-1)}$ (for all $p \in \{, \cdots, D\}$, where $D$ is dimension of spacetime), KS criteria defined the domain of all ``allowable'' complex metrics as those satisfying \cite{Kontsevich:2021dmb, Witten:2021nzp, Lehners:2021mah}
\beq
\label{eq:KScri}
{\rm Re} \left(
\sqrt{g} g^{\mu_1 \nu_1} \cdots g^{\mu_p\nu_p}
F_{\mu_1 \cdots \mu_p} F_{\nu_1 \cdots \nu_p}
\right) >0 \, .
\eeq
This is a strong constraint, which means that in the gravitational path integral where the metric field can be complex, one should sum over only those complex geometries, which, besides respecting the boundary conditions, also respect the ``allowability'' criterion stated in Eq. (\ref{eq:KScri}). 

The task is then to systematically study the gravitational path integral under such conditions. This is quite an ambitious goal to achieve as the gravitational path integral for the action stated in Eq. (\ref{eq:grav_act}) is nonrenormalizable \cite{tHooft:1974toh, Deser:1974nb, Deser:1974cz, Goroff:1985sz, Goroff:1985th, vandeVen:1991gw} if higher-derivative terms are not included. Higher-derivative terms can be included but at the cost of unitarity \cite{Stelle:1976gc}, which can be overcome if certain conditions are satisfied \cite{Salam:1978fd, Narain:2011gs, Narain:2016sgk, Buccio:2024hys}. However, one can still make decent progress by somewhat compromising on the ambitions. In this paper, we will focus on the Einstein-Hilbert (EH) gravity and ignore higher-derivative corrections. In any case as pure EH gravity is one-loop renormalizable on shell, it is sufficient for the points addressed in this paper \cite{tHooft:1974toh, Solodukhin:2015ypa}.

We start by considering the spatially homogeneous and isotropic metric [which is the Friedmann–Lemaître–Robertson–Walker (FLRW) metric] in four spacetime dimensions along with the metric fluctuations in the spatial part. In polar coordinates ${\bf x} = \{t_p, r, \ta, \cdots \}$, it is given by
\beq
\label{eq:frwmet}
{\rm d}s^2 = - N_p^2(t_p) {\rm d} t_p^2 
+ 
\underbrace{
a^2(t_p) \left[
\rho_{ij} + h_{ij}(t_p, {\bf x})
\right]
}_{\g_{ij}} {\rm d}x^i \, {\rm d}x^j\, ,
\eeq
where $\rho_{ij}$ can either correspond to a flat or spherical or hyperboloid metric in 3D. The metric perturbations $h_{00}$ and $h_{0i}$ have been gauge fixed to zero and will not be considered in the following. The unknowns appearing in the metric ansatz are lapse $N_p(t_p)$, scale factor $a(t_p)$ and $h_{ij}(t_p, x_m)$. The former two are only time ($t_p$) dependent while the fluctuation $h_{ij}$ is both time ($t_p$) and space ($x_m$) dependent.  

The Feynman path integral in this metric ansatz will be given by 
\beq
\label{eq:Gform_sch}
G[{\rm Bd}_f, {\rm Bd}_i]
= \int_{\bf C} {\rm d}N_p
\int_{{\rm Bd}_i}^{{\rm Bd}_f} 
{\cal D} a(t_p) {\cal D} h_{ij} {\cal D} \bar{c}_i {\cal D} c^i\,\,
e^{i S_{\rm grav}[a, N_p, h_{ij}] + i S_{\rm gf} + i S_{\rm gh}/\hbar} \, ,
\eeq
where ${\bf C}$ refers to contour of integration. The above path integral arises after the integration over the fermionic ghosts and momenta corresponding to each field has been done following the Batalin-Fradkin-Vilkovisky (BFV) formalism \cite{Batalin:1977pb, Feldbrugge:2017kzv}. The time-reparameterization invariance is gauge fixed by working in the proper-time gauge $N^\prime_p=0$ (for a more elaborate discussion on BFV quantization process and ghosts, see \cite{Teitelboim:1981ua, Teitelboim:1983fk, Halliwell:1988wc}). The path integral over the fluctuation field $h_{ij}$ needs to be separately gauge fixed to prevent the overcounting of gauge orbits. This is systematically done by introducing a gauge-fixing action $S_{\rm gf}$ following the Faddeev-Popov procedure \cite{Faddeev:1967fc}, generating ghosts $c_i$ in the process whose action is given by $S_{\rm gh}$. Interestingly, there is a physical interpretation to the path integral done over the components $a(t_p)$ and $h_{ij}(t_p, {\bf x})$. It gives the transition amplitude for the Universe to evolve from one boundary configuration to another in proper time $N_p$ for every proper duration $0<N_p<\infty$ leading to causal propagation from one boundary ${\rm Bd}_i$ to another ${\rm Bd}_f$ \cite{Teitelboim:1983fh}. However, in the following we will be focusing on the object with $N_p$-range $-\infty<N_p<+\infty$. 

Our purpose in this paper is to study the path integral in Eq. (\ref{eq:Gform_sch}) for the gravitational action given in Eq. (\ref{eq:grav_act}), subject to the boundary conditions imposed on the scale factor $a(t_p)$ and $h_{ij}(t_p, \textbf{x})$. Requirements of the general covariance restrict the freedom to choose the boundary conditions for the scale factor $a(t_p)$ and $h_{ij}(t_p, \textbf{x})$ independently, and the two are intimately tied to each other \cite{Brizuela:2023vmb}. It implies that if certain boundary conditions are imposed on the scale factor and/or its conjugate momenta, then requirements of general covariance limit the available boundary choices for the fluctuation field $h_{ij}(t_p, \textbf{x})$. In this paper, while exploring and studying this connection in more detail, we will also investigate the consequences of these on the path integral. 

In the past, the implication of various boundary conditions on the minisuperspace path integral has been investigated (see also \cite{Chou:2024sgk, Nishimura:2023dky} for numerical studies). In particular, the path integral, where the summation is over all possible geometries which are regular at an initial time, has recently gained a lot of interest due to its simplicity. This is commonly known as the``no-boundary'' proposal of the Universe, where we define the path integral over geometries starting with zero sizes (see \cite{Lehners:2023yrj} for a review of the no-boundary proposal). Recently, studies of the no-boundary proposal in Horava-Lifshitz gravity \cite{Matsui:2023hei} and its implications using resurgence \cite{Matsui:2022lfj} have also been explored. It was noticed that imposition of Dirichlet boundary condition (DBC) on the scale factor leads to unsuppressed behavior of fluctuations in the no-boundary universe \cite{Feldbrugge:2017kzv, DiTucci:2018fdg, Lehners:2018eeo, Feldbrugge:2017fcc, Feldbrugge:2017mbc}. Motivated by this, studies involving Neumann boundary condition (NBC) and Robin boundary condition (RBC) \cite{Krishnan:2016mcj, Krishnan:2017bte} have been pursued where favorable behavior of perturbations was noticed \cite{DiTucci:2019bui, Narain:2021bff, Lehners:2021jmv, DiTucci:2020weq, Narain:2022msz, DiTucci:2019dji, Ailiga:2023wzl, Ailiga:2024mmt}. However, as the boundary choices for the scale factor and fluctuation field are intimately tied to each other due to requirements from covariance, our purpose in this paper is to consider the Neumann and Robin boundary conditions for the scale factor and investigate the boundary choices available for the fluctuation field $h_{ij}(t_p, {\bf x})$ in each case, along with their corresponding implications. 

To compute the transition amplitude from one boundary configuration to another, the gravitational path integral is studied for the fixed boundary choices. Ignoring backreaction, the full gauge-fixed path integral is seen to decompose into a path integral over the scale factor and path integral over the fluctuation field. However, the overall ordinary lapse integral still couples the two. The path integral over the scale factor can be done exactly in the case of Einstein-Hilbert gravity as the theory has only two time derivatives and the action is quadratic. The path integral over the fluctuation field is nontrivial as the action of the fluctuation field is no longer quadratic but will contain an infinite number of terms. As a result it is not possible to do an exact computation for the fluctuation path integral. However, up to one loop the path integral over the fluctuation field can be computed in which case one needs to retain terms up to only quadratic order in the fluctuation field in the action. The resulting quantity is one-loop corrected lapse action. It is seen that this quantity is not finite and diverges, needing regularization. Finiteness is obtained after suitable counterterms are added to cancel divergences. 

The lapse integration is performed on this finite lapse action to compute the UV-finite transition amplitude. The lapse integral is done following the methodology of  Picard-Lefschetz \cite{Witten:2010cx, Witten:2010zr, Basar:2013eka, Tanizaki:2014xba} which has proven to be useful in dealing with highly oscillatory integrals. It allows one to find a contour of integration systematically along which the integrand becomes well behaved leading to a convergent integral. These deformed contours of integration termed {\it Lefschetz thimbles} generalize the notion of standard Wick's rotation to curved spacetime. Working in the WKB approximation, the lapse integral eventually leads to a finite transition amplitude. 

The outline of this paper is as follows: In Sec. \ref{exp} we do the expansion of the action up to quadratic order. We discuss the gauge transformation and gauge fixing including the Fadeev-Popov ghost in Sec. \ref{gfgh}. In Sec. \ref{boundTR} we explore the consequences of general covariance and how the boundary choices for the scale factor and fluctuation field are intimately tied. In Sec. \ref{EQM_qh}, we compute the generic on shell solution for the scale factor and fluctuation field with corresponding action. In Sec. \ref{bound_act}, we study the possible boundary choices for the fluctuation field for the Dirichlet, Neumann and Robin boundary choice for scale factor. In Sec. \ref{trans_amp}, we compute the path integral over the scale factor and fluctuation field up to one loop. In Sec. \ref{sec:lapseNc}, we systematically analyze the lapse integral in the saddle-point approximation. In Sec. \ref{NB_sad_cor} we compute the one-loop quantum corrected action for lapse which is seen to be UV divergent at the no-boundary saddles. The divergences are systematically extracted, and suitable counterterms are added (Sec. \ref{cont_log}). The lapse integration on the finite lapse action via Picard-Lefschetz and WKB methods is done in Sec. \ref{PL_Nc_sad}. The quantum-corrected UV-finite transition amplitude is computed in Sec. \ref{subs:quant_TA}. We compute corrections to saddles and analyze their stability in Sec. \ref{sec:hij_neq_0}. Conclusions are presented in Sec. \ref{conc}.

%%%%%%%%%%%%%%%%%%%%%%%%%%%%%%%%%%%%%%%%%%%%%%%
\section{Action expansion}
\label{exp}
%%%%%%%%%%%%%%%%%%%%%%%%%%%%%%%%%%%%%%%%%%%%%%%

We plug the metric ansatz given in Eq. (\ref{eq:frwmet}) in the gravitational 
action mentioned in Eq. (\ref{eq:grav_act}) and expand it up to second order in $h_{ij}$. 
This is sufficient if one is interested in computing one-loop 
quantum corrections of the corresponding theory. Higher-order terms are needed 
if one goes beyond one loop which will not be considered in this paper.
To achieve this we make use of the Arnowitt–Deser–Misner (ADM) decompositions.
\footnote{
The convention for the
Riemann tensor for the spacetime metric $g_\mn$ is
$R_{\mn}{}^\rho{}_\sg = \pt_\mu \G_\nu{}^\rho{}_\sg 
- \pt_\nu \G_\mu{}^\rho{}_\sg
+\G_\mu {}^\rho{}_\lam \G_\nu{}^\lam{}_\sg
- \G_\nu {}^\rho{}_\lam \G_\mu{}^\lam{}_\sg
$ and the Christoffel connection is taken as
$\G_\al{}^\mu{}_\bt
= 2^{-1} g^{\mu\rho} \left[
\pt_\al g_{\rho\bt} + \pt_\bt g_{\rho\al} - \pt_\rho g_{\al\bt}
\right] . 
$
}
The determinant $\sqrt{-g} = N_p \sqrt{\g}$, where $g = \det g_{\mn}$ and 
$\g = \det \g_{ij}$. 
The covariant derivative compatible to $\g_{ij}$ is $D_k \g_{ij} =0$.
The Christoffel connection $\Theta_i{}^k{}_j$ and extrinsic curvature 
$K_{ij}$ corresponding to it is given by
\bea
\label{eq:chris_gam}
&&
\Theta_i{}^k{}_j = 2^{-1} \g^{km} \left[
\pt_i \g_{mj} + \pt_j \g_{mi} - \pt_m \g_{ij}
\right]\, , 
\hspace{5mm}
K_{ij} = \frac{\g_{ij}^\prime}{2N_p} \, ,
\notag \\
&&
K = \g^{ij} K_{ij} = \frac{(\ln \sqrt{\g})^\prime}{N_p} \, ,
\eea
respectively. In the last line of the above equation $K$ is the trace of 
the extrinsic curvature. The Ricci-tensor corresponding to $\g_{ij}$ is 
\bea
\label{eq:3Rgamma}
&&
{\cal R}_{ij} = \pt_m \Theta_i{}^m{}_j - \pt_i \Theta_m{}^m{}_j
+ \Theta_m{}^m{}_n \Theta_i {}^n{}_j - \Theta_i{}^m{}_n \Theta_m {}^n{}_j \, ,
\notag \\
&&
R = {\cal R} + K_{ij} K^{ij} + K^2 + 2K^\prime/N_p \, ,
\eea
where $({}^\prime)$ denotes the derivative with respect to $t_p$
and ${\cal R} = \g^{ij} \,\, {\cal R}_{ij}$ is the Ricci-scalar 
of the three hypersurface, while $R$ is the Ricci scalar 
of the full spacetime metric. 
In the following we will work in the proper-time gauge 
$N_p^\prime =0$, implying $N_p = {\rm const.}$
In this situation the gravitational action mentioned in Eq. (\ref{eq:grav_act})
becomes the following:
\bea
\label{eq:EHact_ADM}
&&
S_{\rm grav}[g_\mn] = \frac{1}{16 \pi G}\int_{\cal M} {\rm d}^4x \sqrt{-g} 
\bigl[
-2 \Lam + {\cal R} + K_{ij} K^{ij} + K^2 + 2K^\prime/N_p
\bigr]  + S_{\rm bd} \, ,
\notag \\
&&
= \frac{1}{16 \pi G} \int_{\cal M} {\rm d}^4x \sqrt{-g} 
\bigl[
-2 \Lam + {\cal R} + K_{ij} K^{ij} - K^2 
\bigr] 
+ \frac{2}{16 \pi G} \int_{\pt {\cal M}} {\rm d} {\bf x} \sqrt{\g} K + S_{\rm bd} \, ,
\eea
where we have done integration by parts 
and used the expression of $K$ stated in 
Eq. (\ref{eq:chris_gam}) to rewrite 
\beq
\label{eq:rewrite}
\int_{\cal M} {\rm d}^4x \sqrt{-g} (2K^\prime/N_p)
= -2 \int {\rm d}^4 x  \sqrt{-g} K^2 
+ 2 
\underbrace{
\int {\rm d}t_p {\rm d} {\bf x} (\sqrt{\g} K)^\prime
}_{\rm total \,\, derivative} \, .
\eeq
At this point we do the expansion of various metric dependent quantities 
used in the gravity action 
\bea
\label{eq:exp1}
&&
\g^{mn} = a^{-2} \left[
\rho^{mn} - h^{mn} + h^{mi} h_i{}^n + \cdots
\right] \, ,
\notag \\
&&
\sqrt{\g} = a^{3} \sqrt{\rho} \left[
1 + h + \left(\frac{h^2}{8} - \frac{1}{4} h_{ij} h^{ij} \right) +\cdots
\right] \, ,
\notag \\
&&
\Theta_i{}^k{}_j = \bar{\Theta}_i{}^k{}_j 
+ \underbrace{
\frac{1}{2}\left(\bar{D}_i h^k{}_j + \bar{D}_j h^k {}_i - \bar{D}^k h_{ij} \right)
}_{\Theta^{(1)}_i{}^k{}_j}
\underbrace{
- \frac{1}{2} h^{km} \left(\bar{D}_i h_{mj} + \bar{D}_j h_{mi} - \bar{D}_m h_{ij} 
\right) 
}_{\Theta^{(2)}_i{}^k{}_j}+ \cdots \, ,
\eea
where $\bar{D}_i$ is the covariant derivative with respect to the metric $\rho_{ij}$, 
$h = \rho^{ij} h_{ij}$, and the indices of $h_{ij}$ are raised or lowered 
using $\rho_{ij}$. 
The extrinsic curvature can be expanded by making use of the 
definition of $K_{ij}$ mentioned in Eq. (\ref{eq:chris_gam}). This gives
\bea
\label{eq:K_exp}
&&
K = \frac{3a^\prime}{a N_p} 
+ \frac{1}{2N_p} \left[
h^\prime - h^{im} h^\prime_{im}+ \cdots
\right] \, , 
\notag \\
&&
K_{ij} K^{ij} = \frac{1}{4N_p^2} \left[
\frac{12 a^{\prime 2}}{a^2} + \frac{4 a^\prime h^\prime}{a}
- \frac{4 a^\prime h^{im} h^\prime_{im}}{a}
+ h^\prime_{im} h^{\prime im} \cdots
\right] \, ,
\notag \\
&&
{\cal R} = a^{-2} \biggl[
\bar{\cal R} 
+ \left(\bar{D}_i \bar{D}_j h^{ij} - \bar{\Box} h - \bar{\cal R}_{ij} h^{ij}\right)
+ \rho^{ij} \left(\bar{D}_k \Theta^{(2)}_i {}^k{}_j
- \bar{D}_i \Theta^{(2)}_k {}^k{}_j \right) 
\notag \\
&&
+ \bar{\cal R}_{ijmn} h^{im} h^{jn} 
- h^{ij} \bar{D}_i \bar{D}_k h^k{}_j 
+ \frac{1}{2} h^{ij} \bar{\Box} h_{ij} 
+ \frac{1}{2} h^{ij} \bar{D}_i \bar{D}_j h 
+ \frac{1}{2} \bar{D}^i h \bar{D}^j h_{ij}  
\notag \\
&& - \frac{1}{4} \bar{D}^i h \bar{D}_i h 
+ \frac{1}{4} \bar{D}^k h^{ij} \bar{D}_k h_{ij} 
- \frac{1}{2}\bar{D}^k h^{ij} \bar{D}_i h_{jk} 
\biggr]
\, ,
\eea
where $\bar{\cal R}_{ijmn}$, $\bar{\cal R}_{ij}$, $\bar{\cal R}$ 
are the Riemann tensor, Ricci tensor and Ricci scalar corresponding to metric 
$\rho_{ij}$ respectively. These expansions are needed in the 
construction of the suitable boundary term for the fluctuation field 
$h_{ij}$ for a given choice of boundary conditions for the 
scale factor.

%%%%%%%%%%%%%%%%%%%%%%%%%%%%%%%%%%%%%%%%%%%%%%%
\subsection{Gauge-fixing and Faddeev-Popov ghosts}
\label{gfgh} 
%%%%%%%%%%%%%%%%%%%%%%%%%%%%%%%%%%%%%%%%%%%%%%%

As the path integral is diffeomorphism invariant, it needs to be 
gauge fixed. In this subsection we will systematically introduce the 
gauge-fixing condition and compute the corresponding 
Faddev-Popov ghost action. 

We study the gauge-invariant action by making use of the 
background field methods \cite{DeWitt:1980jv, Abbott:1980hw}.
This procedure manifestly preserves background gauge
invariance. The field when decomposed into fixed background and 
fluctuation leads to a path integral over the fluctuation field. 
The invariance of the full field is transmuted to invariance 
of the fluctuation field. Overcounting of the gauge orbits 
in the measure is avoided by constraining the field variable (gauge fixing),
which in turn leads to the generation of auxiliary fields called {\it ghosts}
\cite{Faddeev:1967fc}.

The diffeomorphism invariance of the action in Eq. (\ref{eq:grav_act})
implies that for arbitrary vector field $V^\rho$, the action should be
invariant under the following transformation of the metric field variable,
\beq 
\label{eq:gaugetrgamma}
\de_{\rm D} g_\mn = {\cal L}_V g_\mn = V^\rho \partial_\rho
g_\mn + g_{\mu\rho} \partial_\nu V^\rho +
g_{\nu\rho}\partial_\mu V^\rho \, ,
\eeq
where ${\cal L}_V g_\mn$ is the Lie derivative of the metric
$g_\mn$ along the vector field $V^\rho(t_p, {\bf x})$. 
This would imply that the components of the ADM metric 
mentioned in Eq. (\ref{eq:frwmet}) will transform as
\bea
\label{eq:gaugeTR}
&&
\de_{\rm D}(N_p) 
= \pt_0(N_p V^0) + V^k \pt_k N_p \, ,
\notag \\
&&
\de_{\rm D} (\g_{ij}) = V^0 \pt_0 \g_{ij}
+ V^k \pt_k \g_{ij} + \g_{ik} \pt_j V^k + \g_{jk} \pt_i V^k \, .
\eea
The gauge transformation will also generate a shift vector even if 
the metric ansatz does not have it. However, this can be gauge fixed to 
zero. This is achieved by considering only gauge transformation 
of type $V^\rho (t_p, {\bf x})= \{V^0 (t_p), V^i ({\bf x}) \}$.
Moreover, if $N_p$ is constant ($\pt_0 N_p=0$ and $\pt_i N_p=0$)
and $\pt_0V^0(t_p) =0$, then $N_p$ does not transform under 
gauge transformation and has been gauge fixed 
to a constant (proper time gauge).
\footnote{
The gauge transformation of the shift vector is $\de_{\rm D} N_i = \pt_0(N_i V^0) + V^j \pt_j N_i 
+ N_j \pt_i V^j + (-N^2 + N_k N^k) \pt_i V^0 + \g_{ij} \pt_0 V^j$. In the proper-time gauge, $N_i=0$ implies that rhs vanishes leading to a consistent gauge fixing. 
} 
One is then left only with the gauge transformation of the 
spatial metric $\g_{ij}$. Inserting $\g_{ij} = a^2(t_p)(\rho_{ij}
+ h_{ij})$, one can extract the gauge transformation
of the fluctuation field $h_{ij}$ to be
\beq
\label{eq:hij_trans}
\de_{\rm D} h_{ij} = \frac{2a^\prime}{a} V^0 \rho_{ij}
+ \bar{D}_i V_j + \bar{D}_j V_i
+ \frac{2a^\prime}{a} V^0 h_{ij} + V^0 \pt_0 h_{ij}
+ V^k \bar{D}_k h_{ij}
+ h_{ik} \bar{D}_j V^k + h_{jk} \bar{D}_i V^k \, ,
\eeq
where $\bar{D}_i$ is the covariant derivative 
with respect to $\rho_{ij}$. In QFT when the quantum field 
$h_{ij}$ is considered to be small then the terms independent 
of $h_{ij}$ are the leading ones. Moreover, these are the terms which 
will also contribute at one loop while higher-loop contributions come 
from terms of order ${\cal O}(h)$. The invariance of action for the 
$h_{ij}$ field is broken by choosing an appropriate gauge-fixing 
condition which is implemented following the Faddeev-Popov procedure \cite{Faddeev:1967fc}.
The gauge-fixing condition chosen to break the invariance 
of the $h_{ij}$ field is 
\beq
\label{eq:gaugecond_h}
F_i = \bar{D}^m h_{mi} = 0 \, ,
\hspace{5mm}
F = h_i{}^i=0 \, .
\eeq
The former is the Landau gauge choice with the latter requiring vanishing trace. The Landau gauge choice suppresses the propagation of longitudinal modes while only the transverse mode can propagate. This gauge fixing
will lead to a ghost determinant which can be exponentiated using 
fermionic auxiliary fields (Faddeev-Popov ghosts). 
The ghost determinant corresponding to the gauge-fixing 
condition mentioned in Eq. (\ref{eq:gaugecond_h}) is
\bea
\label{eq:ghost_det}
&&
{\rm Ghost \,\, determinant} \,\, 
\Rightarrow \,\, 
(\det \D_{jm}) \, \notag \\
&&
\D_{jm} = \rho_{jm} \nb^2 + \bar{D}_m \bar{D}_j 
+ \bar{D}_m h_{ij} \bar{D}^i 
+ \bar{D}^i \bar{D}_m h_{ij}
+ h_{im} \bar{D}^i \bar{D}_j 
+ \bar{D}^i h_{jm} \bar{D}_i
+ h_{jm} \nb^2 \, ,
\eea
where $\nb^2 = \bar{D}_i \bar{D}^i$. 
At one loop only the terms independent of the $h_{ij}$ contributes while ${\cal O}(h)$ terms contribute at higher loops. The gauge condition of $h_i{}^i=0$ does not lead to additional ghosts \cite{Ohta:2015zwa, Lin:2017ool}.

%%%%%%%%%%%%%%%%%%%%%%%%%%%%%%%%%%%%%%%%%%%%%%%
\subsection{Boundary action and covariance}
\label{boundTR} 
%%%%%%%%%%%%%%%%%%%%%%%%%%%%%%%%%%%%%%%%%%%%%%%

General covariance brings in strong constraints on the choices of 
boundary conditions available for the fluctuation field $h_{ij}$ 
for the given choice of boundary condition satisfied by the 
background. This is in a sense expected as the action of theory 
for the metric field $g_\mn$ is diffeomorphic. In this subsection 
we explore this connection in some detail to find the set of choices 
for boundary condition available for the fluctuation field $h_{ij}$. 

To set the stage for extracting these constraints, we start by first 
making the following transformation for the scale factor and 
lapse:
\beq
\label{eq:rescale}
N_p(t_p) {\rm d} t_p = \frac{N(t)}{a(t)} {\rm d} t \, ,
\hspace{5mm}
q(t) = a^2(t) \, .
\eeq
This transformation is beneficial as it casts the minisuperspace 
action in a simpler form \cite{Feldbrugge:2017kzv, Narain:2022msz, Ailiga:2023wzl}.
The line element after this coordinate transformation is given by
\beq
\label{eq:frwmet_changed}
{\rm d}s^2 = - \frac{N^2}{q(t)} {\rm d} t^2 
+ \underbrace{
q(t) \left[
\rho_{ij} + h_{ij}(t, {\bf x})
\right]
}_{\g_{ij}} {\rm d}x^i \, {\rm d}x^j \, .
\eeq
Here we will work obeying the Landau gauge condition 
mentioned in Eq. (\ref{eq:gaugecond_h}) along with the 
proper-time gauge, which keeps $N =$ constant ($N_c$). We then insert the 
above metric mentioned in Eq. (\ref{eq:frwmet_changed}) in the 
gravity action given in Eq. (\ref{eq:grav_act}) and 
expand it in powers of $h_{ij}$ keeping terms up to 
second order. In doing so we make use of the set of expansions mentioned 
in Eqs. (\ref{eq:3Rgamma}), (\ref{eq:exp1}), \ref{eq:K_exp}) and utilize the 
transformations given in Eq. (\ref{eq:rescale}). This 
gives the following expansion of the action 
\bea
\label{eq:EHact_exp}
&&
S_{\rm grav}[g_\mn] 
= S_{\rm grav}^{(q)} + S_{\rm grav}^{(h)} = 
\frac{1}{16\pi G}\int_{\cal M} {\rm d} t {\rm d} {\bf x} \sqrt{\rho}
\biggl[
(6k -2 \Lam q )N_c - \frac{3\dot{q}^2}{2N_c} 
\biggr] 
\notag \\
&&
+ \frac{1}{16\pi G}\int_{\cal M} {\rm d} t {\rm d} {\bf x} \sqrt{\rho}
\biggl[ 
\frac{(\Lam q -2 k) N_c}{2}h_{ij} h^{ij}
+ \frac{2q^2 \dot{h}_{ij} \dot{h}^{ij} 
-  (\dot{q}^2 + 4 q \ddot{q} )h_{ij}h^{ij}}{8N_c} 
+ \frac{N_c}{4} h_{ij} \bar{D}^2 h^{ij} 
\biggr] 
\notag \\
&&
+ \frac{1}{16\pi G} \int_{\pt {\cal M}} {\rm d} {\bf x} 
\sqrt{\rho} \left(\frac{3q \dot{q}}{N_c}\right) \biggr \rvert_0^1
+ \frac{1}{16\pi G}\int_{\pt {\cal M}} {\rm d} {\bf x} \sqrt{\rho}
\biggl(
-\frac{q^2}{N_c} h_{ij} \dot{h}^{ij}
- \frac{q \dot{q}}{4N_c} h_{ij} h^{ij}
\biggr) \biggr\rvert_0^1 + S_{\rm bd}\, ,
\eea
where besides using the gauge-fixing conditions, 
we have done integration by parts in some places. Here, $\pt {\cal M}$ refers to temporal boundaries. Note that some of the surface terms generated during these integrations vanish as the field $h_{ij}$ vanishes at spatial infinity. The action $S_{\rm bd}$ will be suitably computed to set up a consistent variational problem for the field variables. It will consist of two parts:
\beq
\label{eq:Sbd_break}
S_{\rm bd} = S_{\rm bd}^{(q)} + S_{\rm bd}^{(h)} \, .
\eeq 
The action $S_{\rm grav}^{(q)}$ and $S_{\rm grav}^{(h)}$ are given by
\bea
\label{eq:Act_sq_sh_begEXP}
&&
S_{\rm grav}^{(q)}
= \frac{1}{16\pi G}\int_{\cal M} {\rm d} t {\rm d} {\bf x} \sqrt{\rho}
\biggl[
(6k -2 \Lam q )N_c - \frac{3\dot{q}^2}{2N_c} 
\biggr] +  \frac{1}{16\pi G} \int_{\pt {\cal M}} {\rm d} {\bf x} 
\sqrt{\rho} \left(\frac{3q \dot{q}}{N_c}\right) \biggr \rvert_0^1 + S_{\rm bd}^{(q)} \, ,
\notag \\
&&
S_{\rm grav}^{(h)}
= \int_{\cal M} 
\frac{{\rm d} t {\rm d} {\bf x} \sqrt{\rho}}{16\pi G}
\biggl[ 
\frac{(\Lam q -2 k) N_c}{2}h_{ij} h^{ij}
+ \frac{2q^2 \dot{h}_{ij} \dot{h}^{ij} 
-  (\dot{q}^2 + 4 q \ddot{q} )h_{ij}h^{ij}}{8N_c} 
+ \frac{N_c}{4} h_{ij} \bar{D}^2 h^{ij} 
\biggr] 
\notag \\
&& + \frac{1}{16\pi G}\int_{\pt {\cal M}} {\rm d} {\bf x} \sqrt{\rho}
\biggl(
-\frac{q^2}{N_c} h_{ij} \dot{h}^{ij}
- \frac{q \dot{q}}{4N_c} h_{ij} h^{ij}
\biggr) \biggr\rvert_0^1 + S_{\rm bd}^{(h)} \, ,
\eea
respectively.
In this section, we will systematically compute the form of 
$S_{\rm bd}^{(h)}$ for a given choice of boundary condition for the 
$q(t)$ resulting in a choice of $S_{\rm bd}^{(q)}$. A generic form 
for $S_{\rm bd}$ respecting general covariance can be taken to be
\beq
\label{eq:Sbd_genform}
S_{\rm bd} = \frac{1}{8\pi G} \int_{\pt {\cal M}} {\rm d} {\bf x}
\sqrt{\g} \,\, F_{\pt {\cal M}}({\cal R}, K) \biggr \rvert_0^1\, ,
\eeq
where the subscript $_{\pt {\cal M}}$ of the function $F$ 
means that the function $F$ can be different for different boundaries. 
We do an expansion of the $S_{\rm bd}$ in powers of $h_{ij}$
up to second order in $h_{ij}$. To achieve this we make use of the 
expansion of $K$ and ${\cal R}$ listed in Eq. (\ref{eq:K_exp}).
For the gauge-fixing choice considered here, neither  
$K$ nor ${\cal R}$ has a linear term in $h_{ij}$. This means that 
the expansion of $F$ is given by
(we define background quantities 
${\cal R}_b=\bar{{\cal R}}/q = {\cal R}_{h=0}$, 
$K_b=3\dot{q}/(2N_c \sqrt{q}) = K_{h=0}$)
\bea
\label{eq:Fexp}
S_{\rm bd}
= &&
\frac{1}{8\pi G} \int_{\pt {\cal M}} {\rm d} {\bf x}\sqrt{\rho}
\biggl[
q^{3/2} F({\cal R}_b, K_b)
- \frac{q^{3/2}}{4} F({\cal R}_b, K_b) h_{ij} h^{ij}
-\frac{q^2}{2N_c} \pt_{K} F({\cal R}_b, K_b)
h_{ij} \dot{h}^{ij}
\notag \\
&&
+ \sqrt{q} \pt_{\cal R} F({\cal R}_b, K_b)
\left\{\frac{k}{2} h_{ij} h^{ij}
+ \frac{1}{4} h_{ij} \bar{D}^2 h^{ij}
\right\} + \cdots
\biggr] \biggr \rvert_0^1\, .
\eea
It is seen that the boundary term also consists of two pieces:
one based on the scale factor entirely while the other depends on the 
fluctuation. These are computed by demanding consistency of the 
full variational problem of the various fields entering the system. 
If one varies the fluctuation independent action 
$S_{\rm grav}^{(q)} = S_{\rm EH}^{(q)} + S_{\rm bd}^{(q)}$ with respect to the
field by writing $q(t) = \bar{q}(t) + \de q(t)$ and expand in powers of 
$\de q(t)$, we get at linear order in $\de q(t)$:
\beq
\label{eq:Sexp_qvar}
\de S^{(q)} = \frac{1}{16\pi G} \int_{0}^{1} {\rm d}t \biggl[
\left(-2 \Lam N_c + \frac{3 \ddot{\bar{q}}}{N_c} \right) \de q
\biggr]
+ \frac{1}{16\pi G} \biggl[
\frac{3\bar{q} \de \dot{q}}{N_c} 
+ 3 \sqrt{\bar{q}} \bar{F} \de q 
+ 2 \bar{q}^{3/2} \de F({\cal R}_b, K_b)
\biggr] \biggr\rvert_0^1\, ,
\eeq
where $\bar{F} = F(\bar{\cal R}_b, \bar{K}_b)$,
$\bar{\cal R}_b = \bar{\cal R}/\bar{q}$ and 
$\bar{K}_b = 3 \dot{\bar q}/(2N_c \sqrt{\bar q})$. 
The equation of motion follows from the 
term proportional to $\de q$:
\beq
\label{eq:dyn_q_eq}
\ddot{\bar{q}} - \frac{2}{3} \Lam N_c^2 =0 \, .
\eeq
However, this is obtained under the assumption that no contribution comes from the second line of Eq. (\ref{eq:EHact_exp}). This is in accordance with the requirements of ignoring backreaction \cite{Feldbrugge:2017mbc,Barvinsky:1992dz} where the behavior of $h_{ij}$, does not affect the dynamical evolution of the scale factor $q(t)$. 
For a consistent variational problem the set of boundary terms must 
vanish. This is achievable either by making a choice of the boundary condition 
or by choosing the function $F$ such that the square bracket 
vanishes or by satisfying both criteria. The same is implemented 
for the fluctuation action $S_{\rm grav}^{(h)}$. 

The fluctuation action when varied 
gives the following: 
\bea
\label{eq:Sexp_hvar}
\de S^{(h)} = 
&&
\int {\rm d}{\bf x} {\rm d}t \frac{\sqrt{\rho}}{16\pi G}
\biggl[
-\frac{\bar{q}^2 \ddot{\bar h}_{ij}}{2N_c} 
- \frac{\bar{q} \dot{\bar{q}} \dot{\bar h}_{ij}}{N_c}
+ \frac{N_c}{2} \bar{D}^2 \bar{h}_{ij} 
+ \biggl\{
(\Lam \bar{q} -2k)N_c - \frac{\dot{\bar{q}}^2}{4N_c} - \frac{\bar{q} \ddot{\bar{q}}}{N_c}
\biggr\} \bar{h}_{ij}
\biggr] \de h^{ij}
\notag \\
&&
+ \frac{1}{16\pi G} \int_{\pt {\cal M}} {\rm d}{\bf x} \sqrt{\rho}
\bigl[
A_{ij} \de h^{ij} + B_{ij} \de \dot{h}^{ij}
\bigr] \biggr \rvert_0^1\, ,
\eea
where
\bea
\label{eq:ABexp}
&&
A_{ij} = -\biggl[
\frac{\bar{q} \dot{\bar{q}}}{2N_c}
+\bar{q}^{3/2} {\bar F} 
-2k \sqrt{\bar{q}} \pt_{\cal R} {\bar F}
\biggr]\bar{h}_{ij} 
+ \sqrt{\bar{q}} \pt_{\cal R} {\bar F} \bar{D}^2 \bar{h}_{ij}
- \frac{\bar{q}^2 (1 + 2\pt_K {\bar F})}{2N_c} \dot{\bar h}_{ij} \, ,
\notag \\
&&
B_{ij} = - \frac{\bar{q}^2 \{1 + \pt_K {\bar F}\}}{N_c} \bar{h}_{ij} \, ,
\eea
where $\pt_{\cal R} {\bar F} = \pt_{\cal R}F(\bar{\cal R}_b, \bar{K}_b)$,
$\pt_K {\bar F} = \pt_KF(\bar{\cal R}_b, \bar{K}_b)$ and 
$\bar{h}_{ij}$ satisfies the equation of motion for the fluctuation field 
\beq
\label{eq:hijEQM}
\bar{q}^2 \ddot{\bar h}_{ij}
+2\bar{q} \dot{\bar{q}} \dot{\bar h}_{ij}
- N_c^2 \bar{D}^2 \bar{h}_{ij} 
+ N_c^2 \biggl[
2k + \left(\frac{\dot{\bar{q}}^2}{2N_c^2}+\frac{6k - 2 \Lam \bar{q}}{3} \right)
\biggr] \bar{h}_{ij} = 0
 \, .
\eeq
To simplify the above we have used the equation of motion for $\bar{q}$.
The expression appearing inside the braces is the constraint 
which vanishes at the saddle points. For a consistent 
variational problem the boundary conditions should be such that 
\beq
\label{eq:BD_h_gen}
\bigl(A_{ij} \de h^{ij} + B_{ij} \de \dot{h}^{ij} \bigr) \biggr \rvert_0^1= 0 \, .
\eeq
This is achieved by either choosing a boundary condition such 
that it identically vanishes or the choice of $F$ makes it vanish. 
However, the function $F$ gets fixed while requiring the consistency of the 
variational problem of the background $\bar{q}$. This in turn
implies that the freedom in making a choice 
for the boundary condition for the fluctuation is constrained and 
cannot be done arbitrarily. In the next section we will explore 
the consequences of this in the cosmology.

%%%%%%%%%%%%%%%%%%%%%%%%%%%%%%%%%%%%%%%%%%%%%%%
\subsection{On shell}
\label{EQM_qh} 
%%%%%%%%%%%%%%%%%%%%%%%%%%%%%%%%%%%%%%%%%%%%%%%
Having set up the variational problem consistently it is crucial to 
find the solution to the equation of motion. The equation of motion for 
$q(t)$ and $h_{ij}(t, {\bf x})$ is given in Eqs. (\ref{eq:dyn_q_eq}) 
and \ref{eq:hijEQM}) respectively. The equation of motion for $q(t)$ is 
easy to solve with a generic solution given by
\beq
\label{eq:qsol_gen}
\bar{q}(t) = \frac{\Lam N_c^2}{3} t^2 + c_1 t + c_2 
= \frac{\Lam N_c^2}{3} (t-r_1)(t-r_2)\, ,
\eeq
where $c_{1,2}$ are constants to be fixed based on the 
requirement of boundary conditions, while $r_{1,2}$ are the 
roots of the quadratic. The equation of motion 
for $h_{ij}(t, {\bf x})$ is a bit more complicated to solve as it 
involves both time and space derivatives, while $\bar{q}(t)$
is given by the above solution. In order to solve it we first do
mode decomposition of the fluctuation field $h_{ij}(t, {\bf x})$. 
For the gauge-fixing choice we have employed 
here, the tensor perturbations are transverse and traceless (TT).
Respecting the background symmetry the 
field TT $h_{ij}(t, {\bf x})$ can be written in terms of 
harmonics \cite{Gerlach:1978gy} as
\beq
\label{eq:h_ij_Sph_Har}
h_{ij}(t, {\bf x}) 
%\sum_{l=2}^\infty h_l(t) G^{(l)}_{ij} ({\bf x})
=\sum_{l=2}^\infty 
\sum_{n=2}^{l} \sum_{m=-n}^{n}
h^{l}_{nm}(t) (G_{ij})^l_{nm} ({\bf x}) \, ,
\eeq
where $h^l_{nm}(t)$ can be thought of as Fourier coefficients
at any given time $t$. The harmonics 
$G^{l}_{ij} ({\bf x})$ are the eigenfunctions satisfying 
\beq
\label{eq:eigenG}
\bar{D}^2 G^{l}_{ij} ({\bf x}) = - [l(l+2) -2] G^{l}_{ij} ({\bf x}) \, , 
\hspace{5mm} l \geq 2 \, .
\eeq
As the Lagrangian of the fluctuation field $h_{ij}$ mentioned 
in Eq. (\ref{eq:EHact_exp}) is seen to respect isotropy, 
it is therefore independent of $n$ and $m$ when the mode decomposition 
is plugged. In this situation we can suppress the 
$n$ and $m$ indices and write $h^{l}_{nm}(t) \equiv h_l(t)$.
Owing to $S^3$ symmetry, there exists degeneracy in the system:
for each $l$ there are 
\beq
\label{eq:degen_h}
g_l =2 (l+3)(l-1) \, ,
\eeq
modes having the same action. The factor $2$ appearing above is due to the contributions from the two propagating polarizations of the $h_{ij}$ field in the Landau gauge choice. 

This mode decomposition when utilized in the 
equation of motion for $\bar{h}_{ij}$ mentioned in 
Eq. (\ref{eq:hijEQM}) it is seen that each component $\bar{h}_l(t)$
follows the following equation: 
\beq
\label{eq:htL_eqm}
\ddot{\bar h}_l + \frac{2 \dot{\bar{q}}}{\bar{q}} \dot{\bar h}_l
+ \frac{N_c^2}{\bar{q}^2}
\biggl[
l(l+2) -2 + 
2k + \left(\frac{\dot{\bar{q}}^2}{2N_c^2}+\frac{6k - 2 \Lam \bar{q}}{3} \right)
\biggr]
\bar{h}_l = 0 \, ,
\eeq
which gets simplified for $k=1$. This equation can be solved 
by using special functions. The most general solution to it is given by
\beq
\label{eq:hl_sol_gen}
\bar{h}_l(t) = \frac{1}{\sqrt{\bar{q}}} 
\bigl[
d^{(l)}_{p} \mathbb{P}_1^{\xi_l} (\tau_t) + d^{(l)}_{q} \mathbb{Q}_1^{\xi_l} (\tau_t) 
\bigr] \, ,
\eeq
where $\mathbb{P}_1^{\xi_l}$ and $\mathbb{Q}_1^{\xi_l}$ are Legendre-P
and Legendre-Q functions, respectively, while
\beq
\label{eq:albt_form}
\xi_l = i \sqrt{1
+ \frac{
36\{l(l+2) + 4k -2\}}{N_c^2 \Lam^2 (r_1 - r_2)^2}} \, ,
\hspace{5mm}
\tau_t = \frac{2t - (r_2+r_1)}{r_1 - r_2} \, .
\eeq
The constants $d^{(l)}_{p}$ and $d^{(l)}_q$ are fixed based on the choice of
boundary conditions for the fluctuation field $h_{ij}$. 
For later use and ease, we define the following function: 
\beq
\label{eq:W_l10_func}
W_l(t,0) = \mathbb{P}_1^{\xi_l}[\tau_t] \mathbb{Q}_1^{\xi_l}[\tau_0] 
- \mathbb{P}_1^{\xi_l}[\tau_0] \mathbb{Q}_1^{\xi_l}[\tau_t] \, .
\eeq
For completeness we also mention the on shell action obtained after 
the on shell solution for $\bar{q}(t)$ and $\bar{h}_l(t)$ is plugged into the 
gravitational action mentioned in Eq. (\ref{eq:EHact_exp}). 
This is given by
\bea
\label{eq:EHact_exp_onshell}
&&
S_{\rm grav}^{\rm on-shell} 
= S_{\rm grav}^{(\bar{q})} + S_{\rm grav}^{(\bar{h})} = 
\frac{V_3}{16\pi G} \Bl[
\frac{2 \Lam^2 N_c^3}{9} + (6k + \Lam c_1) N_c + \frac{3c_1^2}{2N_c}
\Br]
\notag \\
&&
+ \frac{1}{16\pi G}\int_{\pt {\cal M}} {\rm d} {\bf x} \sqrt{\rho}
\Bl(
-\frac{3{\bar q}^2}{4N_c} \bar{h}_{ij} \dot{\bar h}^{ij}
- \frac{\bar{q} \dot{\bar q}}{4N_c} \bar{h}_{ij} \bar{h}^{ij}
\Br) \biggr\rvert_0^1
+ S_{\rm bd} \, ,
\eea
where the on shell $h_{ij}$ is obtained by utilizing the 
mode decomposition stated in Eq. (\ref{eq:h_ij_Sph_Har}) and making 
use of the on shell solution for each mode 
given in Eq. (\ref{eq:hl_sol_gen}). It is interesting to note 
that the on shell action of the fluctuation is purely a surface term,
which is expected once the $t$ integration is performed
in the original action after the on shell solution is plugged. 
The total on shell action also contains contributions from the 
$S_{\rm bd}$, which gets determined once the choice of the 
boundary condition for the field $q(t)$ is stated. This not only  
restricts the available choices of boundary condition for 
$h_{ij}$ but also restricts the possibilities for $S_{\rm bd}$. 
In the next section, we will determine the boundary conditions for 
$h_{ij}$ and $S_{\rm bd}$ in some special scenarios.

%%%%%%%%%%%%%%%%%%%%%%%%%%%%%%%%%%%%%%%%%%%%%%%
\section{Boundary choices and boundary action}
\label{bound_act} 
%%%%%%%%%%%%%%%%%%%%%%%%%%%%%%%%%%%%%%%%%%%%%%%

In this section, we explore the consequences of the restricted choices available 
for the boundary condition for the fluctuation $h_{ij}$ given the 
fixed boundary choice for the background. We first note that the conjugate 
momenta corresponding to field $q(t)$ is given by $\pi_q = -3\dot{q}/(2 N_c)$.
This implies that $K_b=-\pi_q/\sqrt{q}$.
For variational problem consistency as the 
boundary term appearing in Eq. (\ref{eq:Sexp_qvar}) is supposed to 
vanish, this implies
\beq
\label{eq:BDpart_qexp}
\biggl[
\biggl(
3 \sqrt{\bar{q}} \bar{F} 
- \frac{2\bar{\cal R}}{\sqrt{\bar{q}}} \pt_{\cal R} \bar{F} 
+ \bar{\pi}_q \pt_K \bar{F} 
\biggr) \de q
- 2 \bar{q} \biggl(
1+ \pt_K \bar{F}
\biggr) \de \pi_q 
\biggr] \biggr \rvert_0^1= 0 \, , 
\eeq
where as before $\bar{F} = F(\bar{\cal R}_b, \bar{K}_b)$.

%%%%%%%%%%%%%%%%%%%%%%%%%%%%%%%%%%%%%%%%%%%%%%%
\subsection{Dirichelet boundary condition}
\label{DBC_bg} 
%%%%%%%%%%%%%%%%%%%%%%%%%%%%%%%%%%%%%%%%%%%%%%%

Imposing Dirichlet boundary condition at the two end points 
implies that we fix $q$ at the end points. This means $\de q_{1,0}=0$. 
If the momentum is arbitrary then Eq. (\ref{eq:BDpart_qexp}) immediately leads to 
\beq
\label{eq:dbc_qbd_F}
1 + \pt_K \bar{F} = 0
\hspace{5mm}
\Rightarrow 
\hspace{5mm}
F = - K_b  + f({\cal R}_b)\, ,
\eeq
where $f({\cal R}_b)$ is an arbitrary function of ${\cal R}_b$
which for simplicity can be fixed to zero.
For the Dirichlet boundary choice, the parameters 
$c_{1,2}$ appearing in the on shell solution mentioned in Eq. (\ref{eq:qsol_gen}),
and the on shell action including the boundary contribution 
are given by
\bea
\label{eq:onsh_dbc_qbar}
&&
c_1 = q_f - q_i - \Lam N_c^2/3 \, , 
\hspace{5mm} 
c_2 = q_i \, ,
\notag \\
&&
S_{\rm grav}^{(\bar{q})} \br \rvert_{\rm DBC}
= \frac{V_3}{16\pi G}
\Bl[\frac{\Lam^2}{18}N_{c}^3 
+ \{6k - \Lam(q_f + q_i)\} N_c - \frac{3 (q_f - q_i)^2}{2 N_c}\Br] \, .
\eea
Once the functional form of $F$ is known, one can use it to 
compute the boundary condition expected to be satisfied 
by the fluctuation field $h_{ij}$ for the consistent variational problem. 
This is achieved by exploiting Eqs. (\ref{eq:ABexp}) and (\ref{eq:BD_h_gen}). 
For the DBC case this gives
\beq
\label{eq:DBC_hcase}
\frac{1}{2N_c} \bigl(
\bar{q}^2 \dot{\bar h}_{ij} + 2 \bar{q} \dot{\bar q} \bar{h}_{ij}
\bigr) \de h^{ij} \biggr \rvert_0^1 = 0 \, .
\eeq
This means that either $h_{ij}$ is also fixed at the two boundaries
which implies $\de h_{ij} \rvert_0^1 =0$, or we expect the fluctuation 
field to satisfy the condition 
\beq
\label{eq:robinH_dbc_case}
 \bigl(
\bar{q}^2 \dot{h}_{ij} + 2 \bar{q} \dot{\bar q} h_{ij}
\bigr) \rvert_0^1= 0 \, .
\eeq
This is a Robin boundary condition for the fluctuation.
This simple example shows that imposing DBC for the 
background scale factor $q(t)$ restricts our available 
choices of boundary conditions for the fluctuation field $h_{ij}$.
Only two boundary conditions for the fluctuation are allowed 
if we require the whole gravitational system to be
generally covariant. In this paper, we will focus on the 
simplest of the two boundary choices: $\de h_{ij} \rvert_0^1 =0$
($h_{ij}$ is fixed at the two end points). The on shell solution for $h_l(t)$ is mentioned 
in Eq. (\ref{eq:hl_sol_gen}). For the boundary 
condition $\de h_{ij} \rvert_0^1 =0$, this will 
mean that $d^{(l)}_{p}$ and $d^{(l)}_{q}$ are given by
\beq
\label{eq:hl_dpdq_DBC}
d^{(l)}_{p} = \frac{h_1^{(l)}\sqrt{q_f} \mathbb{Q}_1^{\xi_l}[\tau_0] }
{W_l(1,0)}\, , 
\hspace{5mm}
d^{(l)}_{q} = - \frac{h_1^{(l)}\sqrt{q_f} \mathbb{P}_1^{\xi_l}[\tau_0] }{W_l(1,0)} \, ,
\eeq
where $h_1^{(l)}$ is the value of the fluctuation field at the final hypersurface at $t=1$, i.e., $h_l(t=1)$.
The total on shell action including 
the boundary action is given by
\bea
\label{eq:hh_onSH_DBC}
&&
S^{(l)}_{\rm grav} [\bar{q}, \bar{h}_l, N_c] \Br \rvert_{\rm DBC} 
=\frac{1}{16\pi G}\Bl[
\frac{\bar{q}^2}{4N_c}\bar{h}_l\dot{\bar{h}}_l
+\frac{\bar{q}\dot{\bar{q}}}{2N_c} \bar{h}_l^2
\Br]\Br \rvert_{0}^{1} \, ,
\notag \\
&&
= \frac{q_f \bl(h^{(l)}_1\br)^2}{(16\pi G) (8N_c)}
\Bl[
3(q_f-q_i) - \Lam N_c^2 +2q_f \bl\{ \ln W(t,0) \br\}^\prime \br\rvert_{t=1}
\Br] \, ,
\eea
where $({}^\prime)$ denotes the derivative with respect to $t$
and there is no contribution coming from the boundary condition at $t=0$.

%%%%%%%%%%%%%%%%%%%%%%%%%%%%%%%%%%%%%%%%%%%%%%%
\subsection{Neumann boundary condition}
\label{NBC_bg} 
%%%%%%%%%%%%%%%%%%%%%%%%%%%%%%%%%%%%%%%%%%%%%%%

If the Neumann boundary condition is imposed at one of the 
end points then it means that the conjugate momenta 
corresponding to $q(t)$ is fixed at that end point: 
$\de \pi_q =0$. In this case if $q$ is arbitrary at that end point 
then using Eq. (\ref{eq:BDpart_qexp}) we have 
\beq
\label{eq:nbc_bg_q}
\biggl(
3 \sqrt{\bar{q}} \bar{F} 
- \frac{2\bar{\cal R}}{\sqrt{\bar{q}}} \pt_{\cal R} \bar{F} 
+ \bar{\pi}_q \pt_K \bar{F} 
\biggr) = 0 
\hspace{5mm}
\Rightarrow
\hspace{5mm}
2 {\cal R} \pt_{\cal R} F + K \pt_K F = 3F \, . 
\eeq
To solve this we can make an ansatz that $F({\cal R}, K)
\sim {\cal R}^n K^{2m}$. This gives $2n + 2m =3$ leading to an infinite 
set of choices for $F({\cal R}, K)$:
\beq
\label{eq:Fchoice_nbc}
F({\cal R}, K) = c {\cal R}^{3/2-\om} K^{2\om} \, ,
\eeq 
where $c$ is a constant. For each value of $\om$ there is a choice of $F({\cal R}, K)$
giving an infinite series of available choices. All of these different $F$ will lead to the
same boundary condition for the background $q(t)$ (Neumann in this case) 
but will lead to different boundary conditions for the fluctuation field. 
This is an infinite set of covariant boundary action choices available, 
\beq
\label{eq:Sbd_nbc_cov}
S^{\rm NBC}_{\rm bd}
= 
\frac{c}{8\pi G} \int_{\pt {\cal M}} {\rm d} {\bf x}
\sqrt{\g} \,\, {\cal R}^{3/2-\om} K^{2\om} 
\hspace{5mm}
\forall \om \in \mathbb{R} \, ,
\eeq
with the simplest available choice being $c=0$ leading to $F({\cal R}, K) = 0$. 
The choice $c=0$ implies that there is 
no boundary term added. While this boundary choice of $c=0$
is not new to literature, the boundary action for $c\neq0$ and 
various values of $\om$ are new.   

In this paper we will work with this simplest 
choice $F = 0$, and will leave the other possible choices for future exploration. 
For this choice of $F=0$, the boundary condition that is expected to 
be satisfied by the fluctuation field $h_{ij}$ can be obtained
using Eqs. (\ref{eq:ABexp}) and (\ref{eq:BD_h_gen}). This gives
\beq
\label{eq:hij_nbc_F0}
\bigl(\bar{q} \dot{\bar{q}}\bar{h}_{ij}
+ \bar{q}^2 \dot{\bar h}_{ij}\bigr) \de h^{ij}
+2\bar{q}^2 \bar{h}_{ij}\de \dot{h}^{ij} = 0 \, .
\eeq
This gives rise to two possible scenarios: (1) having 
$h_{ij}=0$ which will satisfy the above equation identically 
and (2) requiring 
\beq
\label{eq:hl_nbc_2ndcond}
\de \bigl[
\bar{q} \dot{\bar q} h_l^{3/2} + 3 \bar{q}^2 \dot{h}_l \sqrt{h_l}
\bigr] = 0 
\hspace{5mm} \Rightarrow \hspace{5mm}
\dot{\bar q} h_l^{3/2} + 3 \bar{q} \dot{h}_l \sqrt{h_l} 
= \al 
\, ,
\eeq
where in obtaining the above we have used the harmonic decomposition
of $h_{ij}$ mentioned in Eq. (\ref{eq:h_ij_Sph_Har}). In this paper 
we will focus on the trivial boundary condition for the fluctuation field:
imposing $h_{ij}=0$ at t=0. This means in the case when 
NBC is imposed on the scale factor 
$q(t)$ at initial time and DBC at the final time, then out of the 
four possible choices for the boundary conditions for the fluctuation field, 
we will consider the simplest possibility:
$h_{ij}(t=0, {\bf x})=0$ and $h_{ij}(t=1, {\bf x}) \neq 0$.

The on shell solution for scale factor $\bar{q}(t)$ in the Neumann BC at the initial 
boundary and Dirichlet BC at the final boundary is given by
\beq
\label{eq:qsol_nbc}
\bar{q}(t) = \frac{\Lam N_c^2}{3} (t^2-1) - \frac{2 N_c \pi_i}{3} (t-1) + q_f \, ,
\eeq
where $\pi_i$ is the initial conjugate momenta corresponding to 
$q(t)$ and $q_f$ is the final scale factor. The on shell action for the 
same $S_{\rm grav}^{(\bar{q})}$ is given by
\beq
\label{eq:stot_onsh_nbc}
S_{\rm grav}^{(\bar{q})}[N_c] = \frac{V_3}{8\pi G}
\Bl[\frac{\Lam^2}{9} N_c^3 
- \frac{\Lam \pi_i}{3} N_c^2 
+ \left(3k - \Lam q_f + \frac{\pi_i^2}{3} \right) N_c
+ q_f \pi_i  \Br]\, .
\eeq
The on shell solution for $h_l(t)$ is mentioned 
in Eq. (\ref{eq:hl_sol_gen}), with $d^{(l)}_{p}$ and $d^{(l)}_{q}$ given by
\bea
\label{eq:hl_dpdq_NBC}
d^{(l)}_{p} =  \frac{ h^{(l)}_1 \sqrt{q_f} \mathbb{Q}_1^{\xi_l}[\tau_0] }{W_l(1,0)} \, ,
\hspace{5mm}
d^{(l)}_{q} = - \frac{ h^{(l)}_1 \sqrt{q_f} \mathbb{P}_1^{\xi_l}[\tau_0] }{W_l(1,0)} \, .
\eea
The corresponding on shell action including the contributions 
from the boundary term is given by
\bea
\label{eq:hh_onSH_NBC}
&&
S^{(l)}_{\rm grav} [\bar{q}, \bar{h}_l, N_c] \Br \rvert_{\rm NBC} 
= 
\frac{1}{16\pi G}
\Bl[ 
\Bl\{\frac{\bar{q}^2\bar{h}_l\dot{\bar{h}}_l}{4N_c}
+\frac{\bar{q}\dot{\bar{q}}\bar{h}_l^2}{2N_c}\Br\} 
\Br\rvert_1
+\Bl\{\frac{3\bar{q}^2\bar{h}_l\dot{\bar{h}}_l}{4N_c}
+\frac{\bar{q}\dot{\bar{q}}\bar{h}_l^2}{4N_c}\Br\}
\Br\rvert_0 \Br] \, ,
\notag \\
&&
= \frac{q_f \bl(h^{(l)}_1\br)^2}{(16\pi G) (8N_c)}
\Bl[2\Lam N_c^2 - 2N_c\pi_i +2q_f \bl\{ \ln W(t,0) \br\}^\prime \br\rvert_{t=1}
\Br] \, .
\eea
%

%%%%%%%%%%%%%%%%%%%%%%%%%%%%%%%%%%%%%%%%%%%%%%%
\subsection{Robin boundary condition}
\label{RBC_bg} 
%%%%%%%%%%%%%%%%%%%%%%%%%%%%%%%%%%%%%%%%%%%%%%%

We next consider the case of imposing the Robin boundary condition 
for the background $q(t)$ at one of the end points. 
In this case, we fix a 
linear combination of the scale factor $q$ and the corresponding
conjugate momentum $\pi_q$ at the end point. 
This means 
\beq
\label{eq:robin_cond}
\pi_q + \bt q = P_i= {\rm fixed} 
\hspace{5mm} \Rightarrow \hspace{5mm}
\de \pi_q + \bt \de q = 0 \, .
\eeq
Combining Eq. (\ref{eq:BDpart_qexp}) with the expression for 
$K$ in terms of $\pi_q$ mentioned previously and 
utilizing the requirement of the boundary condition 
given in Eq. (\ref{eq:robin_cond}) we have
\beq
\label{eq:rbc_bg_q}
\biggl(2\bt \sqrt{\frac{6k}{\cal R}} - K \biggr) \pt_K F
- 2 {\cal R} \pt_{\cal R} F + 3 F = - 2\bt \sqrt{\frac{6k}{\cal R}} \, .
\eeq
This is a nonhomogeneous partial differential equation (PDE) which will have a homogeneous 
solution $F_H ({\cal R}, K)$ and a particular solution 
$F_P({\cal R}, K)$. The homogeneous solution can be found by putting 
the rhs of the PDE to zero implying 
\beq
\label{eq:rbc_bg_q_homo}
\biggl(2\bt \sqrt{\frac{6k}{\cal R}} - K \biggr) \pt_K F_H
- 2 {\cal R} \pt_{\cal R} F_H + 3 F_H = 0 \, .
\eeq
To solve this we make an ansatz: 
\beq
\label{eq:rbcF_ansatz}
F_H \sim \biggl(2\bt \sqrt{\frac{6k}{\cal R}} - K \biggr)^{2n} {\cal R}^m \, .
\eeq
This gives $2n+2m =3$ as the condition for the ansatz to be the solution 
of the homogeneous equation. The particular solution can be 
found by making the ansatz $F_P \propto {\cal R}^{-1/2}$. This 
fixes the proportionality constant giving the full solution for $F$ as
\beq
\label{eq:Fchoice_rbc}
F({\cal R}, K) = c  \biggl(2\bt \sqrt{\frac{6k}{\cal R}} - K \biggr)^{2\om}
{\cal R}^{3/2-\om} - \bt \sqrt{\frac{3k}{2 {\cal R}}} \, ,
\eeq 
where $c$ is an arbitrary constant. 
For each value of $\om$ there is a corresponding boundary function 
$F({\cal R}, K)$, giving an infinite set of boundary functions leading to 
the same Robin boundary condition for the scale factor $q(t)$. 
These will however lead to different boundary choices for the 
fluctuation field $h_{ij}$. The infinite set of covariant 
boundary actions in the case of the Robin boundary condition is given by
\beq
\label{eq:Sbd_rbc_cov}
S^{\rm RBC}_{\rm bd}
= 
\frac{1}{8\pi G} \int_{\pt {\cal M}} {\rm d} {\bf x} \sqrt{\g} 
\biggl[
c  \biggl(2\bt \sqrt{\frac{6k}{\cal R}} - K \biggr)^{2\om}
{\cal R}^{3/2-\om} - \bt \sqrt{\frac{3k}{2 {\cal R}}}
\biggr]
\hspace{5mm}
\forall \om \in \mathbb{R} \, ,
\eeq
with the simplest available choice being $c=0$ implying that only 
the particular solution survives. This boundary term reduces to the 
known boundary action that has been studied in the past in the context of 
Einstein-Hilbert gravity with Robin boundary condition in mini-superspace
\cite{Ailiga:2023wzl, Ailiga:2024mmt}. Moreover, in the limit
$\bt\to0$ this boundary action goes to the boundary 
action in the case of Neumann boundary condition. 
The boundary action with $c\neq0$ for various values of $\om$ is new, 
and has not been reported earlier to the best of our knowledge. 

In this paper we will work with the simplest choice ($c=0$)
and will consider the other nontrivial choices in future studies. For the choice of 
$F$, the boundary condition satisfied by the fluctuation field $h_{ij}$
is given by
\beq
\label{eq:rbc_hij_c0}
\biggl[
-\biggl(\frac{\bar{q} \dot{\bar q}}{2N_c} - \frac{7 \bt \bar{q}^2}{12} \biggr) \bar{h}_{ij}
+ \frac{\bt \bar{q}^2}{24} \bar{D}^2 \bar{h}_{ij}
- \frac{\bar{q}^2}{2N_c} \dot{\bar h}_{ij}
\biggr] \de h^{ij}
- \frac{\bar{q}^2 \bar{h}_{ij}}{N_c}\de \dot{h}^{ij} = 0 \, .
\eeq
From this it is seen that the boundary condition $h_{ij}=0$ 
trivially satisfies the above equation as $\de h_{ij} =0$. Besides this 
there are other sets of possible boundary choices for $h_{ij}$,
one corresponding to each value of $\bt$. After plugging the 
harmonic decomposition of $h_{ij}$ mentioned in Eq. (\ref{eq:h_ij_Sph_Har})
in the above equation we get the following condition for each mode: 
\beq
\label{eq:hl_rbc_2ndcond}
\biggl(\dot{\bar q} + \frac{\bt l(l+2) N_c \bar{q}}{12}
- \frac{4 \bt N_c \bar{q}}{3} \biggr) h_l^{3/2}
+ 3 \bar{q} \dot{h}_l \sqrt{h_l} 
= \al \, .
\eeq
The on shell solution 
for scale factor $\bar{q}(t)$ with the Robin BC at the initial 
boundary and Dirichlet condition at the final boundary is given by
\beq
\label{eq:qsol_RBC}
\bar{q}(t) = \frac{\Lam N_c^2}{3} t^2 
+ \frac{P_i}{\bt}
+ \left(1 + \frac{3}{2 \bt N_c} \right)^{-1}
 \left(t + \frac{3}{2 \bt N_c} \right)
 \left(q_f - \frac{P_i}{\bt} - \frac{\Lam N_c^2}{3} \right) \, ,
\eeq
The on shell action for the 
same $S_{\rm grav}^{(\bar{q})}$ is given by
\bea
\label{eq:stot_onsh_rbc}
S_{\rm grav}^{\rm (\bar{q})}[N_c] &=& 
\frac{V_3}{16\pi G}\biggl\{\frac{1}{(27 + 18 N_c \bt)} \biggl[
\bt \Lam^2 N_c^4 + 6 \Lam^2 N_c^3 +
N_c^2 \{108 \bt  k-18 \Lam  (P_i+\bt q_f)\}
\notag \\
&&
+18 N_c \left\{9
k+P_i^2 -  3 q_f \Lambda \right\}
+ 54 P_i q_f - 27 \beta q_f^2
\biggr\}\biggr] \, .
\eea
The on shell solution for $h_l(t)$ is mentioned 
in Eq. (\ref{eq:hl_sol_gen}), with $d^{(l)}_{p}$ and $d^{(l)}_{q}$ given by
\bea
\label{eq:hl_dpdq_RBC}
d^{(l)}_{p} =  \frac{ h^{(l)}_1 \sqrt{q_f} \mathbb{Q}_1^{\xi_l}[\tau_0] }{W_l(1,0)} \, ,
\hspace{5mm}
d^{(l)}_{q} = - \frac{ h^{(l)}_1 \sqrt{q_f} \mathbb{P}_1^{\xi_l}[\tau_0] }{W_l(1,0)} \, ,
\eea
Although this is structurally same as in the NBC case, as
$\xi_l$ depends on $c_1$ and $c_2$ which in turn depend 
on the choice of boundary condition, therefore the 
constants $d^{(l)}_{p}$ and $d^{(l)}_{q}$ are different. 
The corresponding on shell action including the contributions 
from the boundary term is given by
\bea
\label{eq:hh_onSH_RBC}
&&
S^{(l)}_{\rm grav} [\bar{q}, \bar{h}_l, N_c] \Br \rvert_{\rm RBC} 
= \frac{1}{16\pi G} 
\Bl[
\Bl(\frac{\bar{q}^2}{4 N_c} \bar{h}_{l} \dot{\bar{h}}_{l} 
+ \frac{\bar{q} \dot{\bar{q}}}{2 N_c} \bar{h}^2_{l} \Br) \Br \rvert_{1} 
\notag \\
&& 
+ \Bl\{\frac{3\bar{q}^2}{4 N_c} \bar{h}_{l} \dot{\bar{h}}_{l} 
+ \frac{\bar{q} \dot{\bar q}}{4 N_c} \bar{h}^2_{l} 
- \bt \bar{q}^2 \bar{h}^2_{l}\Bl( \frac{7}{24}
- \frac{l(l+2)-2}{48 k}  
\Br)\Br\}\Br\rvert_{0} \Br] 
\notag \\
= && 
\frac{q_f (h_{1}^{(l)})^2}{(16\pi G)(8 N_c)}
\Bl[\frac{6 \Lam N_c^2 - 6 N_c P_i 
+ \beta (6 q_f N_c + 2 \Lam N_c^3 )}{3 + 2 N_c \beta} 
+ 2 q_f \bl\{\ln{W}_l(t,0)\br\}'\big|_{t=1}\Br] \, .
\eea
This will be later utilized while computing the quantum-corrected 
action for the lapse $N_c$.

%%%%%%%%%%%%%%%%%%%%%%%%%%%%%%%%%%%%%%%%%%%%%%%
\section{Path integral over $q(t)$ and $h_{ij}$}
\label{trans_amp} 
%%%%%%%%%%%%%%%%%%%%%%%%%%%%%%%%%%%%%%%%%%%%%%%

After having setup the consistent variational problem and computed the 
relevant boundary action for the various boundary condition choices,
we direct our attention to the gravitational path integral 
mentioned in Eq. (\ref{eq:Gform_sch}). For the metric ansatz stated in 
Eq. (\ref{eq:frwmet_changed}) this path integral acquires the following form:
\beq
\label{eq:grav_path_qform}
G[{\rm Bd}_f, {\rm Bd}_i]
= \int_{-\infty}^{\infty} {\rm d} N_c 
\int_{{\rm Bd}_i}^{ {\rm Bd}_f} {\cal D} q(t) {\cal D} h_{ij}(t, {\bf x})
\exp \left(\frac{i}{\hbar} S_{\rm grav}[q, h_{ij}, N_c] \right) \, ,
\eeq
where ${\rm Bd}_i$ and  ${\rm Bd}_f$ are the initial and final 
boundary configurations respectively; the action $S_{\rm grav}[q, h_{ij}, N_c]$ is mentioned in 
Eq. (\ref{eq:EHact_exp}). The above path integral has been suitably 
gauge fixed to prevent overcounting 
of the gauge orbits, and Faddeev-Popov ghost has been accordingly introduced
as discussed in Sec. \ref{gfgh}. 
For the purpose of this work and in following we will not consider 
ghost contributions as they will not be relevant for the issues to be 
addressed in this paper. 

The gravitational action mentioned in Eq. (\ref{eq:EHact_exp}) is quadratic in 
the fields $q(t)$ and $h_{ij}$, with interaction terms between them giving rise to non-linear effects.
In this paper we will work in the one-loop approximation 
and compute the path integral at one loop. To proceed in this direction 
we make use of background field formalism \cite{Abbott:1980hw} and write 
the fields as 
\beq
\label{eq:qh_exp_bg}
q(t) = \bar{q}(t) + \ep_q Q(t) \, ,
\hspace{5mm}
h_{ij} = \bar{h}_{ij} + \ep_h H_{ij} \, ,
\eeq
where $\ep_q$ and $\ep_h$ are perturbation parameters to keep track 
of the order of terms in the expansion. On plugging Eq. (\ref{eq:qh_exp_bg}) in the gravitational action mentioned in Eq. (\ref{eq:EHact_exp}) and expanded up to order in $\mathcal{O}(\ep^2_{q}), \mathcal{O}(\ep_{q}\ep_{h})$, and $\mathcal{O}(\ep^2_{h})$, we get the following for the terms which are second order:
\bea
\label{eq:Sgrav_exp_QH}
&&
S_{\rm grav} = 
S_{\rm grav}[N_c, \bar{q}, \bar{h}_{ij}]
+ \int_{\cal M} \frac{{\rm d} t {\rm d} {\bf x} \sqrt{\rho}}{16\pi G}
\biggl[
\ep_q N_c \biggl\{
-2 \Lam Q - \frac{3 \dot{\bar q} \dot{Q}}{N_c^2}
+ \frac{\Lam Q}{2}\bar{h}_{ij} \bar{h}^{ij}
+ \frac{Q \bar{q} \dot{\bar h}_{ij} \dot{\bar h}^{ij}}{2N_c^2}
\notag \\
&&
- \frac{(\dot{\bar q} \dot{Q} + 2 \bar{q} \ddot{Q} 
+ 2 Q \ddot{\bar q}) {\bar h}_{ij} \bar{h}^{ij}}{4N_c^2}
\biggr\} 
+ \ep_h
\biggl\{
(\Lam {\bar q} -2 k) N_c {\bar h}_{ij} H^{ij}
+ \frac{2 \bar{q}^2 \dot{\bar h}_{ij} \dot{H}^{ij}
-(\dot{q}^2 + 4{\bar q} \ddot{\bar q}){\bar h}_{ij} H^{ij}}{4N_c}
\notag \\
&&
+ \frac{N_cH_{ij}}{2}  {\bar D}^2 \bar{h}_{ij}
\biggr\} 
+ \ep_q^2 \biggl\{
-\frac{3\dot{Q}^2}{2N_c}
+ \frac{Q^2 \dot{\bar h}_{ij} \dot{\bar h}^{ij}}{4N_c}
-\frac{(\dot{Q}^2 + 4 Q \ddot{Q} ) \bar{h}_{ij}\bar{h}^{ij}}{8N_c} 
\biggl\}
\notag \\
&&
+ \ep_h \ep_q N_c
\biggl\{
\Lam Q {\bar h}_{ij} H^{ij}
+ \frac{\bar{q} Q \dot{\bar h}_{ij} \dot{H}^{ij}}{N_c^2}
- \frac{(\dot{\bar q} \dot{Q} + 2 \bar{q} \ddot{Q} + 2 \ddot{\bar q} Q)\bar{h}_{ij} H^{ij}}{2N_c^2}
\biggr\}
\notag \\
&&
+\ep_h^2 N_c
\biggl\{
\frac{(\Lam {\bar q} -2 k)}{2}H_{ij} H^{ij}
+ \frac{\bar{q}^2 \dot{H}_{ij} \dot{H}^{ij}}{4N_c^2} 
- \frac{(\dot{\bar q}^2 + 4 \bar{q} \ddot{\bar q} )H_{ij}H^{ij}}{8N_c^2}
+ \frac{H_{ij}{\bar D}^2 H^{ij}}{4}  
\biggr\}
\biggr] + \cdots \, ,
\eea
where $S_{\rm grav}[N_c, \bar{q}, \bar{h}_{ij}]$ refers to on shell action of the field $q$ and $h_{ij}$. It should be mentioned that the above is the full expansion of the action up to second order in the field perturbations. Terms linear in $\ep_q$ and $\ep_h$ will lead to the equation of motion for the background $\bar{q}$ and $\bar{h}_{ij}$. It will be noticed that the equation of motion for $\bar{q}$ will also contain terms dependent on $\bar{h}_{ij}$ beside those already mentioned in Eq. (\ref{eq:dyn_q_eq}). These are backreaction terms where the behavior of $\bar{h}_{ij}$ starts to affect the dynamical evolution of $\bar{q}(t)$. This is an artifact of the nonlinear nature of gravity. In what follows we will assume throughout the paper that the behavior of $\bar{h}_{ij}$ will not source the evolution of $\bar{q}(t)$. These are in accordance with the requirements of ignoring backreaction effects at leading order. This also implies that under this approximation, the equation of motion for $\bar{q}(t)$ will remain same as stated in Eq. (\ref{eq:dyn_q_eq}). Moreover, in the spirit of respecting the ideology of background-field formalism we assume that the set of quantum fields $Q(t)$ and $H_{ij}(t, {\bf x})$ don't affect the set of background fields $\bar{q}(t)$ and $\bar{h}_{ij}(t, {\bf x})$. 

Under these set of assumptions it implies that in the second order terms, $\bar{q}(t)$ and $\bar{h}_{ij}(t, {\bf x})$ satisfies Eqs. (\ref{eq:dyn_q_eq}) and (\ref{eq:hijEQM}) respectively. For the various boundary choices considered for the scale factor, it is noticed that the boundary choice $h_{ij}=0$ is very well allowed by the requirements of covariance. In the special case when $h_{ij}$ vanishes at the two boundaries, the solution following from Eq. (\ref{eq:hijEQM}) gives $\bar{h}_{ij}(t, {\bf x})=0$. In this situation, the nondiagonal terms present in the second variation given in Eq. (\ref{eq:Sgrav_exp_QH}) vanish, while the diagonal entries get appropriately simplified. Under such requirements, the path integral simplifies quite a bit. Most importantly, the path integral over $q(t)$ and $h_{ij}$ gets separated, implying the path integral of each can be done independently. This will mean that 
Eq. (\ref{eq:grav_path_qform}) becomes the following:
\bea
\label{eq:grav_path_qhform_1}
G[{\rm Bd}_f, {\rm Bd}_i] \Br \rvert_{\bar{h}_{ij}(t, {\bf x})=0}
= &&
\int_{-\infty}^{\infty} {\rm d} N_c 
\int_{{\rm Bd}_i}^{ {\rm Bd}_f} {\cal D} q(t) 
\exp\left(\frac{i}{\hbar} S^{(q)}_{\rm grav}[q, N_c] \right)
\notag \\
&& \times
\int_{{\rm Bd}_i}^{ {\rm Bd}_f} 
{\cal D} H_{ij}(t, {\bf x})
\exp \left(\frac{i}{\hbar} S^{(h)}_{\rm grav}[\bar{q}, H_{ij}, N_c] \right) \, ,
\eea
where 
\bea
\label{eq:act_SH_begexp}
S^{(h)}_{\rm grav}[\bar{q}, H_{ij}, N_c]
=&& 
\int_{\cal M} \frac{{\rm d} t {\rm d} {\bf x} \sqrt{\rho}}{16\pi G}
N_c
\biggl[
\frac{(\Lam {\bar q} -2 k)}{2}H_{ij} H^{ij}
+ \frac{\bar{q}^2 \dot{H}_{ij} \dot{H}^{ij}}{4N_c^2} 
\notag \\
&&
- \frac{(\dot{\bar q}^2 + 4 \bar{q} \ddot{\bar q} )H_{ij}H^{ij}}{8N_c^2}
+ \frac{H_{ij}{\bar D}^2 H^{ij}}{4}  
\biggr] \, .
\eea
Although, the two path integrals can be done independently, they are however still coupled to each other via lapse integration. In this paper, we will be studying this up to one loop where we remember that in such studies the background perturbation $\bar{h}_{ij}(t, {\bf x})=0$, while the quantum field $H_{ij}(t, {\bf x})$ propagates virtually. In the case when $\bar{h}_{ij}(t, {\bf x})$ is nonvanishing, nondiagonal terms will contribute, which will be studied in future publication.

%%%%%%%%%%%%%%%%%%%%%%%%%%%%%%%%%%%%%%%%%%%%%%%
\subsection{Path integral over $q(t)$}
\label{SubS:PI_qt}
%%%%%%%%%%%%%%%%%%%%%%%%%%%%%%%%%%%%%%%%%%%%%%%

We first study the path integral over $q(t)$ which is mentioned in the first line of 
Eq. (\ref{eq:grav_path_qhform_1}) (without the lapse integration). 
Here, the action appearing above is the action for the scale factor 
$q(t)$ including the suitable boundary action depending on the 
choice of the boundary condition employed. This, however, is a 
one-dimensional path integral but still captures some bits of the 
nontrivialities of the gravitational systems. This path integral 
is exactly doable for the Einstein-Hilbert gravity with various 
boundary conditions and has been computed previously 
in \cite{Feldbrugge:2017kzv, Narain:2021bff, Narain:2022msz, Ailiga:2023wzl}
for Dirichlet, Neumann and Robin boundary conditions.
Here we do not repeat the details of the computation 
but rather given the end result of the path integral 
over the scale factor $q(t)$. These are given by
\bea
\label{eq:q_pathInt_BC}
\int_{{\rm Bd}_i}^{ {\rm Bd}_f} {\cal D} q(t) \,\, 
&&
\exp\bl[\frac{i}{\hbar} S_{\rm grav}^{(q)}\br]
= \bl\{\D_q(N_c) \br\}^{-1/2} \exp\Bl[\frac{i}{\hbar} S_{\rm grav}^{(\bar{q})}(N_c) \Br] 
= e^{i {\cal A}_{q}(N_c)/\hbar}\, ,
\eea
where 
\beq
\label{eq:DeltaN_Q}
\D_q^{\rm DBC}(N_c) = N_c \, ,
\hspace{3mm}
\D_q^{\rm NBC}(N_c) = 1 \, ,
\hspace{3mm}
\D_q^{\rm RBC}(N_c) = 1+ \frac{2N_c\bt}{3} \, ,
\eeq
and $S_{\rm grav}^{(\bar{q})}(N_c)$ for Dirichlet, Neumann and 
Robin boundary conditions 
is mentioned in Eqs. (\ref{eq:onsh_dbc_qbar}), (\ref{eq:stot_onsh_nbc}),
and (\ref{eq:stot_onsh_rbc}) respectively. The factor $\D_q(N_c)$ can 
be exponentiated to obtain the quantum corrected action for the lapse
$N_c$. This is given by
\beq
\label{eq:quant_Nc_act_q}
{\cal A}_{q}(N_c) = S_{\rm grav}^{(\bar{q})}(N_c)
+ \frac{i \hbar}{2} \ln \D_q(N_c) \, . 
\eeq
The quantum corrected action for the lapse leads to corrections 
of order ${\cal O}(\hbar)$ to the saddle-point geometries. In the next 
subsection, we will compute the quantum correction coming from the 
$h_{ij}$.

%

%%%%%%%%%%%%%%%%%%%%%%%%%%%%%%%%%%%%%%%%%%%%%%%
\subsection{Path integral over $H_{ij}$}
\label{SubS:PI_hij}
%%%%%%%%%%%%%%%%%%%%%%%%%%%%%%%%%%%%%%%%%%%%%%%

Now we focus our attention to the path integral of $H_{ij}$, mentioned in the 
second line of Eq. (\ref{eq:grav_path_qhform_1}), where the 
action $S^{(h)}_{\rm grav}[\bar{q}, H_{ij}, N_c]$ is mentioned 
in Eq. (\ref{eq:act_SH_begexp}). The surface terms disappear, as $H_{ij}$ vanishes at the endpoints.
This path integral is already gauge fixed with suitably taken Faddeev-Popov ghost having been suitably taken into account.
The complete ghost operator has been computed and is mentioned in Eq. (\ref{eq:ghost_det}). Note that the operator does not explicitly depend on the scale factor $q(t)$ and lapse $N_c$. In the background field formalism, the term proportional to $h_{ij}$ becomes $\bar{h}_{ij} + H_{ij}$, where $\bar{h}_{ij}$ satisfies Eq. (\ref{eq:hijEQM}). For the specific choice of boundary condition where metric fluctuations vanish at the boundaries, we have $\bar{h}_{ij}(t, {\bf x})=0$. It implies that the ghost operator will contain terms dependent and independent of $H_{ij}$. In the one-loop study, the terms proportional to $H_{ij}$ are irrelevant. They however contribute at two loops or higher. At the one-loop approximation, the ghost operator, although it does not depend on couplings or additional fields, still contributes to the path integral, though by an overall numerical scaling (the ghost-determinant when computed will consist of a divergent and a finite part, both being pure numbers). Since it does not affect the qualitative features of the path integral or alter the conclusions, it can be safely ignored.

To evaluate the $H_{ij}$ path integral we make use of the mode decomposition 
of the field mentioned in Eq. (\ref{eq:h_ij_Sph_Har}). This means 
that the path integral measure will become
\beq
\label{eq:PI_hij_mea}
{\cal D} H_{ij}  \hspace{3mm} \Rightarrow \hspace{3mm}
\prod_{l=2}^\infty \,\, \prod_{n=2}^l \,\, \prod_{m=-n}^n \,\, 
{\cal D} H^l_{nm}(t)
\eeq
while the fluctuation action will become a sum over action for 
various modes. As for each $l$ there are $g_l$ degenerate modes 
having the same action, so the path integral over all these modes 
will contribute in same manner. 
The action for each mode including the contribution 
coming from the boundary is given by
\beq
\label{eq:mode_hL_act}
S^{(h)}_{\rm grav}
= \frac{1}{16\pi G} \sum_{l=2}^\infty \,\, \sum_{n=2}^l \,\, \sum_{m=-n}^n
\bl(S^{(l)}_{\rm grav} \br)_{nm} \, ,
\eeq
where $\bl(S^{(l)}_{\rm grav} \br)_{nm}$ is the action 
corresponding to each mode and is given by
\bea
\label{eq:SL_hL_act}
\bl(S^{(l)}_{\rm grav} \br)_{nm} &&=  \int_0^1 {\rm d}t
\Bl[
\frac{\bar{q}^2 }{4N_c}\bl(\dot{H}^l_{mn}\br)^2
- N_c\Bl\{
\frac{l(l+2) - 2}{4}  
+ \frac{\dot{\bar q}^2 + 4 \bar{q}\ddot{\bar{q}}}{8N_c^2} 
- \frac{ \Lam \bar{q} - 2k}{2}
\Br\} \bl(H^l_{mn}\br)^2
\Br]
\, .
\eea
Given that $\bar{q}(t)$ is quadratic polynomial in $t$, the action 
$\bl(S^{(l)}_{\rm grav} \br)_{nm}$ has a very simple structure. Moreover, in the one-loop studies, the quadratic nature of the action for $H^l_{mn}(t)$ leads to an exactly doable path integration over each mode $H^l_{mn}(t)$. As the action for each mode respects spherical symmetry and is independent of $n$ and $m$, therefore 
while doing path integral these indices will be suppressed in the following. However, appropriate degeneracy contribution will be 
included wherever needed. For a given $l$, the path integral 
over all the modes with various $n$ and $m$ will have same contribution. 
This implies that only path integral over the various modes 
with different $l$ needs to be done, with the contribution of each 
being raised by power $g_l$ (the degeneracy). This implies 
\bea
\label{eq:h_PI_mode_degen}
\int_{{\rm Bd}_i}^{ {\rm Bd}_f} 
{\cal D} H_{ij}(t, {\bf x}) 
&&
\exp \left(\frac{i}{\hbar} S^{(h)}_{\rm grav}[\bar{q}, H_{ij}, N_c] \right) 
\notag \\
&&
= \prod_{l=2}^\infty
\biggl[
\int_{{\rm Bd}_i}^{ {\rm Bd}_f} 
{\cal D} H_l(t) \exp\left(\frac{i}{\hbar} S^{(l)}_{\rm grav}[\bar{q}, H_l, N_c] \right) 
\Br]^{g_l} 
= e^{i {\cal A}_{h}(N_c)/\hbar}\, ,
\eea
where $n$ and $m$ indices are no longer needed as their contribution 
has been taken into account by including degeneracy factor $g_l$,
and $S^{(l)}_{\rm grav}$ is the action in Eq. (\ref{eq:SL_hL_act}) 
with indices $n$ and $m$ suppressed. ${\rm Bd}_i$ and ${\rm Bd}_f$
are the boundary condition imposed at the initial and final boundary 
for the fluctuation field $H_{ij}$ respectively, and 
$\bar{q}(t)$ is the on shell solution for the scale factor 
which is quadratic in $t$ and mentioned in Eq. (\ref{eq:qsol_gen}).

The path integral over $H_l(t)$ mentioned in Eq. (\ref{eq:h_PI_mode_degen}) although quadratic (as we are working in one-loop approximation) still needs care in its evaluation. Structurally, the path integral will have the following form:
\beq
\label{eq:hL_PI_stru}
\int_{{\rm Bd}_i}^{ {\rm Bd}_f} 
{\cal D} H_l(t) \exp\Bl(\frac{i}{\hbar} 
S^{(l)}_{\rm grav}[\bar{q}, H_l, N_c] \Br)
= \bl\{\D^{(l)}_h(N_c) \br\}^{-1/2}\, ,
\eeq
where $H_{l}$ vanishes on the boundaries ${\rm Bd}_i$ and ${\rm Bd}_f$ and the action for the mode $H_l(t)$ is mentioned in Eq. (\ref{eq:SL_hL_act}). This being a Gaussian path integral is easily evaluated for each mode giving the rhs of Eq. (\ref{eq:hL_PI_stru}). The action of the system can be thought of 
as a harmonic oscillator with time-dependent mass and frequency. 
Such systems lack time-translational invariance and hence 
do not suffer from complications involving zero modes. In such cases,
one can make use of the results stated in \cite{Khandekar:1986ib, Sabir:1991PrJ}
to compute the path integral exactly. 
These can be derived alternatively via Wronski's construction 
described in \cite{McKane:1995vp, Kleinert:1998rz, Kleinert:1999Cher}. 

However, naively implementing the results of 
\cite{Khandekar:1986ib, Sabir:1991PrJ, Kleinert:1998rz, Kleinert:1999Cher}
to compute $\D^{(l)}_h(N_c)$ does not guarantee that the 
functional determinant will be devoid of problems. 
For example, as will be seen later for some 
$N_c$ it can diverge! In such cases it is worth asking about 
the nature of divergences present in the functional determinant 
which arises in the one-loop computation.
For generic boundary 
choices $\D^{(l)}_h(N_c)$ structurally is given by
\beq
\label{eq:Del_struct_hl}
\D^{(l)}_h(N_c) = -
\bl(2\pi i \hbar\br)
\Bl(
\frac{\Lam N_c^2 c_2}{3} + c_1c_2 + c_2^2
\Br)^{-1/2} W_l(1,0) \mathbb{M}(\xi_{l},N_c)^{-1}\, ,
\eeq
where $\xi_l$ and $\tau_t$ are mentioned in Eq. (\ref{eq:albt_form}),
$W_l(1,0)$ can be obtained from Eq. (\ref{eq:W_l10_func}),
$c_1$ and $c_2$ depend on the choice of boundary condition 
imposed on the scale factor $q(t)$.   
It should be highlighted that the $N_c$ dependence 
is embedded in $c_1$, $c_2$, $\xi_l$ and the expression for $\tau_t$.
The factor $2\pi i \hbar$ arises due to normalization and is 
standard in the path integral computations, it can be ignored for the 
purpose of this paper as it does not play role in the following.

The quantum corrected lapse action is given by 
\bea
\label{eq:QT_Nc_act_h_divST}
{\cal A}_{h}(N_c) = \sum_{l=2}^\infty 
\Bl[
-\frac{i g_l \hbar}{4} \ln \Bl(
\frac{\Lam N_c^2 c_2}{3} + c_1c_2 + c_2^2
\Br)
+ \frac{i g_l \hbar}{2} \ln W_l(1,0)-\frac{i g_l \hbar}{2} \ln \mathbb{M}(\xi_{l},N_c) \Br]  \, .
\eea
This is the contribution of the quantum fluctuation field $H_{ij}$
at one loop to the action of the lapse $N_c$. Structurally, it will remain 
same for generic boundary choices. 
Note, that the dependence on $c_1$ and $c_2$ 
is also present in $\xi_l$ via the relation mentioned in Eq. (\ref{eq:albt_form}).
This expression for $W_l(1,0)$ mentioned in Eq. (\ref{eq:W_l10_func}) depends on $c_1$ and $c_2$ indirectly via $\xi_l$. The choice of boundary conditions imposed on the scale factor eventually translates into determining $c_1$ and $c_2$.
The same boundary choice, due to covariance further restricts the 
choices available for the boundary condition for the fluctuation field $h_{ij}$.

%%%%%%%%%%%%%%%%%%%%%%%%%%%%%%%%%%%%%%%%%%%%%%%
\section{Analysing $N_c$-integral}
\label{sec:lapseNc}
%%%%%%%%%%%%%%%%%%%%%%%%%%%%%%%%%%%%%%%%%%%%%%%

After having computed the path integral over the scale factor $q(t)$
and fluctuation field $h_{ij}$ to first order in $\hbar$ (up to 
one loop), we turn our attention to the final step in the computation 
of the transition amplitude mentioned in Eq. (\ref{eq:grav_path_qform}). 
This implies that we now focus our interests in studying 
\beq
\label{eq:grav_path_NC_form}
G[{\rm Bd}_f, {\rm Bd}_i]
= \int_{-\infty}^{\infty} {\rm d} N_c \,\,
\exp \bl\{i {\cal A}(N_c)/\hbar \br\} \, ,
\eeq
where the total action for the lapse $N_c$, including the contribution from 
both the path integral over $q(t)$ and $H_{ij}$ is given by
\bea
\label{eq:Nc_act_total}
&&
{\cal A}(N_c) = {\cal A}_{q}(N_c) + {\cal A}_{H}(N_c)
= S_{\rm grav}^{(\bar{q})}(N_c)
+  \frac{i \hbar}{2} \ln \D_q(N_c) -\frac{i\hbar}{2} \sum_{l=2}^\infty 
g_l  \ln \mathbb{M}(\xi_{l},N_c)  
\notag \\
&&
- \frac{2i \hbar}{3} \ln \Bl(
\frac{\Lam N_c^2 c_2}{3} + c_1c_2 + c_2^2
\Br)
+ \frac{i \hbar}{2} \sum_{l=2}^\infty 
g_l  \ln W_l(1,0) 
= {\cal A}_0(N_c) + i \hbar {\cal A}_1(N_c) 
\, ,
\eea
where ${\cal A}_0(N_c)$ is the classical action and ${\cal A}_1(N_c) $ is the one-loop quantum correction. Here we have utilized the following regularized summation 
via the usage of $\zeta$-function
\beq
\label{eq:reg_sum_gl}
\sum_{l=2}^\infty g_l = 2 \zeta(-2) + 4 \zeta(-1) - 6\zeta(0) = 8/3 \, .
\eeq
This is possible as $N_c$, $c_1$, and $c_2$ do not depend on the mode number $l$. In principle, the summation over $l$ should be cutoff at $l_{\rm max}$, which arises due to the ultraviolet cutoff of the theory. For example, the frequency corresponding to each $l$-mode at the final hypersurface is like $\nu_l \sim l/\sqrt{q_f}$, thereby implying a corresponding energy to be roughly $E_l \sim l/\sqrt{q_f}$. The theory is principally well defined for $E_l < \Lam_{\rm cutoff}$, which puts a bound on $l$, giving $l_{\rm max} \sim \sqrt{q_f} \Lam_{\rm cutoff}$. However, as $l_{\rm max}$ is expected to be quite large one can take it to be infinite wherever it does not lead to divergences. 

The action stated in Eq. (\ref{eq:Nc_act_total}) includes ${\cal O}(\hbar)$ corrections from both the path integral of $q(t)$ and $H_{ij}$ (these are one-loop corrections to the lapse action). 
The corresponding quantum correction $\ln \D_q(N_c)$
are known in literature \cite{Lehners:2021jmv, DiTucci:2020weq, 
Narain:2022msz, Ailiga:2023wzl, Ailiga:2024mmt}. 
The contribution coming from the $H_{ij}$ is the quantum correction mentioned in Eq. (\ref{eq:Nc_act_total}). This also incorporates the effects of the boundary actions that have been computed in the earlier section. One-loop studies in cosmological systems similar to the one considered in this paper have been attempted in \cite{Barvinsky:1992dz}. However, their study was done with Euclidean path integral in the background field formalism with metric perturbations over a fixed background spacetime given by the Hartle-Hawking Universe. Our study in this paper goes beyond by considering the path integral over all the geometries respecting $\mathbb{R} \times S^3$ symmetry along with the perturbations over $S^3$. In this case, the path integral done over the scale factor allows for all possibilities respecting boundary choices, where the lapse $N_c$ is not fixed to unity but can take any value in the complex $N_c$ plane. The path integral over the fluctuation is performed at one loop, taking these things into consideration.  

The $N_c$ integral in general is quite complicated for various 
boundary conditions. These complications come from the path integral over the $h_{ij}$ field. 
However, some progress can be made using the saddle-point 
approximation, which we will utilize to compute $N_c$ integral.
It must be highlighted, though, that the $N_c$ integral can be 
done exactly using $Airy$ functions when $h_{ij}$ contribution is not taken 
into account. This has been studied extensively in past in 
\cite{Lehners:2021jmv, DiTucci:2020weq, Narain:2022msz, Ailiga:2023wzl, Ailiga:2024mmt}. 
In this paper, we go beyond and make an attempt 
to address the transition amplitude incorporating the effects 
from the fluctuation field $h_{ij}$ up to one loop.

%%%%%%%%%%%%%%%%%%%%%%%%%%%%%%%%%%%%%%%%%%%%%%%
\subsection{Picard-Lefschetz methodology}
\label{PL_int} 
%%%%%%%%%%%%%%%%%%%%%%%%%%%%%%%%%%%%%%%%%%%%%%%

The integral in Eq. (\ref{eq:grav_path_NC_form}) is highly oscillatory 
if the integration contour is along the real line. This can be made 
convergent if the contour of integration can be deformed 
to a new contour along which the integrand is well behaved and 
convergent. Picard-Lefschetz (Pl) methodology systematically describes 
the procedure for obtaining this. Here, we do a quick recap of the 
key points of the PL methods, while the details can be seen in 
\cite{Witten:2010cx, Witten:2010zr, Basar:2013eka, Tanizaki:2014xba, Lehners:2018eeo, Narain:2021bff, Ailiga:2024mmt}.

In the PL methodology, one allows the $N_c$ to be a complex variable and 
analytically continues the exponent $\mathcal{A}(N_c)$ to a holomorphic 
function of $N_c$, which satisfies the following Cauchy's equation
\begin{align}
\label{eq:CRfunc}
\frac{\partial { \mathcal{A}}}{\partial \bar{N_c}} = 0 
\Rightarrow
\begin{cases}
\pt{\rm Re} {\cal A}/ \pt y_1
&= \pt {\rm Im} {\cal A}/\pt y_2 \, , \\
\pt{\rm Re} {\cal A}/ \pt y_2
&= - \pt{\rm Im} {\cal A} / \pt y_1 \, .
\end{cases}
\end{align}
where, $N_c=y_1(\lambda)+i y_2(\lambda)$. Let us write the whole 
exponent as $i\mathcal{A}(N_c)=\mathfrak{h}+i\mathcal{H}$. 
Then the equation for the \emph{thimbles} (also called downward flow) is given by
\beq
\label{eq:downFlowDef}
\frac{{\rm d} y_i}{{\rm d} \lam}
= - \mathfrak{g}_{ij} \frac{\pt \mathfrak{h}}{\pt y_j} \, ,
\eeq
where, $\mathfrak{g}_{ij},(i,j=1,2)$ is the Riemannian metric in the 
complex $N_c$ plane and $\lambda$ is a flow parameter along the thimbles. 
These are the steepest descent trajectories denoted by 
$\mathcal{J}_\sigma$; $\sigma$ denotes the saddle point associated 
with the \emph{thimble}. Along these lines, the Morse function ($\mathfrak{h}$) 
decreases monotonically as one moves away from the saddle point, with the 
maximum being at the saddle point and satisfies
\beq
\label{eq:flowMonoDec_h}
\frac{\rm d \mathfrak{h}}{{\rm d} \lam}
= \mathfrak{g}_{ij} \frac{{\rm d} y^i}{{\rm d} \lam} \frac{\pt \mathfrak{h}}{\pt y_j} 
= - \left(\frac{{\rm d} y_i}{{\rm d}\lam}\frac{{\rm d} y^i}{{\rm d}\lam}\right)
\leq 0 \, ;
\eeq
 also the phase of the integrand remains constant, i.e., 
$\frac{d\mathcal{H}}{d\lambda}=0$ along these thimbles,
\beq
\label{eq:consHflow}
\frac{{\rm d}\mathcal{H}}{{\rm d} \lam} 
= \frac{1}{2i} \frac{{\rm d} ({\cal A} - {\cal \bar{A}})}{{\rm d} \lam} 
= \frac{1}{2i} \left(
\frac{\pt {\cal A}}{\pt N_c} \frac{{\rm d} N_c}{{\rm d} \lam} 
- \frac{\pt \bar{\cal A}}{\pt \bar{N_c}}\frac{{\rm d}\bar{N_c}}{{\rm d} \lam}
\right) = 0 \, .
\eeq
These two facts make the integral convergent along 
the \emph{thimble}. Instead of a minus sign in Eq .(\ref{eq:downFlowDef}), 
if one would have defined it with a plus sign, then it would be the steepest 
ascent lines ($\mathcal{K}_\sigma$) along which the Morse function ($\mathfrak{h}$) 
increases as one moves away from the saddles. Integral becomes 
divergent along these lines.

Once the saddles and their corresponding thimbles are found,
one can analyze the behavior 
of the Morse function to identify the {\it allowed} and {\it forbidden} regions in the 
complex $N_c$ plane. The allowed region, where the integral is well-behaved, 
is denoted by $J_\sigma$, while the forbidden region, where the integral diverges, 
is denoted by $K_\sigma$. The {\it allowed} and {\it forbidden} regions are given by 
$\mathfrak{h}(J_\sigma)<\mathfrak{h}(N_\sigma)$, 
whereas $\mathfrak{h}(K_\sigma)>\mathfrak{h}(N_\sigma)$ respectively.

Once all the saddles, corresponding thimbles and the region of convergence are known, 
the crucial question comes: what is the correct deformed contour along which the 
integral absolutely converges? According to the PL method, given the original 
contour ($\mathbb{D}$), the deformed contour will be
\begin{equation}
\mathcal{C}=\sum_\sigma n_\sigma \mathcal{J}_\sigma,
\hspace{5mm} \,n_\sigma={\rm Int} (\mathbb{D}, {\cal K}_\sg),
\end{equation}
where $\rm Int (. , .)$ counts the intersection between two curves 
and satisfies $\rm Int (\mathcal{J}_\sigma,\mathcal{K}_{\sigma'})=\delta_{\sigma\sigma'}$. 
$n_\sigma$ takes values $(0,\pm 1)$ and the sign of $n_\sigma$ signifies the orientation.
The above equation implies that the deformed contour of integration will be 
the sum (with appropriate orientation) of those \emph{thimbles} 
($\mathcal{J}_\sigma$) for which the associated steepest ascent line 
($\mathcal{K}_\sigma$) intersects the original contour of integration. 
Saddle points attached to those thimbles are the only \emph{relevant} 
saddles which contribute to the integral. While deforming the original contour, 
one should be careful so that the endpoints remain fixed so that there is no 
end-point contribution coming from the arc at infinity and the deformation 
encounters no singularities of the integrand, if any. 

After all these are settled, the original oscillatory integral now becomes 
the sum of integrals over each thimble:
\beq
\label{eq:sumOthim}
\int_{-\infty}^{\infty} {\rm d} N_c\, e^{i \mathcal{A}(N_c)/\hbar}
\Rightarrow \sum_\sg n_\sg \int_{{\cal J}_\sg} {\rm d}N_c
e^{i \mathcal{A}(N_c)/\hbar}\, .
\eeq
Now, since along each $\mathcal{J}_\sigma$, the integral is absolutely 
convergent, one can perform this integral in the saddle-point approximation retaining all terms relevant at the one loop. For this we first note that the quantum corrected lapse action can be written in the following form:
\beq
\label{eq:ANc_1lp_form}
{\cal A}(N_c) = {\cal A}_0(N_c) + \hbar {\cal A}_1(N_c) + \cdots \, ,
\eeq
where $\cdots$ contain higher-loop contribution, and ${\cal A}_0(N_c)$ and ${\cal A}_1(N_c)$ can be obtained from Eq. (\ref{eq:Nc_act_total}). If saddle-point is denoted by $N_\sg$, which receives $\hbar$-corrections, then it can be written in the form
\beq
\label{eq:Nsg_1lp_form}
N_\sg = N_\sg^{(q)} + \hbar \bl(
{\cal N}_\sg^{(q)} + {\cal N}_\sg^{(h)}
\br) + \cdots \, ,
\eeq
where $N_\sg^{(q)}$ is the saddle point for the zeroth-order lapse action ${\cal A}_0(N_c)$, while $\bl(
{\cal N}_\sg^{(q)} + {\cal N}_\sg^{(h)}
\br)$ corresponds to one-loop corrections (note there is no contribution coming from the $h_{ij}$ field at the zeroth order in $\hbar$, which is due to boundary choice of vanishing perturbation at the boundaries). Expanding ${\cal A}(N_c)$ around the saddle point $N_\sg$ gives [ keeping terms up to $\mathcal{O}(\hbar)$]
\bea
\label{eq:ANc_exp_sad_form}
{\cal A}(N_c) 
&& = {\cal A}(N_\sg) + \frac{1}{2}(N_c - N_\sg)^2 {\cal A}^{\prime\prime}(N_\sg) + \cdots
\notag \\
&&
= {\cal A}_0(N_\sg^{(q)})
+ \hbar {\cal A}_0^\prime(N_\sg^{(q)}) 
\bl(
{\cal N}_\sg^{(q)} + {\cal N}_\sg^{(h)}
\br)
+ \hbar {\cal A}_1(N_\sg^{(q)}) 
\notag \\
&&
+ \frac{1}{2}(N_c - N_\sg)^2 \bl[
{\cal A}_0^{\prime\prime} (N_\sg^{(q)})
+ \hbar {\cal A}_0^{\prime\prime\prime} (N_\sg^{(q)})\bl(
{\cal N}_\sg^{(q)} + {\cal N}_\sg^{(h)}
\br)
+ \hbar {\cal A}_1^{\prime\prime} (N_\sg^{(q)})
+ \cdots
\br] + \cdots 
\notag \\
&& 
= {\cal A}_0(N_\sg^{(q)})
+ \hbar {\cal A}_1(N_\sg^{(q)}) 
+ \frac{1}{2} (N_c - N_\sg)^2 \bl[
{\cal A}_0^{\prime\prime} (N_\sg^{(q)})
\notag \\
&&
+ \hbar {\cal A}_0^{\prime\prime\prime} (N_\sg^{(q)})\bl(
{\cal N}_\sg^{(q)} + {\cal N}_\sg^{(h)}
\br)
+ \hbar {\cal A}_1^{\prime\prime} (N_\sg^{(q)})
+ \cdots
\br] + \cdots \, ,
\eea
where some terms vanish as ${\cal A}^\prime(N_\sg)=0$ and ${\cal A}_0^\prime(N_\sg^{(q)})=0$.
In the saddle-point approximation and retaining terms relevant at one loop, we will get (up to some irrelevant numerical factor)
\beq
\label{eq:LDordI}
\int_{{\cal J}_\sg} {\rm d}N_c\,
e^{i \mathcal{A}(N_c)/\hbar} 
= 
\frac{e^{i\theta_\sg}}{\sqrt{\bl\vert {\cal A}_0^{\prime\prime} (N_\sg^{(q)}) \br\rvert}}
\times 
\exp\Bl[ 
\frac{i}{\hbar} {\cal A}_0(N_\sg^{(q)})
+ i {\cal A}_1(N_\sg^{(q)}) + \cdots
\Br] \, ,
\eeq
Here, $\theta_\sigma$ is the direction of the flow lines (steepest descent) at the corresponding saddle point ($N_\sigma$). It is
\begin{equation}
\label{eq:theta_sg_exp}
    \theta_\sigma =\frac{(2k-1)\pi}{4}-\frac{\delta_{\sigma}}{2},\hspace{5mm}k\in \mathbb{Z}_{\rm odd},
\end{equation} 
where $\delta_{\sg}$ is defined via $\mathcal{A}''(N_\sg)=|\mathcal{A}''(N_\sg)|e^{i\delta_\sg}$. In principle, there are corrections to $\theta_\sg$ and the denominator in Eq. (\ref{eq:LDordI}). However, they will play a role at two or higher loops. We will utilize the expression mentioned in Eq. (\ref{eq:LDordI}) later to compute the $N_c$ integration.

%%%%%%%%%%%%%%%%%%%%%%%%%%%%%%%%%%%%%%%%%%%%%%%
\subsubsection{$N_\sg^{(q)}$: saddles of ${\cal A}_0(N_c)$}
\label{subsubS:Nsq}
%%%%%%%%%%%%%%%%%%%%%%%%%%%%%%%%%%%%%%%%%%%%%%%

Past studies done in the literature dealing with Dirichlet BC showed that imposing DBC on the background field $q(t)$, corresponding defining the path integral starting with zero size, leads to unsuppressed behavior of gravitation fluctuations in the no-boundary proposal of the Universe \cite{Feldbrugge:2017kzv,Lehners:2018eeo,Feldbrugge:2017fcc,Feldbrugge:2017mbc}. These studies shows perhaps imposing Dirichlet BC at initial boundary is not the ``right'' choice and motivates one to investigate Neumann and Robin boundary conditions on the scale factor $q(t)$ at the initial boundary. These have been shown to lead to stable universe where perturbations are well behaved \cite{DiTucci:2019bui, Narain:2021bff, Lehners:2021jmv, DiTucci:2020weq, Narain:2022msz, DiTucci:2019dji, Ailiga:2023wzl, Ailiga:2024mmt}. Extremizing the on shell action of $q(t)$ in either boundary choices lead to zeroth-order saddles. These follow from the solution of ${\rm d} S_{\rm grav}^{(\bar{q})}/{\rm d} N_c=0$, where $S_{\rm grav}^{(\bar{q})}$ is given in Eqs. (\ref{eq:stot_onsh_nbc}) and (\ref{eq:stot_onsh_rbc}) for Neumann BC and Robin BC, respectively.

For $\pi_{i}=-3i$ and $P_{i}=-3i$, we have the two Hartle-Hawking no-boundary saddle point solutions given by
\beq
\label{eq:NB_sad}
N_{\pm}^{(\rm nb)}=
\frac{3}{\Lam}\Bl(-i\pm \sqrt{\frac{\Lam q_f}{3}-1}\Br)\, ,
\eeq
for Neumann BC and Robin BC, respectively. Additionally, for Robin BC, we also have two non-no-boundary solutions given by
\beq
\label{eq:NNB_sad}
N_{1,2}^{(\rm \cancel{nb})}=-\frac{3}{\beta}
-\frac{3}{\Lambda}\Bl(- i  \pm\sqrt{\frac{\Lambda  q_f}{3}-1} \Br),
\eeq
where we choose $\beta$ to be purely negative imaginary defined by $\beta=-i\Lambda x/2$, $x>0$, \cite{Ailiga:2023wzl}. This particular choice turns out to be relevant for the present discussion. These constitute $N_\sg^{(q)}$ for these boundary choices.

%%%%%%%%%%%%%%%%%%%%%%%%%%%%%%%%%%%%%%%%%%%%%%
\subsubsection{$\hbar$-correction to saddles: computation of ${\cal N}_\sg^{(q)} + {\cal N}_\sg^{(h)}$}
\label{sad_cor2_qm} 
%%%%%%%%%%%%%%%%%%%%%%%%%%%%%%%%%%%%%%%%%%%%%%% 

For completeness, we compute the first order in $\hbar$ correction to the saddles $N_\sg^{(q)}$. These, however, do not enter in the computation of the one-loop studies but will nevertheless be useful in studies involving degenerate situations. 

Saddles by definition satisfy ${\cal A}^\prime(N_\sg)=0$. On utilizing Eqs. (\ref{eq:Nsg_1lp_form}) and (\ref{eq:ANc_exp_sad_form}), we note that to first order in $\hbar$ we have 
\beq
\label{eq:ANc_sad_exp_1lp}
 {\cal A}_0^\prime(N_\sg^{(q)})
+ \hbar {\cal A}_0^{\prime\prime}(N_\sg^{(q)}) 
\bl(
{\cal N}_\sg^{(q)} + {\cal N}_\sg^{(h)}
\br)
+ \hbar {\cal A}_1^\prime(N_\sg^{(q)}) = 0 \, .
\eeq
At zeroth order by definition we have ${\cal A}_0^\prime(N_\sg^{(q)})=0$. This gives 
\beq
\label{eq:Hbar_sadcor_genform}
\bl(
{\cal N}_\sg^{(q)} + {\cal N}_\sg^{(h)}
\br)
= - {\cal A}_1^\prime(N_\sg^{(q)})/{\cal A}_0^{\prime\prime}(N_\sg^{(q)}) \, .
\eeq
In the special situation, when there is no contribution coming from the fluctuation field $h_{ij}$, then the quantum correction to the saddles comes entirely from the ${\cal N}_\sg^{(q)}$. This can be computed from 
\begin{equation}
\label{eq:sadcor_1lp_onlyQ}
[S_{\rm grav}^{(q)}(N_\sigma^{(q)})]^\prime
+ \hbar [S_{\rm grav}^{(q)}(N_\sigma^{(q)})]^{\prime\prime}\mathcal{N}_\sigma^{(q)}
+(i\hbar/2)\Delta_q'(N_\sigma^{(q)})/\Delta_q(N_\sigma^{(q)})
=0 \, .
\end{equation}
The first term in the above vanishes as $N_\sigma^{(q)}$ satisfies the zeroth order saddle point equation. In that, the first-order log correction to the background saddle is given by
\begin{equation}
\label{eq:logcorrectionto_sad}
\mathcal{N}_\sigma^{(q)}
=-(i\hbar/2)\Delta'_q(N_\sigma^{(q)})[\Delta_q(N_\sigma^{(q)})]^{-1}/\bl[S^{(q)}_{\rm grav}(N_\sigma^{(q)}) \br]^{\prime\prime} \, .
\end{equation}
For the case when Robin boundary condition is applied on the initial hypersurface and Dirichlet boundary condition at the final hypersurface, we have $\Delta_q(N_c)=1+ 2N_c\beta/3$. In this case we get
\begin{equation}
\label{eq:calNs_q_hbarCR}
\mathcal{N}_\sigma^{(q)}
=-(i\beta/3)\bl(1+2\beta N_\sigma^{(q)}/3\br)^{-1}/\bl[S^{(q)}_{\rm grav}(N_\sigma^{(q)}) \br]^{\prime\prime}\, .
\end{equation}
This is the correction received due to the logarithmic term, provided the denominator is nonvanishing. Note that it is proportional to $\beta$. So, for the Neumann ($\beta=0$) the saddles receives no correction due to quantum fluctuations of scale factor. There will be however $\hbar$ correction coming from the quantum fluctuations of $h_{ij}$, which is encoded in $\mathcal{N}_\sigma^{(h)}$ term of Eq. (\ref{eq:Nsg_1lp_form}). We will not be computing it here as it is not only beyond the scope of this work but also does not play a role in the transition amplitude computed to one loop.

%%%%%%%%%%%%%%%%%%%%%%%%%%%%%%%%%%%%%%%%%%%%%%%
\section{No-boundary saddles: one-loop UV divergences}
\label{NB_sad_cor} 
%%%%%%%%%%%%%%%%%%%%%%%%%%%%%%%%%%%%%%%%%%%%%%%

It is crucial to know how the saddles behave once the contribution from $h_{ij}$ is incorporated. This is particularly important in the studies of the no-boundary proposal of the Universe, where the boundary conditions are chosen to ensure that the universe starts with a vanishing initial size. This proposal has been successfully explored for the case when Dirichlet, Neumann or Robin boundary condition is imposed on the scale factor within the minisuperspace approximation \cite{Feldbrugge:2017kzv, Narain:2021bff, Narain:2022msz, Ailiga:2023wzl}. In either case, the existence of ``no-boundary'' saddles and its relevance in the $N_c$ integration following Picard-Lefschetz methodology motivates one to explore the robustness of the proposal in various ways. In particular, it has been noticed that Gaussian perturbations around the no-boundary saddles for the Dirichlet boundary conditions are unstable \cite{Feldbrugge:2017fcc, Lehners:2018eeo, DiTucci:2018fdg, Feldbrugge:2017mbc}, while perturbations are seen to be well behaved and show stable behavior in the case of Neumann and Robin boundary conditions \cite{DiTucci:2019bui, DiTucci:2019dji}. 

Robustness of the no-Boundary universe was also tested under the inclusion of the higher-derivative gravitational corrections and its compatibility with the asymptotic safety paradigm \cite{Lehners:2019ibe, Lehners:2023fud}. These explorations further assures faith in the robustness of the no-boundary proposal of the Universe. These studies while heuristically done, still gives confidence to proceed forward in further analyzing the no-boundary proposal. Recent exact studies involving inclusion of Gauss-Bonnet gravitational term lead to conclusion that boundary conditions corresponding to no-boundary Universe are favored in the path integral as they tend to give most dominant contribution \cite{Narain:2021bff, Narain:2022msz, Ailiga:2023wzl}.

In this paper, we challenge the robustness of the no-boundary proposal of Universe by further incorporating the fluctuation field $h_{ij}$ and quantum corrections. We do a one-loop study of the path integral over the fluctuation field $h_{ij}$, and aim to compute the one-loop corrected action for the lapse $N_c$ which is mentioned in Eq. (\ref{eq:Nc_act_total}). However, as Einstein-Hilbert gravity is known to UV nonrenormalizable, it is therefore expected that UV divergences will show up in some form in the one-loop study of the no-boundary universe. In this section, we will extract these UV divergences systematically and compute the relevant counterterms that are needed to make the quantum lapse action finite. 

%%%%%%%%%%%%%%%%%%%%%%%%%%%%%%%%%%%%%%%%%%%%%%%
\subsection{UV Divergence near no-boundary saddles}
\label{small_q0qf} 
%%%%%%%%%%%%%%%%%%%%%%%%%%%%%%%%%%%%%%%%%%%%%%%

%
At the no-boundary saddles (with various boundary conditions), the scale factor vanishes: $\bar{q}_0 \to 0$ at the initial hypersurface. This immediately implies that $c_2 \to0$. Also, at these saddles $\xi_l$ and $\tau_t$ from Eq. (\ref{eq:albt_form}) is given by
\begin{equation}
\label{eq:xiL_tauT_nbc_nb1}
\xi_l^{\rm s}=1+l \, ,
\hspace{5mm}
\tau_{t}^{\rm nb} =1-i \Lam N_{\pm}^{\rm (nb)}t/3 \, .
\end{equation}
Note that $\tau_0^{\rm nb}=1$. $W_l(1,0)$ can be obtained from Eq. (\ref{eq:W_l10_func}).
At the no-boundary saddles $W_l(1,0)$ gets simplified 
as $\mathbb{P}_1^{l+1}(\tau_0^{\rm nb}) =0$. This means that Eq. (\ref{eq:W_l10_func})
becomes 
\beq
\label{eq:Wl_nb_nbc}
W_l^{\rm nb}(1,0) = 
\mathbb{P}_1^{l+1}(\tau_1^{\rm nb}) \mathbb{Q}_1^{l+1}(\tau_0^{\rm nb}) \, .
\eeq
where, $\mathbb{P}_1^{l+1}(\tau_t^{\rm nb})$ becomes algebraic and is given by
\beq
\label{eq:alt_Pxi_nb_form}
\mathbb{P}_1^{l+1} (\tau_t^{\rm nb}) 
= (-1)_{l+1}\Bl(\frac{1-\tau_{t}^{\rm nb}}{1+\tau_{t}^{\rm nb}}\Br)^{(l+1)/2}
\bl(\tau_{t}^{\rm nb}+l+1\br) \, ,
\eeq
where $(a)_b$ is the Pochhammer symbol. Here, $(-1)_{l+1}$ vanishes when $l$ is positive integer. However, it is cancelled by the $\mathbb{M}(\xi_{l}^{s},N_{nb})^{-1}$ which is infinity leading to a finite quantity, as shown in the Appendix \ref{app:gelfand}.
It is seen that near the no-boundary saddles, the function $W_l^{\rm nb}(1,0)$ behaves as 
\beq
\label{eq:Wl_be_NoB}
W_l(1,0) \mathbb{M}(\xi_{l}^{s},N_{nb})^{-1}\Br \rvert_{N_c \to N_{\rm nb}}
= \bar{q}_0^{-\xi_l^{\rm s}/2} \mathbb{W}_l(1,0) 
\,\, \xRightarrow[N_c \to N_{\rm nb}]{} \,\, 
\bar{q}_0^{-(l+1)/2} \mathbb{W}_l(1,0)
\, ,
\eeq
where $\xi_l^{\rm s}$ is defined in Eq. (\ref{eq:xiL_tauT_nbc_nb1}), while $\mathbb{W}_l(1,0)$ is the finite part depending on the choice of boundary conditions. It should be stressed that $W_l(1,0)$ remains finite at other saddles but behaves singularly in case of ``no-boundary'' saddles. On realizing that $\bl(\Lambda N_c^2/3+c_1+c_2\br)=q_f$, one can express the quantum corrected lapse $N_c$ action near the no-boundary saddles as follows:
\bea
\label{eq:Nc_act_total_nb}
&&
{\cal A}(N_c) \br \rvert_{N_c \to N_{\rm nb}, } 
= S_{\rm grav}^{(\bar{q})}(N_c)
+ \frac{i \hbar}{2} \ln \D_q(N_c)
\notag \\
&&
- \frac{2 i \hbar}{3} \ln q_f 
+ \frac{i \hbar}{2} \sum_{l=2}^\infty 
g_l  \ln \mathbb{W}_l(1,0) 
- 
\underbrace{
\frac{117i \hbar}{80} \ln 
\bar{q}_0 
}_{{\rm Log-div}}
+ {\cal O}(\hbar, N_c - N_{\rm nb})\, .
\eea
Note that the last term in the above is $\log$ divergent, where we have used zeta function to regularize the infinite summation over $l$ and obtain the coefficient of $\ln \bar{q}_0$. In the quantum corrected action for lapse, the last term diverges logarithmically at the ``no-boundary'' saddles. Such $\log$ divergence arises from each mode of the field and needs to be summed up, leading to the above. This is the action computed near the no-boundary saddle. The terms included in ${\cal O}(\hbar, N_c - N_{\rm nb})$ become relevant beyond one loop. In the Picard-Lefschetz computation of the $N_c$ integration at one loop they do not contribute.

%%%%%%%%%%%%%%%%%%%%%%%%%%%%%%%%%%%%%%%%%%%%%%%
\subsection{UV Divergence due to $l$ summation}
\label{small_q0qf} 
%%%%%%%%%%%%%%%%%%%%%%%%%%%%%%%%%%%%%%%%%%%%%%%

Besides the $log$ divergence, the $l$-summation series mentioned in the last line of Eq. (\ref{eq:Nc_act_total_nb}) also leads to UV divergence. This can be extracted systematically as shown in the Appendix \ref{logsum}. At the no-boundary saddles, we first realize that, 
\begin{equation}
\label{eq:Wlbb_nb_10}
    \mathbb{W}_l(1,0)=\mathbb{P
    }_1^{l+1}(\tau_1^{\rm nb})\mathbb{A}_l \, ,
\end{equation}
where, $\mathbb{A}_l$ (which is independent of $q_f$, and same for all saddles) is given by
\begin{equation}
\label{eq:Al_nb_10}
    \begin{split}
        \mathbb{A}_l&=  \frac{ie^{i\pi(l+1)}\Gamma(1-l)}{2(l+1)(l+2)} \left(\frac{12}{\Lambda}\right)^{(l+1)/2},
\hspace{5mm}
\forall \,\, l\geq2 \,\, {\rm and} \,\, l \in \mathbb{I}^+ 
    \end{split} \, .
\end{equation}
The $i$ factor appearing in Eq. (\ref{eq:Al_nb_10}) is canceled by the $(-i)$ present in Eq. (\ref{eq:Del_struct_hl}). Also, $e^{i\pi(l+1)}$ is canceled using the identity in Eq. (\ref{identity}) leading to a real quantity.
Once the $\log$ function is expanded, it will lead to the following:
\bea
\label{eq:log_exp_wl_nb_10}
    \sum_{l=2}^\infty g_l\ln \mathbb{W}_l(1,0)
    && =- \ln 2 \sum_{l=2}^\infty g_l +  \frac{1}{2}\ln\left[\frac{12 (1-\tau_1^{\rm nb})}{\Lam(1+\tau_1^{\rm nb})}\right]\sum_{l=2}^\infty g_l(l+1) 
    \notag \\
   && +\sum_{l=2}^\infty g_l\ln(l+1+\tau_1^{nb}) - \sum_{l=2}^\infty g_l \ln (l+1) - \sum_{l=2}^\infty g_l \ln (l+2)  \,.
\eea
In this, the first and second series can be easily summed using zeta functions, giving $\sum_{l=2}^\infty g_l = 8/3 $ and $\sum_{l=2}^\infty g_l(l+1)=191/60$ respectively. The infinite summation in the second line consists of three parts: one dependent on $\tau_1^{nb}$, and the other two are independent of any parameters of the theory. All of them, although summable, are divergent and need to be regularized. This can be done via the zeta function regularization technique and has been computed in the Appendix \ref{logsum}. Here, we utilize those to extract the divergent and finite parts at the no-boundary saddles. The divergent part is given by
\begin{equation}
\label{eq:Log_wl_nb10_parts_div}
\sum_{l=2}^{\infty} g_l\ln \mathbb{W}_l(1,0) \Biggr \rvert_{\rm div}
   = 
  \frac{2}{ 3\epsilon}\bigl[-22  + 12\tau _1^{\rm nb}  - (\tau _1^{\rm nb})^3 \bigr] \, ,
\end{equation}
while the finite part is given by
\begin{eqnarray}
\label{eq:Log_wl_nb10_parts_fin}
\sum_{l=2}^{\infty} g_l\ln \mathbb{W}_l(1,0) \Biggr \rvert_{\rm fin}
&&= \frac{191}{120}\ln\left[\frac{1-\tau_1^{\rm nb}}{\Lam(1+\tau_1^{\rm nb})}\right]  + \frac{1}{6}\left[-136 - 7 \tau_{1}^{nb} + 9 (\tau_{1}^{nb})^2 - 2 (\tau_{1}^{nb})^3 \right]   
\notag \\
&& +2[ 4 - (\tau_{1}^{nb})^2 ] \text{ln[$\Gamma $}\left(\tau _1^{\rm nb}+3\right)]- 4 \psi ^{(-3)}\left(\tau _1^{\rm nb}+3\right)  
\notag \\
&&
+ 4 \tau _1^{\rm nb} \psi ^{(-2)}\left(\tau _1^{\rm nb}+3\right)+ \ln( \pi^{12} A^{16})+\frac{391 \log (2)}{60}\notag \\
&&+\frac{191 \log (3)}{120}
-\frac{\zeta (3)}{2 \pi ^2}+ {\cal O}(N_c - N_{\rm nb}) \, .
\end{eqnarray}

In this expression $\gamma$ is Euler-Gamma, $A$ is the Glaisher constant and $\psi^{(n)}$ is Polygamma. This is the finite contribution computed at the no-boundary saddle with corrections coming in for $N_c$ away from the saddle points, which is summed up in ${\cal O}(N_c - N_{\rm nb})$. Again, at one-loop study these ${\cal O}(N_c - N_{\rm nb})$ terms do not contribute in the Picard-Lefschetz computation of the $N_c$-integration, and they become relevant beyond one loop.  

It should be highlighted that these divergent contribution mentioned in Eq. (\ref{eq:Log_wl_nb10_parts_div}) arises when $N_c$ is at the no-boundary saddles only.

%%%%%%%%%%%%%%%%%%%%%%%%%%%%%%%%%%%%%%%%%%%%%%%
\subsection{UV Divergence for $q_f<3/\Lam$}
\label{small_q0qf} 
%%%%%%%%%%%%%%%%%%%%%%%%%%%%%%%%%%%%%%%%%%%%%%%

Along with this, there is an additional UV divergence that arises for small $q_f<3/\Lam$. This arises only at the no-boundary saddle, and can be extracted by investigating the small $q_f$ behavior of the finite part of $\mathbb{W}_l(1,0)$ mentioned in Eq. (\ref{eq:Log_wl_nb10_parts_fin}). For $q_f<3/\Lambda$, only one of the two no-boundary saddles contribute in the path integral, as that is relevant. This contributing relevant saddle is $N_+^{\rm (nb)}$. At this saddle, in the limit $q_f\rightarrow 0$,
\begin{equation}
\label{eq:Wmathbb_qf_sm}
\mathbb{W}_l(1,0)\Br \rvert_{N_c \to N_+^{\rm (nb)}}
\xRightarrow[\text{$q_f\rightarrow 0$}] {} \,\,\,\,  
q_f^{(l+1)/2}\times\mathcal{W}_l
\end{equation}
where
\begin{equation}
\label{eq:calWl_qf_sm}
    \mathcal{W}_l=\frac{i}{2(l+1)} \, ,
\end{equation}
is a pure constant independent of $\Lambda$. The $i$ factor appearing in Eq. (\ref{eq:calWl_qf_sm}) is canceled by the $(-i)$ present in Eq. (\ref{eq:Del_struct_hl}) leading to a real quantity. Note that the power of $q_f$ is exactly equal and opposite to the power of $\bar{q}_0$ in $W_l(1,0)$ as mentioned in Eq. (\ref{eq:Wl_be_NoB}). So, the finite part in Eq. (\ref{eq:Log_wl_nb10_parts_fin}) in small $q_f$ limit becomes the following [note that the divergent part mentioned in Eq. (\ref{eq:Log_wl_nb10_parts_div} is finite in $q_f\rightarrow 0$ limit, whereas the finite part is not]:
\begin{equation}
\label{eq:mathBBW_qf_sm_fin}
\frac{i\hbar}{2}\sum_{l=2}^\infty g_l\ln \mathbb{W}_l(1,0)\Biggr \rvert_{\rm fin}
\xRightarrow[\text{$q_f\rightarrow 0$}]{N_c \to N_+^{\rm (nb)}}
i \hbar \Bl[
\underbrace{\frac{191}{240}\ln q_f}_{{\rm Log-div}} - \underbrace{\frac{11}{3\ep}}_{div}+\mathcal{C}_+ 
+\mathcal{O}(q_f)
+ {\cal O}(N_c - N_+^{\rm (nb)})\Br] \, , 
\end{equation}
where
\begin{equation}
\label{eq:C+_sm_qf}
   \mathcal{C}_+= \frac{\ln (4\pi^6) - 7}{3}  - \frac{\zeta(3)}{4\pi^2} .
\end{equation}
is a positive constant and $\mathcal{O}(q_f)$ is real for positive $q_f<3/\Lam$. Here again, the first terms on the rhs of Eq. (\ref{eq:mathBBW_qf_sm_fin}) is computed at the no-boundary saddle, with corrections coming in for $N_c$ away from the saddle point. These are summed up in ${\cal O}(N_c - N^+_{\rm nb})$, and they do not contribute to the one-loop study.

%%%%%%%%%%%%%%%%%%%%%%%%%%%%%%%%%%%%%%%%%%%%%%%
\subsection{Counterterms and lapse effective action}
\label{cont_log}
%%%%%%%%%%%%%%%%%%%%%%%%%%%%%%%%%%%%%%%%%%%%%%% 

To ensure the finiteness of the lapse action at the no-boundary saddles, one needs to suitably add counterterms overcoming these UV divergences. The suitable counterterm to be added is given by 
\begin{equation}
\label{eq:Nc_act_total_nb_ct}
\mathcal{A}_{\rm ct}(N_c)= 
\frac{117i \hbar}{80} \ln 
\bar{q}_0
-\frac{i\hbar}{ 3\epsilon}\bigl[-22  + 12\tau _1^{\rm nb}  
- (\tau _1^{\rm nb})^3 \bigr]
-  \frac{31i \hbar}{240}\ln q_f\,\, \theta\Bl(\frac{3}{\Lam}-q_f\Br)
\, ,
\end{equation}
where $\tau_1 = 1-i \Lam N_c /3$ and $\theta\bl(3/\Lam-q_f\br)$ refers to the step function which is introduced to pick only the small $q_f$ divergence. These counterterms precisely cancel the divergent terms arising from the computation of the one-loop determinant at the no-boundary saddle, ensuring the finiteness of the quantum corrected lapse action. We have purposely written the counterterms in this fashion to bring out the UV divergences coming from small $\bar{q}_0$, $l$ summation and small $q_f$. The first two terms arise for all $q_f$ at both the no-boundary saddles, while the later arises only for ``small" $q_f<3/\Lam$ at only the relevant no-boundary saddle ($N_+^{\rm nb}$). 

Once the quantum corrected action for the lapse is supplemented with the counterterm, the action no longer suffers from UV divergences, leading to a finite effective action for the lapse ($N_c$). The finite lapse Effective action is given by
\bea
\label{eq:Nc_act_total_nb_fin}
&&
{\cal A}_{\rm fin}(N_c)
= {\cal A}_0(N_c) + \hbar {\cal A}_1(N_c) 
+ {\cal A}_{\rm ct}(N_c) \, ,
\notag \\
&&
=S^{(\bar{q})}_{\rm grav}(N_c)
+\frac{i\hbar}{2}\ln\Delta_q(N_c) 
-\frac{2i\hbar}{3}\ln q_f
- \frac{31i \hbar}{240}
%\left[1-\theta\Bl(q_f-\frac{3}{\Lam}\Br) \right]
\theta\Bl(\frac{3}{\Lam}-q_f\Br)\ln q_f 
\notag \\
&&
+\frac{i\hbar}{2} \sum_{l=2}^{\infty} g_l\ln \mathbb{W}_l(1,0) \Biggr \rvert_{\rm fin}
= {\cal A}_{\rm fin}^{(0)}(N_c)
+ \hbar {\cal A}_{\rm fin}^{(1)}(N_c)
\eea
where ${\cal A}_0(N_c) + \hbar {\cal A}_1(N_c)$ and ${\cal A}_{\rm ct}(N_c)$ are mentioned in Eqs. (\ref{eq:Nc_act_total}) and (\ref{eq:Nc_act_total_nb_ct}) respectively.

%%%%%%%%%%%%%%%%%%%%%%%%%%%%%%%%%%%%%%%%%%%%%%%
\section{$N_c$ integration via Picard-Lefschetz}
\label{PL_Nc_sad} 
%%%%%%%%%%%%%%%%%%%%%%%%%%%%%%%%%%%%%%%%%%%%%%%

The finite lapse effective action action allows one to carry forward with the computation of the lapse-integral via Picard-Lefschetz methods. This leads to a finite transition amplitude, including the quantum corrections. The finite transition amplitude that we will focus our attention on is given by
\beq
\label{eq:grav_path_NC_form_fin}
G_{\rm fin}[{\rm Bd}_f, {\rm Bd}_i]
= \int_{-\infty}^{\infty} {\rm d} N_c \,\,
\exp \bl\{i {\cal A}_{\rm fin}(N_c)/\hbar \br\} \, ,
\eeq
where ${\cal A}_{\rm fin}(N_c)$ is finite lapse action defined in Eq. (\ref{eq:Nc_act_total_nb_fin}). This is computed by making use of Picard-Lefschetz and WKB methods. In this approximation, we can utilize the expression stated in Eq. (\ref{eq:LDordI}); one needs to compute the saddles of the one-loop finite lapse action ${\cal A}_{\rm fin}(N_c)$. Ignoring the contribution from $H_{ij}$ to the saddles, for the case of NBC, there will be two complex conjugate saddles as shown in Eq. (\ref{eq:grav_path_NC_form_fin}). Both these saddles are seen to be relevant as the steepest ascent emanating from them hits the original integration contour. The original integration contour can then be deformed to lie along the thimbles passing through these saddles, along which the integral becomes convergent.

When $H_{ij}$ contributions are incorporated, these saddles, along with the set of steepest ascent/descent lines, will be modified. The corrections to the background saddles are discussed in the Sec. \ref{sad_cor2}. But to leading order, the qualitative picture remains the same in the sense the relevance of the saddle remains unaffected. In particular, on inserting Eq. (\ref{eq:Nsg_1lp_form}) in the expression of ${\cal A}_{\rm fin}(N_c)$, it is seen we get the following expansion:
\bea
\label{eq:ANc_sad_exp_1st}
&&
{\cal A}_{\rm fin}(N_c)
= {\cal A}_{\rm fin}^{(0)}(N_\sg^{(q)})
+ \hbar {\cal A}_{\rm fin}^{(1)}(N_\sg^{(q)}) 
+ \frac{1}{2}(N_c - N_\sg)^2 \bl[
{\cal A}_{0, {\rm fin}}^{\prime\prime} (N_\sg^{(q)})
\notag \\
&&
+ \hbar {\cal A}_{0, {\rm fin}}^{\prime\prime\prime} (N_\sg^{(q)})\bl(
{\cal N}_\sg^{(q)} + {\cal N}_\sg^{(h)}
\br)
+ \hbar {\cal A}_{1, {\rm fin}}^{\prime\prime} (N_\sg^{(q)})
+ \cdots
\br] + \cdots \, .
\eea
Here $({}^\prime)$, denotes the derivative with respect to lapse $N_c$ computed at the saddle point $N_\sg$. In the saddle-point approximation and retaining terms relevant at one loop, we will get using Eqs. (\ref{eq:sumOthim}) and (\ref{eq:LDordI}):
\beq
\label{eq:LDordI_fin}
G_{\rm fin}[{\rm Bd}_f, {\rm Bd}_i]
= 
\sum_\sg \frac{e^{i\theta_\sg}}{\sqrt{\bl\vert {\cal A}_{0, {\rm fin}}^{\prime\prime} (N_\sg^{(q)}) \br\rvert}}
\times 
\exp\Bl[ 
\frac{i}{\hbar} {\cal A}_{\rm fin}(N_\sg^{(q)})
+ \cdots
\Br] \, ,
\eeq
This is schematically the expression of transition amplitude in saddle-point approximation up to one loop. In the following we will make use of this to compute the transition amplitude for various values of the scale factor.

%%%%%%%%%%%%%%%%%%%%%%%%%%%%%%%%%%%%%%%%%%%%%%%
\subsection{$q_f>3/\Lam$}
\label{PL_Nc_sad_qfLG} 
%%%%%%%%%%%%%%%%%%%%%%%%%%%%%%%%%%%%%%%%%%%%%%%

According to the Picard-Lefschetz theory, for $q_f \geq 3/\Lam$, both the no-boundary corrected saddles are relevant. In the case of imposing Robin and Dirichlet boundary conditions on the background at the initial and final boundary, respectively, we restrict the parameter $0 < x \leq 1$ to ensure that the corrected no-boundary saddle points remain relevant. Under the saddle-point approximation, the transition amplitude is computed using the Picard-Lefschetz method given as \cite{Narain:2022msz,DiTucci:2019dji}. Making use of the expression mentioned in Eq. (\ref{eq:LDordI_fin}) we get
\begin{equation}
\label{eq:Gtrans_qf>3/lam_gen}
\begin{split}
G_{\rm fin}[\rm Bd_f,\rm Bd_i]& \approx \frac{e^{i\theta_{s}^+ }e^{i\mathcal{A}_{\rm fin}(N^{\rm nb}_+)/\hbar} } 
{
\sqrt{\bl\vert {\cal A}_{0,{\rm fin}}^{\prime\prime} (N_{+}^{\rm nb}) \br\rvert}} 
+ \frac{e^{i\theta_{s}^- }e^{i\mathcal{A}_{\rm fin}(N^{\rm nb}_-)/\hbar} } {
\sqrt{\bl\vert {\cal A}_{0,{\rm fin}}^{\prime\prime} (N_{-}^{\rm nb}) \br\rvert}} \, .
\end{split}
\end{equation}
Here, $ \mathcal{A}_{\rm fin}(N^{\rm nb}_\pm) $ is the one-loop finite lapse action (obtained after adding the suitable counterterm) evaluated at the uncorrected no-boundary saddle points. Below, we have written the expressions for $\mathcal{A}_{\rm fin}(N^{\rm nb}_\pm)$, where $ N^{nb}_\pm$ are the zeroth order no-boundary saddles mentioned in Eq. (\ref{eq:NB_sad}), and $\theta^{\pm}_{s}$ for the corresponding saddles given in Eq. (\ref{eq:theta_sg_exp}) respectively. The one-loop corrected {\it finite} lapse action at the no-boundary saddles for $q_f>3/\Lam$ can be obtained from Eq. (\ref{eq:Nc_act_total_nb_fin}). This is given by
\begin{equation}
\begin{split}
\label{eq:fin_pt_Afin_nb_qfLG}
\mathcal{A}_{\rm fin}(N_c) \Br \rvert_{N_c\rightarrow N_{\rm nb}}&
=S^{(\bar{q})}_{\rm grav}(N_c)
+\frac{i\hbar}{2}\ln\Delta_q(N_c) -\frac{2i\hbar}{3}\ln q_f+\frac{i\hbar}{2} \sum_{l=2}^{\infty} g_l\ln \mathbb{W}_l(1,0) \Biggr \rvert_{\rm fin}.
\end{split}
\end{equation}
Note the appearance of the finite part in the last line of Eq. (\ref{eq:fin_pt_Afin_nb_qfLG}). A complete expression for the quantum corrected lapse action at the saddle point, for the case of Robin boundary condition at the initial hypersurface and the Dirichlet boundary condition at the final hypersurface, for $q_f>3/\Lam$ is mentioned in Eq. (\ref{eq:finite_re_img_act_nb_sad_RBC_1}) in Appendix \ref{app:qcLapse_ex}. The first term in Eq. (\ref{eq:fin_pt_Afin_nb_qfLG}) contain classical contributions coming from the scale factor, while the rest contains the one-loop ($\hbar$) corrections. The corresponding finite action for the Neumann boundary condition and Dirichlet boundary condition at the initial and final hyper-surfaces, respectively,  can be obtained by taking \(x \rightarrow 0\). It is worthwhile to note that as the two no-boundary saddles for $q_f>3/\Lambda$ satisfy $N^{\rm nb}_+ = - \bigl(N^{\rm nb}_{-}\bigr)^*$ (where ${}^*$ denotes complex conjugation), this implies that the real and imaginary part of the quantum corrected lapse action at the two no-boundary saddles should satisfy the following relations:
\begin{equation}
\label{identity}
Re[i\mathcal{A}_{\rm fin}(N^{\rm nb}_+)]=Re[i\mathcal{A}_{\rm fin}(N^{\rm nb}_-)]\,,
\hspace{5mm} \text{and}\hspace{5mm} 
Im[i\mathcal{A}_{\rm fin}(N^{\rm nb}_+)]=-Im[i\mathcal{A}_{\rm fin}(N^{\rm nb}_-)] \, .
\end{equation}
It is easy to see that the classical part satisfies the above condition. However, for the $\hbar$-corrected terms this is not easy to see, as it is difficult to separate the real and imaginary parts for arbitrary $q_f$. One can check it explicitly for large $q_f$, where one can write real and imaginary parts separately i.e., $ \mathcal{A}_{\rm fin}(N^{\rm nb}_\pm) = \mathcal{A}_{\rm fin}^{\rm real}(N^{\rm nb}_\pm) + i \mathcal{A}_{\rm fin}^{\rm imag}(N^{\rm nb}_\pm)$. The term proportional to $\hbar$ leads to the one-loop correction to the no-boundary wave function of the universe. This correction term is independent of $h_{_1}^{(l)}$ but depends on $q_f$. In the special case of vanishing metric perturbations at the two hypersurfaces (putting $h_1^{(l)}=0$ at the final hypersurface), the quantum corrections give a nonvanishing contribution which arising due to virtual gravitons. For generic $q_f$, this contribution is quite complicated. However, in the large $q_f$ limit, it simplifies to,
\begin{equation}
\label{eq:Afin_ReIm_LG_qf}
   \begin{split}
        \mathcal{A}_{\rm fin}^{\rm real}(N^{\rm nb}_\pm) \Br\rvert_{\rm one-loop} \approx & \pm \hbar\left[\frac{1}{18}(1+3\ln \bold{q_f})\bold{q_f}^{3/2}+\frac{7}{4}\bold{q_f}^{1/2}\ln \bold{q_f}+\cdots\right] \, ,\\
         \mathcal{A}_{\rm fin}^{\rm imag}(N^{\rm nb}_\pm)\Br\rvert_{\rm one-loop} \approx & \hbar\left[-\frac{\pi}{6}\bold{q_f}^{3/2}+\frac{1}{2}\bold{q_f}+ \cdots\right]\, ,
   \end{split} 
\end{equation}
where we have defined $\bold{q_f}=\Lam q_f/3$. The real part of subsequent subleading terms in the expression falls as $\bold{q_f}$, whereas the imaginary part falls as $\sqrt{\bold{q_f}}$, where $\ln \bold{q_f}$ is taken to be $\mathcal{O}(1)$.
%\textcolor{red}{$\cdots$ include subsubleading terms in $\bold{q_f}$}.
For large $q_f$, one can explicitly verify that the quantum corrected lapse action satisfies Eq. (\ref{identity}).  

It is crucial to note that the terms proportional $\hbar$, indicate a growing behavior for large $q_f$. This is a typical characteristic of the de Sitter geometry where infrared divergences are known to show up in quantum corrections and indicate a secular growth of terms \cite{Akhmedov:2013vka, Akhmedov:2019cfd, Akhmedov:2024npw, Miao:2024shs}.

%%%%%%%%%%%%%%%%%%%%%%%%%%%%%%%%%%%%%%%%%%%%%%%
\subsection{$q_f<3/\Lam$}
\label{PL_Nc_sad_qf_sm} 
%%%%%%%%%%%%%%%%%%%%%%%%%%%%%%%%%%%%%%%%%%%%%%%

Now consider the case when $q_f<3/\Lambda$. In this case for $q_f<3/\Lam$, we have only one relevant no-boundary saddle i.e., $N_{+}^{\rm nb}$, unlike the $q_f \geq 3/\Lam$ case. The corresponding transition amplitude, in this situation, can be computed using the expression mentioned in Eq. (\ref{eq:LDordI_fin}). This is given by \cite{Narain:2022msz,DiTucci:2019dji}
\begin{equation}
\label{eq:Gfin_qf_less_3/lam}
G_{\rm fin}[\rm Bd_f,\rm Bd_i] \approx \frac{e^{i\theta_{s}^+ }e^{i\mathcal{A}_{\rm fin}(N^{\rm nb}_+)/\hbar} } {
\sqrt{\bl\vert {\cal A}_{0,{\rm fin}}^{\prime\prime} (N_{+}^{\rm nb}) \br\rvert}} \, .
\end{equation}
where the complete expression for $\mathcal{A}_{\rm fin}(N^{\rm nb}_+)$ for $q_f<3/\Lam$ is given in Eq. (\ref{eq:finite_re_img_act_nb_sad_RBC_2}) in Appendix \ref{app:qcLapse_ex}. However, this acquires a simplified structure in the small $q_f$ limit. It turns out that in this limit, the finite lapse action at the relevant saddle $N_+^{\rm nb}$ behaves in the following manner:
\begin{equation}
\label{eq:Afin_qf_sm_N+}
\mathcal{A}_{\rm fin}(N_+^{\rm nb})\Br\rvert_{\rm one-loop}
\sim i\hbar\bl[\mathcal{C}_+
+\mathcal{O}\left(\bold{q_f}\right)\br] \, ,
\end{equation}
where ${\cal C}_+$ is mentioned in Eq. (\ref{eq:C+_sm_qf}) leading to a finite quantity for small $q_f$ without further UV divergences. It is interesting to note that the quantum corrected lapse action is purely imaginary, resulting in the one-loop correction being purely real for $\bold{q_f}<1$. We also note that for $\bold{q_f}<1$, at the relevant no-boundary saddles, geometry looks like a sphere ($S^4$). Our finding is in agreement with the previous conclusion where the authors have shown the one-loop correction to the gravity effective action on a sphere is real \cite{Mazur:1989ch}.

The special case of $q_f=3/\Lam$, is discussed in Eq. (\ref{eq:finite_re_img_act_nb_sad_RBC_3}) in Appendix \ref{app:qcLapse_ex}, where the exact expression is given.

%%%%%%%%%%%%%%%%%%%%%%%%%%%%%%%%%%%%%%%%%%%%%%%
\subsection{Degeneracy lift}
\label{subs:deg_lift}
%%%%%%%%%%%%%%%%%%%%%%%%%%%%%%%%%%%%%%%%%%%%%%%
%

\begin{figure}[h]
\centering
\includegraphics[scale=0.5]{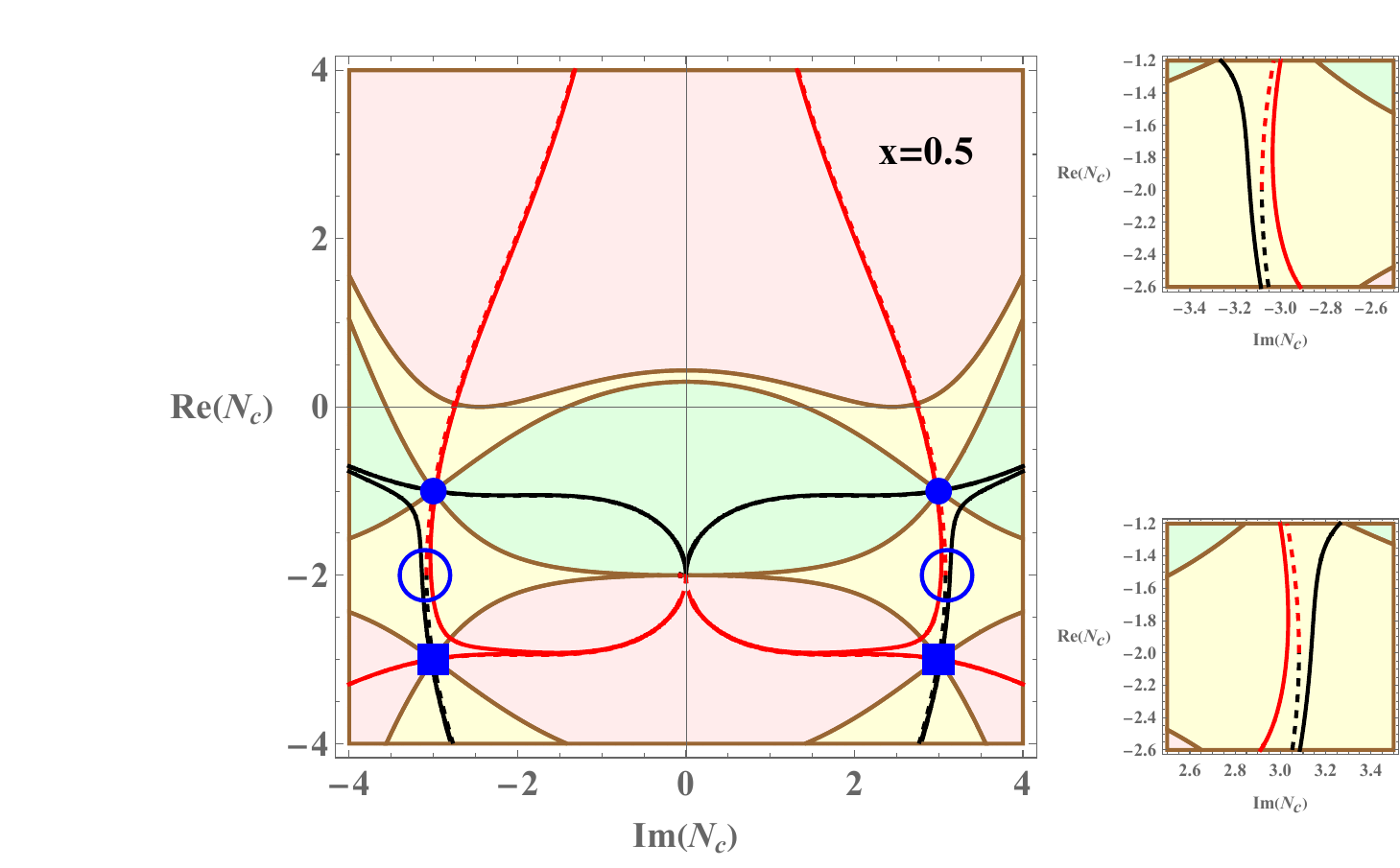}
\caption{We analyze the no-boundary universe scenario, considering cases both with and without quantum corrections. When incorporating quantum corrections, we use the parameter values: $P_i = - 3i, \Lam = 3$, $q_{f} = 10$, $x = 1/2$ and $\hbar = 1$. Here, dotted flow-lines represent without any quantum corrections, while solid lines incorporate quantum corrections from scale factor. The red lines denote the steepest ascent paths, and the black lines indicate the steepest descent paths. The blue dots are the no-boundary saddles: $N_{+}^{(nb)} = 3-i$ and $N_{-}^{(nb)} = -3-i$, whereas the blue squares are the non-no boundary saddles without any loop corrections: $N_{+}^{\cancel{(nb)}} = 3-3i$ and $N_{-}^{\cancel{(nb)}} = -3-3i$. The light green region represents the allowed region, where $h < h(N_{\sg})$ for all values of $\sg$. The light yellow region corresponds to intermediate values of $h$. The light pink is the forbidden region, has $h > h(N_{\sg})$ for all values of $\sg$. These regions are constructed without incorporating any loop corrections.}
    \label{Fig:stokes_0.5}
\end{figure}

An interesting feature of incorporating quantum corrections in the computation of transition amplitude via Picard-Lefschetz is that it helps in lifting degeneracies from the system, which arise due to the presence of symmetries in the theory. In past studies involving various boundary choices but without quantum corrections \cite{Feldbrugge:2017kzv, Narain:2021bff, Narain:2022msz, Ailiga:2023wzl}, quite a few instances involving degenerate scenarios have been witnessed. One of the most common degenerate situations is the overlapping of the steepest ascent of one saddle with the steepest descent of another saddle. In previous works, a suggestion to uplift this degeneracy is to include a small perturbation in the theory. This process, although it works, is a bit {\it ad hoc} in the sense there is no systematic procedure for inserting the perturbation in the lapse action, and the results may depend on the perturbation inserted. 

It is seen that such issues can be resolved when quantum corrections are taken into account. In particular, when the contributions coming from the quantum fluctuations of scale factor are included, then it systematically provides a perturbation thereby lifting the degeneracy cleanly. Here, one does not need to introduce an {\it ad hoc} perturbation to the lapse action as a external input; the quantum fluctuations of the field provides the necessary perturbation needed for breaking the degeneracy. This can be seen from the example depicted in the figure \ref{Fig:stokes_0.5}. Here, from the figure, it can be noticed that without quantum corrections, the steepest ascent of one saddle overlaps with the steepest descent from another. However, when the quantum corrections are included, then the flow lines no longer overlap. This helps in finding the relevant saddles correctly, as the flow lines only intersect at the saddle points. 

A full detailed study of degeneracies and how they get resolved via quantum corrections is beyond the scope of this work and will be treated in future publication. Here, we just highlight one example of how the quantum corrections cleanly offer a solution in resolving the degeneracies.

%%%%%%%%%%%%%%%%%%%%%%%%%%%%%%%%%%%%%%%%%%%%%%%
\section{Quantum corrected transition amplitude}
\label{subs:quant_TA}
%%%%%%%%%%%%%%%%%%%%%%%%%%%%%%%%%%%%%%%%%%%%%%%

Once the quantum corrected lapse action is computed for various $q_f$ along with the set of UV divergences, it is crucial to ask about the transition amplitude in the no-boundary proposal of universe, where quantum fluctuations from the fields have been incorporated. In particular, an interesting object to study here is the transition amplitude with vanishing perturbations ($h_{ij} =0$) at the two hypersurfaces but retaining their contribution of $h_{ij}$ virtually. In this situation, it contributes with $h_{ij}$ propagating in the loops only. A naive expectation in any standard flat spacetime QFT study is that the quantum contribution via loops is suppressed compared to the leading behavior. However, in de Sitter spacetime, it has been seen that such expectations are challenged where secular growth of quantum corrections happens in the deep infrared \cite{ Akhmedov:2013vka, Akhmedov:2019cfd, Akhmedov:2024npw, Miao:2024shs, Starobinsky:1994bd}. This could signal a breakdown of the perturbation theory. While there is no clear agreement over how this can be resolved, several proposals debating on it exist. Some of these involve the resummation of the secularly growing terms \cite{Baumgart:2019clc, Honda:2023unh, Cespedes:2023aal, Starobinsky:1994bd, Huenupi:2024ztu, Kamenshchik:2024ybm}, incorporating an IR-cutoff \cite{Huenupi:2024ksc} and regulating the theory \cite{Xue:2011hm}, or incorporating nonlocality \cite{Narain:2018rif} which also acts as a kind of IR regulator. A notable feature of the secular growth of correlation functions is in massless theories, while massive theories tend to yield IR-finite results.
Recent studies (Refs.\cite{Kamenshchik:2024ybm, Kamenshchik:2025ses}) of free massive scalar field on the de Sitter space show that correlation functions remain IR-finite in the massive theory, meanwhile massless case shows secular growth behavior.

In the present study of computing the transition amplitude of the no-boundary Universe by analyzing the gravitational path integral directly in the Lorentzian signature, we notice that a similar growth coming from the quantum fluctuations of the metric perturbations $h_{ij}$. These secularly growing terms dominate over the leading contribution in the infrared, similar to the one noticed in previous works \cite{Akhmedov:2013vka, Akhmedov:2019cfd, Akhmedov:2024npw, Miao:2024shs}. While such IR behavior is expected in a sense as one is studying quantum metric fluctuations on de Sitter, in the context of Lorentzian quantum cosmology, they have been investigated for the first time using Picard-Lefschetz formalism. 

In the special case when the metric perturbations vanish ($h_{ij}=0$) at the two boundary hypersurfaces, implies that each mode of the $h_{ij}$-field in the mode decomposition vanishes: $h_1^{(l)}=0$. In this simple setting, the metric perturbations only propagate virtually in the loop and vanish at the boundary. For such a situation, the correction to the orientation of thimbles vanishes, i.e., $\delta \theta_s=0$.

%%%%%%%%%%%%%%%%%%%%%%%%%%%%%%%%%%%%%%%%%%%%%%%
\subsection{$q_f>3/\Lam$}
\label{subs:GTR_qf>3/Lam_LG} 
%%%%%%%%%%%%%%%%%%%%%%%%%%%%%%%%%%%%%%%%%%%%%%%

\begin{figure}
    \centering
    \includegraphics[
    width=0.779\textwidth]
    {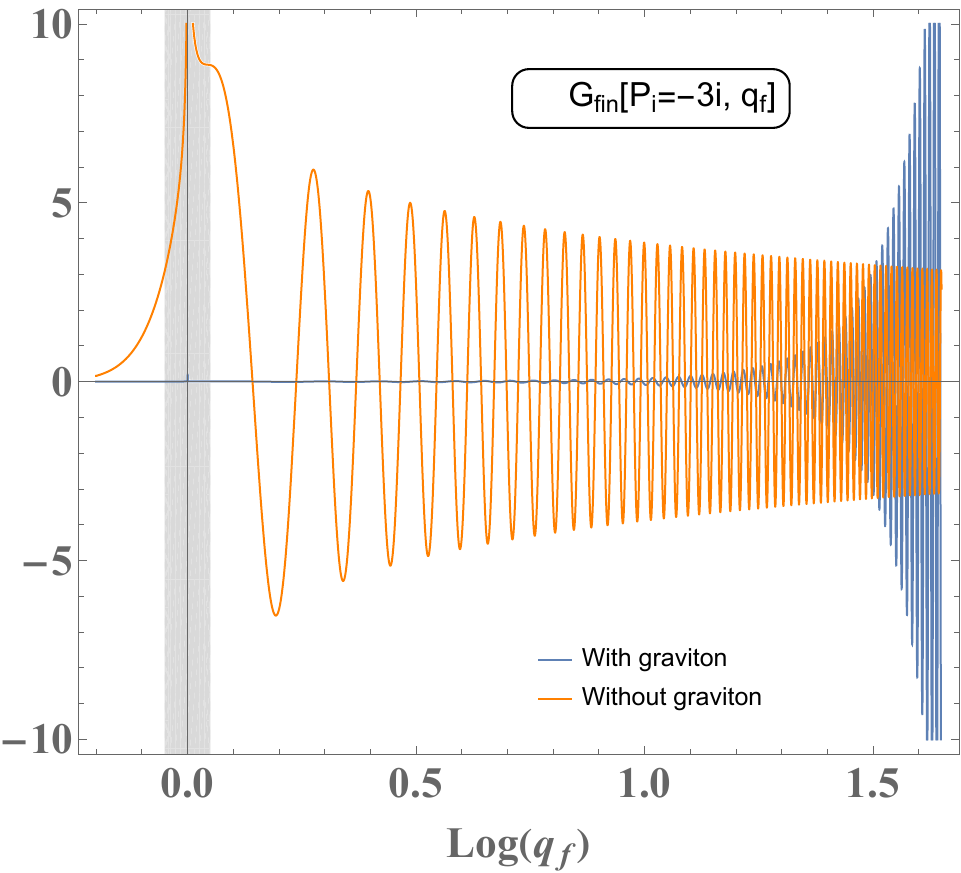}
\caption{The plot shows the transition amplitude $G_{\rm fin}
\bl[P_i=-3i,q_f \br]$ vs $\log(q_f)$ (where $a_f=\sqrt{q_f}$ is the final scale factor of the Universe) in the no-boundary universe, with the initial condition imposed to be Robin choice, while the final boundary condition is Dirichlet type. We have taken $\Lam=3, \hbar = 1/2\pi, 8\pi G = 1, V_3 = 2\pi^2$ and $k=1$ for this plot. To bring out the features elegantly, we have scaled down the $y$-axis by $10^{16}$ and plotted the $x$ axis in log scale. The metric fluctuations $h_{ij}$ vanishes at the two boundaries. The orange curve depicts the transition amplitude in the absence of any metric fluctuations ($h_{ij}$ field). As the Universe expands, the amplitude decays monotonically for $q_f>3/\Lam$. The blue curve depicts the transition amplitude with the metric fluctuation, where $h_{ij}$ contributes virtually through the loops. This amplitude shows a growing behavior for large $q_f>3/\Lam$, leading to infrared divergence coming due to the graviton loops. The blue curve increases monotonically; however, for small $q_f$, it is $10^{3}$ order smaller than the orange curve (without graviton), which leads to a very small amplitude with our scaling of $y$ axis.
The deviation is seen as a breakdown of the WKB approximation near $q_f=3/\Lam$, which jumps near the origin for both curves, depicted by a light gray region. 
}
    \label{fig:IRbehave}
\end{figure}

For $q_f>\frac{3}{\Lambda}$, we have $\theta_s^\pm$ given by the following: 
\begin{equation}
\label{eq:theta_s_qf>3/lam}
    \theta_s^+(x)=\frac{k\pi}{2}-\frac{\pi}{4}-\frac{1}{2}tan^{-1}\left[\frac{x\sqrt{q_f\Lambda-3}}{(1-x)\sqrt{3}}\right],\hspace{5mm} k\in \mathbb{Z}_{odd}
\end{equation}
with $0\leq\theta_s^+(x)\leq\pi/2$, $\theta_s^-(x)=-\theta_s^+(x)$. For Neumann BC case ($x=0$), $\theta_s^+$ is $\pi/4$ independent of $q_f$ and $\Lambda$. For Robin BC with $0<x<1$, one can numerically find $\theta_s^\pm (x)$ using Eq. (\ref{eq:theta_s_qf>3/lam}). With this inclusion, the transition amplitude mentioned in Eq. (\ref{eq:Gtrans_qf>3/lam_gen}) in the WKB approximation becomes the following: 
\begin{equation}
\label{eq:Gtrans_qf>3/Lam}
G_{\rm fin}[q_f>3/\Lambda,P_i=-3i]
\approx 
\frac{e^{i\theta_s^+(x)}\exp\left(i\mathcal{A}_{\rm fin}(N_+^{\rm nb})/\hbar\right)}
{
\sqrt{\bl\vert {\cal A}_{0,{\rm fin}}^{\prime\prime} (N_{+}^{\rm nb}) \br\rvert}}
+\frac{e^{-i\theta_s^+(x)}\exp\left(i\mathcal{A}_{\rm fin}(N_-^{\rm nb})/\hbar\right)}
{
\sqrt{\bl\vert {\cal A}_{0,{\rm fin}}^{\prime\prime} (N_{-}^{\rm nb}) \br\rvert}} \, .
\end{equation}
From Eq. (\ref{identity}), it is easy to see that the transition amplitude acquires the following structure:
\bea
\label{amplitude}
&&
G_{\rm fin}[q_f>3/\Lambda,P_i=-3i]
\approx 
\frac{2\exp\left(-Im[\mathcal{A}_{\rm fin}(N_+^{\rm nb})]/\hbar\right)}
{\bl\{(Re[{\cal A}_{0,{\rm fin}}^{\prime\prime} (N_{+}^{\rm nb})])^2
+(Im[{\cal A}_{0,{\rm fin}}^{\prime\prime} (N_{+}^{\rm nb})])^2 \br\}^{1/4}}
\notag \\
&&
\times 
\cos\left(Re[\mathcal{A}_{\rm fin}(N_+^{\rm nb})]/\hbar+\theta_s^+(x)\right) 
\notag \\
&&
= \frac{2\exp\left(-Im[{\cal A}_{\rm fin}^{(0)}(N_+^{\rm nb})]/\hbar\right)}
{\bl\{(Re[{\cal A}_{0,{\rm fin}}^{\prime\prime} (N_{+}^{\rm nb})])^2
+(Im[{\cal A}_{0,{\rm fin}}^{\prime\prime} (N_{+}^{\rm nb})])^2 \br\}^{1/4}}
\exp\left(-Im[{\cal A}_{\rm fin}^{(1)}(N_+^{\rm nb})]\right)
\notag \\
&&
\times 
\cos\Bl(\frac{1}{\hbar}Re[{\cal A}_{\rm fin}^{(0)}(N_+^{\rm nb})]
+ Re[{\cal A}_{\rm fin}^{(1)}(N_+^{\rm nb})]
+ \theta_s^+(x)\Br)
\, ,
\eea
where in getting the last equality we have used Eq. (\ref{eq:Nc_act_total_nb_fin}). This is the one-loop finite no-boundary transition amplitude for the case of $q_f>3/\Lam$. It is a real quantity as the contributions from the two relevant saddles combine to give a real amplitude. The quantum corrected finite lapse action at the saddle point for $q_f>3/\Lam$ is mentioned in Eq. (\ref{eq:finite_re_img_act_nb_sad_RBC_1}) in Appendix \ref{app:qcLapse_ex}. The one-loop graviton contribution enters at two places in Eq. (\ref{amplitude}): one as an exponential given by $\exp\left(-Im[{\cal A}_{\rm fin}^{(1)}(N_+^{\rm nb})]\right)$ and other inside the cosine function. In the early stages of the Universe, when $q_f$ is not very large, this exponential gives a suppression, but as the Universe grows in size, this term starts to grow exponentially. This is more cleanly appreciated by writing $q_f = 3(1+\de)/\Lam$, and doing a small/large-$\de$ expansion of $\exp\left(-Im[{\cal A}_{\rm fin}^{(1)}(N_+^{\rm nb})]\right)$:
\bea
\label{eq:Exp_Imcal_Afin1}
&&
\exp\left(-Im[{\cal A}_{\rm fin}^{(1)}(N_+^{\rm nb})]\right) \biggr \rvert_{\de \to0}
\sim \frac{
\exp[\zeta(3)/4\pi^2 
+ 29/12 
+2 \psi^{(-3)}(3)]}
{128 \times 3^{139/240} \times 2^{31/120}\pi ^6 A^8}
+ {\cal O}(\de)
\notag \\
&&
\exp\left(-Im[{\cal A}_{\rm fin}^{(1)}(N_+^{\rm nb})]\right) \biggr \rvert_{\de \to \infty}
\sim e^{
\pi\de^{3/2}/6
- \de/2 + 2\pi \sqrt{\de}
- (11/6) \ln \de + \cdots}
[1 +  {\cal O}(1/\de)] \, .
\eea
From this, it is easy to see that for small $\de$, one gets a suppressing contribution leading to a very small amplitude while giving an exponential growth for large $\de$, indicating infrared divergence in the graviton contribution. This behavior is depicted in Fig. \ref{fig:IRbehave}. 

Turning our focus back to the transition amplitude expression mentioned in Eq. (\ref{amplitude}), it is seen that in general it is quite involved. However, for large $q_f \to \infty$, the expression gets reasonably simplified. The denominator appearing in Eq. (\ref{amplitude}) can be further re-expressed as follows:
\bea
\label{oneloopampltitude}
\bl\{(Re[{\cal A}_{0,{\rm fin}}^{\prime\prime} (N_{+}^{\rm nb})])^2
&&
+(Im[{\cal A}_{0,{\rm fin}}^{\prime\prime} (N_{+}^{\rm nb})])^2 \br\}^{1/4} %\Br \rvert_{q_f \to \infty}
\notag \\
&&
=
\left[\frac{\Lambda  V_3}{4 \pi  G }\right]^{1/2}\left(\frac{3}{\Lambda  q_f-3}+\frac{x^2}{(x-1)^2}\right)^{-1/4}
\, .
\eea
This is applicable only for $q_f>3/\Lam$ and $0\leq x <1$. When this is combined with the large $q_f$ behavior of ${\cal A}_{\rm fin}(N_c)$ using Eq. (\ref{eq:Afin_ReIm_LG_qf}), it will lead to the large $q_f$ behavior of the finite transition amplitude for no-boundary universe at one loop. In the large $q_f$ limit, one can obtain a simplified expression for the transition amplitude. This is given by
\bea
\label{eq:Gtr_qf>3/lam_LG}
&&
G_{\rm fin}[q_f>3/\Lambda,P_i=-3i]\Biggr \rvert_{q_f\to \infty}\sim \,\frac{1}{\mathcal{F}_\infty}\times \exp\left(\frac{3V_3}{4\pi G\hbar\Lambda}+\frac{\pi}{6}\bold{q_f}^{3/2}-\frac{1}{2}\bold{q_f}+ \cdots+\mathcal{O}(\hbar)\right)
\notag \\
&&
\times \cos\left[\frac{3V_3}{4\pi G\hbar\Lambda}\bold{q_f}^{3/2}-\frac{1}{18}(1+3\ln \bold{q_f})\bold{q_f}^{3/2}-\frac{7}{4}\bold{q_f}^{1/2}\ln \bold{q_f}+\cdots+\mathcal{O}(\hbar)-\bar{\theta}_s^+\right] \, ,
\eea
where $\mathcal{F}_\infty$ is the contribution coming from the denominator in Eq. \ref{oneloopampltitude}:
\begin{equation}
\label{eq:calF_form_LGqf}
    \begin{split}
        \mathcal{F}_\infty &\sim \bold{q_f}^{1/4}\hspace{5mm}\text{for $x=0$} \, ,\\
        &\sim \text{const}\hspace{5mm}\text{for $0<x<1$} \, ,
    \end{split}
\end{equation}
and $\bar{\theta}_s^+=\pi/4$ for $x=0$, and $\bar{\theta}_s^+=0$ for $0<x<1$ in the large $\bold{q_f}$ limit. The growing behavior in the exponential is consistent with the IR growth seen in the graviton sector in Eq. (\ref{eq:Exp_Imcal_Afin1}). For generic  $q_f>3/\Lam$, the real and imaginary parts in $\mathcal{A}_{\rm fin}(N_+^{\rm nb})$ mix in a nontrivial manner, forbidding one to obtain such a simplified analytic expression. For generic $q_f$, we have however mentioned $\mathcal{A}_{\rm fin}(N_+^{\rm nb})$ in \ref{eq:finite_re_img_act_nb_sad_RBC_1} in Appendix \ref{app:qcLapse_ex}, in which case the transition amplitude can only be analyzed numerically. 

It is worth pointing out that the leading large $\bold{q_f}$ behavior coming from virtual graviton is independent of $x$, which is also expected as it includes the contribution at the no-boundary saddles only. It is interesting to note that for large $q_f$, the amplitude shows a growing behavior coming due to the quantum fluctuations of the $h_{ij}$ field, overpowering the leading $1/\hbar$ contribution. This infrared behavior is depicted in Fig.\ref{fig:IRbehave}, where the transition amplitude with and without graviton contribution is compared. The former grows while the later decays as the Universe expands. This infrared divergent behavior does not seem surprising in a sense, as we are studying QFT on de Sitter, which by now is known to experience infrared divergences and secular growth of quantum corrected terms \cite{Akhmedov:2013vka, Akhmedov:2019cfd, Akhmedov:2024npw, Miao:2024shs}. The leading growth term of $\sim q_f^{3/2}$ was also noticed in Euclidean quantum cosmology studies in \cite{Barvinsky:1992dz}.

%%%%%%%%%%%%%%%%%%%%%%%%%%%%%%%%%%%%%%%%%%%%%%%
\subsection{$q_f<3/\Lam$}
\label{subs:GTR_qf<3/Lam_LG} 
%%%%%%%%%%%%%%%%%%%%%%%%%%%%%%%%%%%%%%%%%%%%%%%

For $q_f<3/\Lambda$, the transition amplitude at the one loop is given by ($\theta_s^+=0$ for all $x$, $q_f$, and $\Lambda$)
\bea
\label{eq:GTfin_qf<3/lam}
&&
G_{\rm fin}[q_f<3/\Lambda,P_i=-3i]
\approx
\frac{\exp\left(i\mathcal{A}_{\rm fin}(N_+^{\rm nb})/\hbar\right)}
{
\sqrt{\bl\vert {\cal A}_{0,{\rm fin}}^{\prime\prime} (N_{+}^{\rm nb}) \br\rvert}} \, ,
\notag\\
&&
\approx 
\exp\left(-Im[{\cal A}_{\rm fin}(N_+^{\rm nb})]/\hbar\right)
\bl\{(Re[{\cal A}_{0,{\rm fin}}^{\prime\prime} (N_{+}^{\rm nb})])^2
+(Im[{\cal A}_{0,{\rm fin}}^{\prime\prime} (N_{+}^{\rm nb})])^2 \br\}^{-1/4} \, .
\eea
As $\mathcal{A}_{\rm fin}(N_+^{\rm nb})$ is purely imaginary for $q_f<3/\Lambda$, leading to a real amplitude, it agrees with previously known results from the literature. The expression for $\mathcal{A}_{\rm fin}(N_+^{\rm nb})$ is given in Eq.  \ref{eq:finite_re_img_act_nb_sad_RBC_3}. The last factor in the last line of Eq. (\ref{eq:GTfin_qf<3/lam}) is given by
\begin{equation}
\label{eq:WKB_fac_qf<3/lam}
\Bl\{(Re[{\cal A}_{0,{\rm fin}}^{\prime\prime} (N_{+}^{\rm nb})])^2
+(Im[{\cal A}_{0,{\rm fin}}^{\prime\prime} (N_{+}^{\rm nb})])^2 \Br\}^{1/4}
= \left[\frac{\Lambda V_3}{4\pi G}\right]^{1/2}\left(\frac{x}{1-x}+\frac{\sqrt{3}}{\sqrt{3-q_f\Lambda}}\right)^{-1/2} \, .
\end{equation}
In the small $q_f$ limit, it takes the following form:
\begin{equation}
\label{eq:GTr_fin_qf<3/lam_small}
G_{\rm fin}[q_f<3/\Lambda,P_i=-3i]\Biggr \rvert_{q_f\to0}
\approx 
\frac{1}{\mathcal{F}_0}\times \exp \Bl[
\frac{1}{\hbar}\Bl\{\frac{9 V_3 \bold{q_f}}{8 \pi  G\Lambda}
+\mathcal{O}(\bold{q}_f^2)\Br\}
-\mathcal{C}_+
+\mathcal{O}_0(\bold{q_f})+\cdots\Br],
\end{equation}
where ${\cal C}_+$ is mentioned in Eq. (\ref{eq:C+_sm_qf}) and 
\begin{equation}
\label{eq:calF_qf<3/lam_small}
   \mathcal{F}_0=\left[(1-x)+\mathcal{O}(\bold{q_f})\right]^{1/2}. 
\end{equation}
It is interesting to note that in the limit $\bold{q_f}\rightarrow 0$ limit, the leading contribution (constant independent of $\bold{q_f}$) comes from the quantum fluctuations of the fields, as the ($\frac{1}{\hbar}$) term is the order of $\bold{q_f}$ and higher. For small ${\bf q}_f$ these $1/\hbar$ contributions vanish, leaving behind the dominant contribution coming from the quantum fluctuations.

%%%%%%%%%%%%%%%%%%%%%%%%%%%%%%%%%%%%%%%%%%%%%%%
\subsection{$q_f=3/\Lam$}
\label{subs:GTR_qf=3/Lam_LG} 
%%%%%%%%%%%%%%%%%%%%%%%%%%%%%%%%%%%%%%%%%%%%%%%
For $q_f=3/\Lambda$, both the no-boundary saddles merge leading to a degenerate situation where $N_+^{(\rm nb)}=N_-^{(\rm nb)}$. In this case, WKB analysis fails, and we needs to go beyond WKB in order to do the $N_c$ integration needed for the computation of the transition amplitude in the no-boundary universe. This implies
\begin{equation}
\label{eq:Afin_exp_3rd_order}
\mathcal{A}_{\rm fin}(N_c)
\approx 
\mathcal{A}_{\rm fin}(N_+^{(\rm nb)})
+\frac{1}{6}\mathcal{A}_{\rm fin}^{\prime\prime\prime}(N_+^{\rm (nb)})(N_c-N_+^{(\rm nb)})^3+\cdots \, ,
\end{equation}
where the first and second derivative term vanishes as $N_+^{(\rm nb)}$ is a degenerate saddle. The transition amplitude is given by ($\delta N_c=(N_c-N_+^{(\rm nb)})$):
\bea
\label{eq:Gtr_fin_qf=3/lam_deg}
G_{\rm fin}[q_f=3/\Lambda,P_i=-3i]
\approx 
e^{\mathcal{A}_{\rm fin}(N_+^{(\rm nb)})/\hbar}\Bl[
&&
\int_{-\infty}^0d[\delta N_c]e^{i\mathcal{A}_{0,\rm fin}^{\prime\prime\prime}(N_+^{\rm (nb)})\delta N_c^3/6\hbar}
\notag \\
&&
+\int_{0}^{\infty}d[\delta N_c]e^{i\mathcal{A}_{0,\rm fin}^{\prime\prime\prime}(N_+^{\rm (nb)})\delta N_c^3/6\hbar}\Br]_{q_f=3/\Lambda} \, ,
\eea
Here, the two integrals need to be performed separately and summed up. This is to be done as not only the saddles after merging become entirely imaginary in nature, but also the $\de N_c$ integral whose integrand is of form $\exp\{k \lvert \delta N_c \rvert^3\}$. On writing, $\delta N_c=n \exp(i\theta_s)$, we have
\bea
\label{eq:Gfin_tra_qf=3/lam_deg1}
&&
G_{\rm fin}
\Bl[q_f=3/\Lambda,P_i=-3i\Br]
\approx e^{\mathcal{A}_{\rm fin}(N_+^{(\rm nb)})/\hbar}
\Bl[e^{-i\pi/6}\int_{-\infty}^0 {\rm d}n \,e^{|\mathcal{A}_{0,\rm fin}^{\prime\prime\prime}(N_+^{\rm (nb)})|n^3/6\hbar}
\notag \\
&&
+e^{i\pi/6}\int_{0}^{\infty} {\rm d}n\,
e^{-|\mathcal{A}_{0,\rm fin}^{\prime\prime\prime}(N_+^{\rm (nb)})|n^3/6\hbar}\Br]
\approx \sqrt{3}e^{\mathcal{A}_{\rm fin}(N_+^{(\rm nb)})/\hbar}\int_{0}^{\infty}dn\,e^{-|\mathcal{A}_{0,\rm fin}^{\prime\prime\prime}(N_+^{\rm (nb)})|n^3/6\hbar}\Biggr \rvert_{q_f=3/\Lambda} \, ,
\notag \\
&&
\approx e^{\mathcal{A}_{\rm fin}(N_+^{(\rm nb)})/\hbar} \frac{\sqrt{3}\,6^{1/3}\Gamma(4/3)}{|\mathcal{A}_{0,\rm fin}^{\prime\prime\prime}(N_+^{\rm (nb)})|^{1/3}}\Biggr \rvert_{q_f=3/\Lambda}\, .
\eea
As this needs to be computed up to one loop, the denominator in the last line of Eq. (\ref{eq:Gfin_tra_qf=3/lam_deg1}) can be suitably evaluated leading to $|\mathcal{A}_{0,\rm fin}^{\prime\prime\prime}(N_+^{\rm (nb)})|=\Lambda ^2 V_3/(12 \pi  G)$. This will lead to,
\begin{equation}
\label{eq:GFin_tr_qf=3/lam_deg2}
G_{\rm fin}
\Bl[q_f=3/\Lambda,P_i=-3i \Br]
\approx 6 (3\pi^2)^{1/6} \Gamma \left(\frac{4}{3}\right) \sqrt[3]{\frac{G}{\Lambda ^2 V_3}}\exp\left(-\mathcal{C}/\hbar\right) \, ,
\end{equation}
where
\begin{equation}
\label{eq:C_Gfin_qf=3/lam_deg}
\begin{split}
\mathcal{C}=-\frac{3V_3}{4\pi G \Lambda}
& +\frac{\hbar}{2} \Bl[
\ln(1 - x)
-\frac{\zeta (3)}{2 \pi ^2} + \ln( \pi^3 A^{4})
-\frac{31}{120}\ln{\Lambda}  - \frac{149}{60}\ln{2} + \frac{37}{20}\ln{3} + \frac{61}{6} 
\Br]\, .
\end{split}
\end{equation}
%

%%%%%
The amplitude is real, as expected. Note, this holds only for $0 \leq x<1$.

%%%%%%%%%%%%%%%%%%%%%%%%%%%%%%%%%%%%%%%%%%%%%%%
%%%%%%%%%%%%%%%%%%%%%%%%%%%%%%%%%%%%%%%%%%%%%%%
\section{$h_{ij}\neq 0$ at the final hypersurface}
\label{sec:hij_neq_0}
%%%%%%%%%%%%%%%%%%%%%%%%%%%%%%%%%%%%%%%%%%%%%%%

For completeness, we consider the situation when the metric perturbation $h_{ij}\neq 0$ at the final hypersurface. In this section, we do not study the one-loop effects. Our main focus in this section will to investigate the correction to the saddles when $h_{ij}\neq 0$, and stability of the background saddles $N_{\sg}^{(q)}$ for various boundary choices, without incorporating backreaction effects.

%%%%%%%%%%%%%%%%%%%%%%%%%%%%%%%%%%%%%%%%%%%%%%%
\subsection{Saddle correction due to $h_{ij}\neq0$}
\label{sad_cor} 
%%%%%%%%%%%%%%%%%%%%%%%%%%%%%%%%%%%%%%%%%%%%%%%

In the saddle-point approximation, we start by computing the saddles of the action ${\cal A}(N_c)$ mentioned in Eq. (\ref{eq:Nc_act_total}). These can be computed by varying ${\cal A}(N_c)$
with respect to $N_c$: ${\rm d} {\cal A}/{\rm d} N_c =0$. The 
saddles will not only have contribution from 
$h_{ij}$ but will also receive quantum corrections of ${\cal O}(\hbar)$.
If we denote a saddle by $N_{\sg}$, then perturbatively, it can be expressed as 
\beq
\label{eq:Nsexp_sad_hij}
N_\sigma = N_\sigma^{(q)} + N_\sigma^{(h)} 
+ \hbar \, {\cal N}_\sigma^{(q)} + \hbar \, {\cal N}_\sigma^{(h)} + \cdots \, ,
\eeq
where the $(\cdots)$ represents higher-loop corrections which we ignore. Previously, we have computed $N_\sigma^{(q)}$ and ${\cal N}_\sigma^{(q)}$; here our main focus will be the computation of $N_\sigma^{(h)}$. This is the correction the saddles will get when the perturbation does not vanish at the final boundary. To compute this, 
we first note that the contribution of the 
$h_{ij}$ path integral to the lapse $N_c$ action is just the on shell action of the fluctuation field, when ${\cal O}(\hbar)$ terms are ignored. This can be written as
\beq
\label{eq:Sl_form_act_hl}
S^{(h)}_{\rm grav} = \sum_{l=2}^\infty 
g_l S^{(l)}_{\rm grav}[\bar{q}, \bar{h}_l, N_c]
= q_f \sum_{l=2}^\infty 
g_l  \,\, \bl(h^{(l)}_1\br)^2 \,\, \mathbb{F}_l(N_c) \, ,
\eeq
where $\mathbb{F}_l(N_c)$ for various choice of boundary condition 
can be inferred from Eqs. (\ref{eq:hh_onSH_DBC} and \ref{eq:hh_onSH_NBC}).
A point in complex $N_c$ plane is a saddle point of the whole 
action [ignoring the ${\cal O}(\hbar)$ terms] if it satisfies 
\beq
\label{eq:sad_cond_full}
\frac{{\rm d} S_{\rm grav}}{{\rm d} N_c} \Br \rvert_{N_{\sg}}
= \frac{{\rm d} S_{\rm grav}^{(\bar{q})}}{{\rm d} N_c}\Br \rvert_{N_{\sg}}
+ q_f \sum_{l=2}^\infty 
g_l  \,\, \bl(h^{(l)}_1\br)^2 \frac{{\rm d} \mathbb{F}_l(N_c)}{{\rm d} N_c} \Br \rvert_{N_{\sg}}= 0 \, .
\eeq
If $N_{\sg} = N_{\sg}^{(q)}$ is the saddle point of the $S_{\rm grav}^{(\bar{q})}$
(the on shell action of scale factor), then the first term vanishes.
As $\bl(h^{(l)}_1\br)$ can be arbitrary so it implies that 
for $N_{\sg}^{(q)}$ to be also the saddle of the full action it must satisfy
for all $l$
\beq
\label{eq:sad_cond_l_gen_h}
\frac{{\rm d} \mathbb{F}_l(N_c)}{{\rm d} N_c} \Br \rvert_{N_{\sg}^{(q)}} = 0 \, ,
\hspace{5mm}
\forall \,\, l\geq2 \,\, {\rm and} \,\, l \in \mathbb{I}^+ \, .
\eeq
This leads to an infinite set of algebraic equations that need to be individually
satisfied for all $l\geq 2$! 
If we have only a finite set of $l$ modes to consider 
$2 \leq l \leq M$, then one can adjust the boundary conditions imposed 
on the scale factor such that the saddles of the $S_{\rm grav}^{(\bar{q})}$
are also the saddles of the full action $S_{\rm grav}$, where now the lapse action 
[excluding the ${\cal O}(\hbar)$ quantum corrections] is given by
\beq
\label{eq:lapseAct_cut}
S_{\rm grav} = S^{(q)}_{\rm grav}
+ q_f \sum_{l=2}^M 
g_l  \,\, \bl(h^{(l)}_1\br)^2 \,\, \mathbb{F}_l(N_c) \, .
\eeq
This can be easily verified numerically for $l=2$. 
The saddle of the $S_{\rm grav}^{(\bar{q})}$ is not  
the saddle of $S_{\rm grav}$. One can appropriately adjust the 
boundary condition so that two have a common saddle. 
As we increase $M \to M + 1$ and incorporate an additional mode participating 
in the path integral, then the system of saddle equations that need to be 
satisfied increases by 1. The only way these can be satisfied 
if an additional adjustment is done on the boundary condition so that 
the saddle of $S_{\rm grav}^{(\bar{q})}$ also is the saddle of 
$S_{\rm grav}$. As $M$ increases further, the amount of adjustment 
needed to the boundary condition so that saddles of the 
$S_{\rm grav}^{(\bar{q})}$ also becomes the saddle of 
$S_{\rm grav}$, increases further. This eventually leads to an 
infinite amount of adjustment as $M\to \infty$. This is 
analogous to fine-tuning of the boundary condition, 
leading to conclusion that the saddles of $S_{\rm grav}^{(\bar{q})}$
will not be saddles of the full action $S_{\rm grav}$. However, 
the new saddles can be thought of as corrected old saddle. 

%%%%%%%%%%%%%%%%%%%%%%%%%%%%%%%%%%%%%%%%%%%%%%%
\subsubsection{Semi-rigorous proof for saddles getting corrected}
\label{sad_cor1} 
%%%%%%%%%%%%%%%%%%%%%%%%%%%%%%%%%%%%%%%%%%%%%%%

For generic boundary condition, it is hard to have a 
proof for the condition mentioned in Eq. (\ref{eq:sad_cond_l_gen_h}).
However, it can explicitly checked whether it is indeed true for the
special case of Neumann boundary condition. To do so, we use the 
on shell action for the fluctuation field mentioned in Eq. (\ref{eq:hh_onSH_NBC})
to compute the expression for $\mathbb{F}_l(N_c)$. 
This is given by
\beq
\label{eq:FNl_NBC}
\mathbb{F}_l(N_c)=\bl\{ 2\Lambda N_c-2\pi_i\br\}
+2\frac{q_f}{N_c}\bl\{\ln W_l(t,0) \br\}^\prime\br \rvert_{t=1} \, .
\eeq
On plugging this in the condition mentioned in Eq. (\ref{eq:sad_cond_l_gen_h})
we get the following:
\beq
\label{eq:NBC_Fnl_cond_sad}
\Lam-\frac{q_f}{N_c^2}\bl\{\ln W_l(t,0)\br\}^\prime\br \rvert_{t=1} \Br \rvert_{N_{\sg}^{(q)}}
+\frac{q_f}{N_c}\frac{\rm d}{{\rm d}N_c}
\bl\{\ln W_l(t,0)\br\}^\prime\br \rvert_{t=1} \Br \rvert_{N_{\sg}^{(q)}}=0 \, .
\eeq
The expression of $W_l(t,0)$ is mentioned in Eq. (\ref{eq:W_l10_func}).
This however, gets simplified at the no-boundary saddles 
as $\xi_l =l+1$ and $\mathbb{P}_1^{\xi_l}[\tau_0] = 0$. This means 
that we have the following:
\bea
\label{eq:Fnl_nbc_Der_sad}
&&
\bl\{\ln W_l(t,0)\br\}^\prime\br \rvert_{t=1} \Br \rvert_{N_{\sg}^{(q)}}
= 
\frac{\dot{\mathbb{P}}_1^{\xi_l}(\tau_1)}{\mathbb{P}_1^{\xi_l}(\tau_1)}\Br \rvert_{N_{\sg}^{(q)}} \, ,
\notag \\
&&
\frac{\rm d}{{\rm d}N_c}
\bl\{\ln W_l(t,0)\br\}^\prime\br \rvert_{t=1} \Br \rvert_{N_{\sg}^{(q)}}
=
\frac{\rm d}{{\rm d} N_c} \Bl(
\frac{\dot{\mathbb{P}}_1^{\xi_l}(\tau_1)}{\mathbb{P}_1^{\xi_l}(\tau_1)}
\Br)
\Br \rvert_{N_{\sg}^{(q)}} \, .
\eea
At the no-boundary saddles these can be computed exactly for the 
Neumann BC. These are given by
\beq
\label{eq:F_Sad}
\frac{\dot{\mathbb{P}}_1^{\xi_l}(\tau_1)}{\mathbb{P}_1^{\xi_l}(\tau_1)}\Br \rvert_{N_{\sg}^{(q)}}
= \frac{i N_{\sg}^{(q)}}{q_f} \Bl[ (l+1) 
- \frac{q_f \Lam}{3+3l \mp i \sqrt{3\Lam q_f - 9}}\Br] ,
\eeq
and
\bea
\label{eq:Deriv_F_Sad}
\frac{\rm d}{{\rm d} N_c} &&
\Bl(
\frac{\dot{\mathbb{P}}_1^{\xi_l}(\tau_1)}{\mathbb{P}_1^{\xi_l}(\tau_1)}
\Br)
\Br \rvert_{N_{\sg}^{(q)}} 
= 
\frac{1}{q_f (l+1) \bl[3(l+1) \pm i \sqrt{3\Lam q_f - 9} \br]^2} 
\Bl[ -9i \bl(l^4 + 4 l^3 + 6l^2 + 2l-2\br)  
\notag \\
&&
+ 3iq_f\Lam (l^2-l-3) \pm \sqrt{3q_f\Lam - 9} 
\bl\{6l^3 + 15 l^2 + 6l - 6 + 2q_f \Lam (l+1) \br\} \Br] \, .
\eea
When all the terms are put together in the condition stated in 
Eq. (\ref{eq:NBC_Fnl_cond_sad}), it is seen that two scenarios arises 
on comparing the wavelength of mode on the 
final three geometry $\lam_l \sim \sqrt{q_f} l^{-1}$ with the Hubble radius given by
$\bold{H} = \sqrt{3/\Lam}$:
modes with $\lam_l \gg \bold{H}$, and modes $\lam_l \ll \bold{H}$.
We define $\eta_l = \lam_l/ \bold{H}$. 
In former case ($\eta_l\gg1$) the modes lie outside the Hubble horizon and 
are not expected to affect the structure inside the Hubble radius. 
In the later case ($\eta_l \ll 1$) it affects the structure inside the Hubble radius. 
In these regimes the lhsof Eq. (\ref{eq:NBC_Fnl_cond_sad}) reduces 
to the following for $\lam_l \gg \bold{H}$ and $\lam_l \ll \bold{H}$:
\bea
\label{eq: NBC_frozen_limt}
&&
\Lam-\frac{q_f}{N_c^2}\bigl\{\ln 
W_l(t,0)\bigr\}^\prime\bigr \rvert_{t=1} \Bigr \rvert_{N_{\sg}^{(q)}}
+\frac{q_f}{N_c}\frac{\rm d}{{\rm d}N_c}
\bigl\{\ln W_l(t,0)\bigr\}^\prime\bigr \rvert_{t=1} \Bigr \rvert_{N_{\sg}^{(q)}}  
\notag \\
&&
= \mp \frac{4i}{\bold{H}\sqrt{q_f}} + \mathcal{O}\Bigl(\frac{1}{\eta_l} \Bigr) \, , 
\hspace{10mm}
\forall \,\,\,\, \eta_l \gg 1 \, ,
\notag \\
&&
= 
\mp \frac{2i \Lam}{3 \eta_l }  
+ \mathcal{O}\bigl(\eta_l^0 \bigr) \, ,
\hspace{10mm}
\forall \,\,\,\, \eta_l \ll 1 \, 
\eea
respectively. It is immediately noticed in the large $q_f$ limit, that the condition 
stated in Eq. (\ref{eq:NBC_Fnl_cond_sad}) is satisfied for 
$\eta_l \gg 1$, while it is not satisfied for $\eta_l \ll 1$.
For a fixed $q_f$, modes with higher $l$ will have smaller $\lam_l$
hence smaller $\eta_l$,
and will be inside the Hubble radius, but as $q_f$ becomes large 
then they will go outside the Hubble radius after a while.
However, as $l= 2$ to $\infty$, so there will always be modes 
lying inside the Hubble radius and hence will not satisfy the 
condition stated in Eq. (\ref{eq:NBC_Fnl_cond_sad}). This means 
there will always a finite set of modes which will satisfy the 
condition in Eq. (\ref{eq:NBC_Fnl_cond_sad}), thereby implying 
the no-boundary saddles will be corrected. Similarly, one can 
show that no-boundary saddles will be corrected in the Robin BC case too. 

%%%%%%%%%%%%%%%%%%%%%%%%%%%%%%%%%%%%%%%%%%%%%%%
\subsubsection{Correction to saddles: Computation of $N_{\sg}^{(h)}$}
\label{sad_cor2} 
%%%%%%%%%%%%%%%%%%%%%%%%%%%%%%%%%%%%%%%%%%%%%%%  

After having established that the background saddles indeed do get corrected, it is natural to ask about the correction they receive. This can be computed analytically, at least to first order in the perturbation. To compute the first order correction to the saddle we start with the lapse action stated in Eq. (\ref{eq:Nc_act_total}) at $\hbar=0$. The saddle-point equation is given in (\ref{eq:sad_cond_full}). In this we plug Eq. (\ref{eq:Nsexp_sad_hij}) and solve it order by order. 
To zeroth order (ignoring all contributions coming from the fluctuation field), one gets the following equation:
\beq
\label{eq:0thSad_eq_per}
\bl[S^{(q)}_{\rm grav}(N_{\sg}^{(q)}) \br]^\prime=0\,. 
\eeq
This will give the standard background saddles for various boundary conditions (NBC or RBC) and are known in the literature. Note, here $(')$ represents the derivative with respect to lapse variable $N_c$. To first order we get the following equation that need to be satisfied: 
\beq
\label{eq:1st_Sad_eq_per}
\bl[S^{(q)}_{\rm grav}(N_{\sg}^{(q)}) \br]^\prime
+ \bl[S^{(q)}_{\rm grav}(N_{\sg}^{(q)}) \br]^{\prime\prime} N_{\sg}^{(h)}
+ q_f \sum_{l=2}^\infty 
g_l \,\, \bl(h^{(l)}_1\br)^2 \,\, 
\mathbb{F}_l^\prime(N_{\sg}^{(q)})
= 0 \, .
\eeq
The first term in the above vanishes as $N_{\sg}^{(q)}$ satisfies the zeroth-order saddle point equation. In that, the first-order correction to the background saddle is given by
\beq
\label{eq:1st_nh_sad_per}
N_{\sg}^{(h)}
= - \frac{q_f}{\bl[S^{(q)}_{\rm grav}(N_{\sg}^{(q)}) \br]^{\prime\prime}}
\sum_{l=2}^\infty 
g_l \,\, \bl(h^{(l)}_1\br)^2 \,\, 
\mathbb{F}_l^\prime(N_{\sg}^{(q)}) \, .
\eeq
This is the correction received due to the fluctuation field, provided the second derivative in the denominator is nonzero. However, in degenerate cases this vanishes and needs a separate treatment, which we do not cover here. 

%%%%%%%%%%%%%%%%%%%%%%%%%%%%%%%%%%%%%%%%%%%%%%%
\subsection{Stability of saddles}
\label{hij_beh_sad}
%%%%%%%%%%%%%%%%%%%%%%%%%%%%%%%%%%%%%%%%%%%%%%%
%
A natural question that immediately arises at this point is the 
issue of stability of saddles under the metric perturbations 
which in the present scenario are given by $h_{ij}$. Such an exploration is not
new and has been done previously for the case of Dirichlet, Neumann and 
Robin boundary conditions in \cite{Feldbrugge:2017fcc, Lehners:2018eeo,
DiTucci:2018fdg, Feldbrugge:2017mbc,DiTucci:2019bui, DiTucci:2019dji}.
Rigorous studies done in the context of imposing Dirichlet boundary 
condition on the scale factor showed that perturbations were not stable 
at the saddle points which correspond to no-boundary proposal of 
Universe \cite{Feldbrugge:2017fcc, Lehners:2018eeo,
DiTucci:2018fdg, Feldbrugge:2017mbc, Honda:2024aro, Matsui:2024bfn}. 
This discouragement and 
subsequent motivation led the community to seek stable saddles 
with Neumann and Robin choice of boundary condition 
\cite{DiTucci:2019bui, DiTucci:2019dji}. It was noticed that for 
standard boundary choices for the fluctuation $h_{ij}$, the 
perturbations are well behaved at the no-boundary saddles. 
In this paper we readdress stability issues systematically 
for those choices of boundary condition imposed on the 
perturbation $h_{ij}$ which are 
motivated due to covariance of the gravitational system
as discussed in Sec. \ref{boundTR}. 
%
%%%%%%%%%%%%%%%%%%%%%%%%%%%%%%%%%%%%%%%%%%%%%%%
\subsubsection{Neumann No-boundary saddles}
\label{nbs_nbc}
%%%%%%%%%%%%%%%%%%%%%%%%%%%%%%%%%%%%%%%%%%%%%%%
%
If a Neumann BC is imposed at the initial boundary then as discussed in Sec. \ref{NBC_bg}, initial conjugate momenta corresponding to scale factor is fixed to $\pi_i$, while the initial scale factor is left arbitrary. The set of boundary action respecting covariance and NBC are mentioned in Eq. (\ref{eq:Sbd_nbc_cov}). Working with the simplest choice of the vanishing boundary term still imposes constraints on boundary choices of the fluctuation field $h_{ij}$, implying that the boundary condition on $h_{ij}$ cannot be arbitrarily chosen. This gives two possible boundary choices for $h_{ij}$ at $t=0$: one is trivial boundary choice of $h_{ij}(t=0) = 0$ and another nontrivial boundary choice is mentioned in Eq. (\ref{eq:hl_nbc_2ndcond}. Focusing on the trivial boundary choice of vanishing fluctuation field at the initial boundary we analyze its behavior at the no-boundary saddles.

It was noticed in \cite{Ailiga:2023wzl, Ailiga:2024mmt} that the gravitational path integral gets the most dominant contribution for the boundary configuration corresponding to $\pi_i = -3i$. This was also seen to correspond to stable configuration when Gauss-Bonnet 
corrections were taken into account.  
The lapse action from the background field given in 
Eq. (\ref{eq:stot_onsh_nbc}), which when varied with respect to $N_c$ 
gives the saddle-point solutions giving dominant contributions in the path integrals. For $\pi_i = -3i$, these are mentioned in Eq. (\ref{eq:NB_sad}). 

A crucial observation is made that at the saddle point, the configuration $\pi_i = -3i$ corresponds to a universe starting with zero initial size ($\bar{q}(t=0) =0$) with a smooth rounding off at initial times. At these Neumann no-boundary saddles, 
$\xi_l$ and $\tau_t$ from Eq. (\ref{eq:albt_form}) reduces to Eq. (\ref{eq:xiL_tauT_nbc_nb1}).
Note that $\tau_0^{\rm nb}=1$ and $\xi_l^{\rm s} = l+1$ at all saddles. A generic solution for the $\bar{h}_l(t)$ is mentioned in Eq. (\ref{eq:hl_sol_gen}) with $d^{(l)}_{p}$ and $d^{(l)}_{q}$ given in Eq. (\ref{eq:hl_dpdq_NBC}) respectively. $W_l(1,0)$ can be obtained from Eq. (\ref{eq:W_l10_func}), but at the no-boundary saddles $W_l(1,0)$ gets simplified as $\mathbb{P}_1^{l+1}(\tau_0) =0$. This is mentioned in Eq. (\ref{eq:Wl_nb_nbc}). Using this one can immediately work out the expression for $d^{(l)}_{p}$ and $d^{(l)}_{q}$ using Eq. (\ref{eq:hl_dpdq_NBC}) thereby giving 
\beq
\label{eq:hl_dpdq_NBC_nb}
d^{(l)}_{p} =  \frac{ h^{(l)}_1 \sqrt{q_f}}{\mathbb{P}_1^{l+1}[\tau_1]} \, ,
\hspace{5mm}
d^{(l)}_{q} = 0 \,  .
\eeq
At the no-boundary saddle as $\xi^{\rm nb}_l$ is an integer, the function $\mathbb{P}_1^{l+1} (\tau_t)$ has an alternative form mentioned in Eq. (\ref{eq:alt_Pxi_nb_form}). Making use of this form of $\mathbb{P}_1^{l+1} (\tau_t)$, one can express 
$\bar{h}_l^{\rm nb}(t)$ in the following manner: 
\beq
\label{eq:hLt_nbc_nb_form}
\bar{h}_l^{\rm nb}(t) =h^{(l)}_1 \sqrt{\frac{q_f}{\bar{q}_{\rm nb}(t)}}
\Bl(
\frac{1-\tau_{t}^{\rm nb}}{1+\tau_{t}^{\rm nb}}
\Br)^{(l+1)/2}
\Bl(
\frac{1+\tau_{1}^{\rm nb}}{1-\tau_{1}^{\rm nb}}
\Br)^{(l+1)/2}
\Bl(\frac{\tau_{t}^{\rm nb}+l+1}{\tau_{1}^{\rm nb}+l+1}
\Br) \, .
\eeq
For the NBC, at the no-boundary saddles
\beq
\label{eq:q0ATnb_nbc}
\bar{q}^\pm_{\rm nb}(t) = t \bl\{
q_f t - 2(t-1) i N_{\pm}^{(\rm nb)}
\br\} \, .
\eeq
This implies that for NBC case the perturbations near the initial boundary 
at the no-boundary saddles will behave like $\bar{h}_l^{\rm nb}(t) \sim t^{l/2}$,
which is finite for all $l\geq2$ and vanishes as $t\to0$. The on shell action 
at the Neumann no-boundary saddles 
can be computed using Eq. (\ref{eq:hh_onSH_NBC})
and making use of the Eq. (\ref{eq:Wl_nb_nbc}). This gives
\begin{equation}
\label{eq:hh_onSH_NBC_nb}
i S^{(l)}_{\rm grav} [\bar{q}, \bar{h}_l, N_{\pm}^{(\rm nb)}] 
=\frac{q_f \bl(h^{(l)}_1\br)^2}{(64\pi G)}
\Bl[
\frac{l (l+2)\sqrt{3}}{-(l+1)\sqrt{3} \pm i \sqrt{\Lam q_f-3}}
\pm 4 i \sqrt{\frac{\Lambda q_f}{3}-1}
\Br] \, .
\end{equation}
From the above one can compute that for $\Lambda > 0$ and large $q_f$,
perturbations behave like  
$e^{-3l(l+1)(l+2)(h_1^{(l)})^2/(4\Lambda\hbar)}$, 
indicating suppressed perturbations for each mode. 
This behavior corresponds to the Bunch-Davies vacuum of de Sitter 
space, where the perturbation is known to be Gaussian. The $l^3$ dependence 
shows the scale invariance of the perturbation. For each mode, the phase in the 
perturbation grows as $q_f^{1/2}$. Comparing it to the background, 
we find that the phase in perturbation grows slower than the background, 
where it grows with $q_f^{3/2}$. The additional $l$ independent phase terms 
come from the requirement of covariance. These are the overall phase 
terms same for all modes. This is in agreement with the conclusion obtained 
in the past \cite{Lehners:2021jmv}. However, unlike in the previous works, 
here, the choice of boundary conditions imposed on the fluctuations 
is constrained due to requirements of covariance. And the stability at the 
saddles is analyzed keeping the covariance into account.
%
%
%%%%%%%%%%%%%%%%%%%%%%%%%%%%%%%%%%%%%%%%%%%%%%%
\subsubsection{Robin BC saddles}
\label{nbs_rbc}
%%%%%%%%%%%%%%%%%%%%%%%%%%%%%%%%%%%%%%%%%%%%%%%
%
If Robin BC is imposed at the initial boundary, then as discussed in Sec. \ref{RBC_bg}, the linear combination of scale factor and conjugate momenta corresponding to the scale factor is fixed to $P_i$. The set of boundary actions respecting covariance and RBC are mentioned in Eq. (\ref{eq:Sbd_rbc_cov}). Working with the simplest choice, still imposes constraints on the boundary condition of the fluctuation field $h_{ij}$, thereby implying that the boundary condition for $h_{ij}$ cannot be arbitrarily chosen. This gives two possible boundary choices for $h_{ij}$ at $t=0$: one is trivial boundary choice of $h_{ij}(t=0) = 0$, and another nontrivial boundary choice is mentioned in Eq. (\ref{eq:hl_rbc_2ndcond}). 

Focusing on the trivial boundary choice of vanishing fluctuation field at the initial boundary, we analyze its behavior both at the no-boundary (nb) and other saddles, which we denote as $N^{(\cancel{nb})}$ (non-no-boundary ones). It was noticed in \cite{Ailiga:2023wzl, Ailiga:2024mmt} that the gravitational path integral gets the most dominant contribution to the boundary configuration corresponding to $P_i = -3i$. This was also seen to correspond to the stable configuration when Gauss-Bonnet corrections were taken into account.  

The lapse action from the background field is given in Eq. (\ref{eq:stot_onsh_rbc}), which, when varied with respect to $N_c$ gives four saddle point solutions. For $P_i = -3i$, these are given in Eq. (\ref{eq:NB_sad}) and in Eq. (\ref{eq:NNB_sad}). $N_{\pm}^{(\rm nb)}$ are the no-boundary saddles which are exactly the same as for the NBC case.
A crucial observation is made that at these saddle 
points, the configuration $P_i = -3i$
correspond to a universe starting with zero initial size ($\bar{q}(t=0) =0$) with a smooth rounding off at initial times. The other two saddles are not no-boundary saddles where the initial size cannot be zero and depends on the initial parameter $\beta$. In the limit, $\beta\rightarrow 0$, $N_{1,2}^{(\rm \cancel{nb})}$ saddles go to infinity, and we recover the Neumann limit.

For Robin boundary condition, at the no-boundary saddles $\xi_l$ and $\tau_t$ can be obtained from Eq. (\ref{eq:albt_form}) and they are same as in case of Neumann BC mentioned in Eq. (\ref{eq:xiL_tauT_nbc_nb1}). Just like in NBC case, the $\tau_0^{\rm nb}=1$ and $\xi_l^{\rm s} = l+1$ at all saddles for the RBC too. A generic solution for the $\bar{h}_l(t)$ is mentioned in Eq. (\ref{eq:hl_sol_gen}) with $d^{(l)}_{p}$ and $d^{(l)}_{q}$ given in Eq. (\ref{eq:hl_dpdq_RBC}) respectively. $W_l(1,0)$ can be obtained from Eq. (\ref{eq:W_l10_func}). 
Interestingly, $\tau_{t}^{\rm nb}$ is independent of $\beta$.
So, the solution $h_l(t)$ will be exactly similar to the NBC case, which we will not repeat here.
For the RBC, at the no-boundary saddles
\beq
\label{eq:q0ATnb_rbc}
\bar{q}^\pm_{\rm nb}(t) = t \bl\{
q_f t - 2(t-1) i N_{\pm}^{(\rm nb)}
\br\} \, .
\eeq
This is the same as the NBC case. So, the perturbation analysis 
at these no-boundary saddles for RBC is exactly similar to the NBC case. 
Perturbation is Gaussian at no-boundary saddles, irrespective of the boundary condition. 
For the other two saddles, i.e., $N^{(\rm \cancel{nb})}_{1,2}$, with $\beta = - i \Lam x/2$,
\begin{equation}
\label{eq:tauT_nnb}
\tau_t^{\rm (\cancel{nb})} = 1 - 2/x + i \Lam N_{\pm}^{\rm (\cancel{nb})} t/3.
\end{equation}
So, $\tau_t^{\rm (\cancel{nb})}$ depends on $\beta$ (or $x$) in this case. 
At $t=0$ and finite $x$, $\tau_t^{\rm (\cancel{nb})}=1-2/x \neq 1$. 
As a result, neither $\mathbb{P}^{l+1}_{1}[\tau_0^{\rm (\cancel{nb})}]$ 
nor $\bar{q}(0)$ vanishes at these saddles. The equation of motion 
for $\bar{h}_l(t)$ mentioned in Eq. (\ref{eq:htL_eqm}) can be solved 
giving Eq. (\ref{eq:hl_sol_gen}), with $d^{(l)}_{p}$ and $d^{(l)}_{q}$ 
given in Eq. (\ref{eq:hl_dpdq_RBC}) respectively,
except that all the quantities need to be computed 
at these non-no-boundary saddles. 
The on shell action in Eq. (\ref{eq:hh_onSH_RBC}) becomes
\begin{equation}
\label{eq:hh_onSH_RBC_nnb}
iS^{(l)}_{\rm grav}\bl[\bar{q},\bar{h}_l,N_{1,2}^{(\rm \cancel{nb})} \br]
=\frac{q_f h_1^{(l)}}{64\pi G}\Bl[-\sqrt{3 \Lambda  q_f-9}
+\frac{q_f}{N_{1,2}^{\rm (\cancel{nb})}}\{\ln W_l(t,0)\}^\prime \br \rvert_{t=1}\Br] \, . 
\end{equation}
The above expression is usually quite algebraically involved 
making it difficult to analyze it. However, our numerical studies 
shows that similar to the 
no-boundary saddles, ${\rm Re} \bl(iS^{(l)}_{\rm grav} \br)$ varies with $q_f$ and saturates 
to a constant value in the limit $q_f$ becomes large. Depending on 
the value of $x$, it may either saturate to a positive or a negative value, 
unlike the no-boundary saddles. Similar to the no-boundary saddles, 
for each $l$ mode in the large $q_f$ limit, ${\rm Im} \bl(iS^{(l)}_{\rm grav} \br)$ grows 
with $\sqrt{q_f}$. 

%%%%%%%%%%%%%%%%%%%%%%%%%%%%%%%%%%%%%%%%%%%%%%%
\section{Conclusions and outlook}
\label{conc}
%%%%%%%%%%%%%%%%%%%%%%%%%%%%%%%%%%%%%%%%%%%%%%%

Motivated by the findings of our past works in \cite{Narain:2022msz, Ailiga:2023wzl, Ailiga:2024mmt}, we have been intrigued by the importance of boundary choices and the significant role they play in dictating the behavior of the saddles in the gravitational path integral. In particular, we noticed that the presence of the higher derivative Gauss-Bonnet term naturally favors boundary choices which align with the ideology of the no-boundary proposal of the Universe, advocating a universe starting with vanishing size. It further shows a universe having a Euclideanised beginning, transitioning to Lorentzian as it evolves \cite{DiTucci:2019bui, Narain:2021bff, Lehners:2021jmv, DiTucci:2020weq, DiTucci:2019dji, Narain:2022msz}. These have been verified in exact computations of the path integral of the minisuperspace gravitational system \cite{Narain:2022msz, Ailiga:2023wzl, Ailiga:2024mmt}.

Usually, in such studies, the boundary conditions are imposed in canonical form. For example, in the case of the Dirichlet boundary problem, the field is fixed at the boundary, while in the Neumann boundary problem, the corresponding conjugate momenta of the field is fixed, and in the Robin BC case, a combination of the two is fixed. In the minisuperspace approximations, the gravitational system reduces to just a dynamically evolving scale factor and lapse $N_c$. The gravitational path integral becomes a path integral over scale factor and an ordinary integral over lapse $N_c$ after gauge fixing. For Einstein-Hilbert gravity (with/without Gauss-Bonnet), the path integral over the scale factor is exactly doable as the action is quadratic with not more than two time derivatives \cite{Narain:2022msz, Ailiga:2023wzl, Ailiga:2024mmt}. The ordinary integral over the lapse is doable exactly only in some special boundary choices, thereby implying that even for such simple systems of minisuperspace, where the gravitational degrees of freedom have been significantly reduced, exact results are only available for special choices of boundary conditions \cite{DiTucci:2019bui}. However, implementing WKB methods to compute the ordinary lapse integral offers extra freedom and allows us to go beyond some of the standard choices of boundary conditions.  

In achieving this, an important role is played by the PL methods while dealing with the lapse integration, which, in general, is highly oscillatory. PL methodology helps in not only finding the right convergent contour of integration systematically (integration done along the thimbles of relevant saddles in the complex plane) but also dictates the procedure of deforming the original integration contour to deformed contour (see \cite{Feldbrugge:2017kzv, Narain:2021bff, Lehners:2023yrj, Ailiga:2024mmt} for review of Picard-Lefschetz methods). However, whenever boundary conditions change, the qualitative picture in the complex plane changes, and the saddles and corresponding thimbles change, thereby implying that the system is sensitive to the boundary choices. 

Moreover, the stability of such saddles is also affected once boundary choices are changed. For example, the relevant no-boundary saddle with Dirichlet boundary condition at initial time has unsuppressed perturbations. However, when the no-boundary Universe is studied with either Neumann or Robin BC, then relevant saddles have well-behaved perturbations leading to a stable universe. An immediate question that arises here is, what are the boundary conditions for the perturbations? Past studies involved imposing Dirichlet boundary conditions for the perturbations $h_{ij}$, a standard choice though not a well-motivated one except for simplicity. One may wonder if it is possible to determine the available boundary choices for the perturbation for the given boundary condition for the scale factor.  

In this paper, we have explored this question deeply using the requirements coming from covariance. It turns out that the boundary condition for the fluctuation $h_{ij}$ cannot be arbitrarily chosen and is intimately tied to the boundary choice imposed on the scale factor. This is expected as the full theory respects diffeomorphism invariance. We use this intricate connection to find the allowed boundary choice for the fluctuation field for the given boundary choice of the scale factor. In this process, we also construct a family of boundary actions that sets up a consistent variational problem. The details of this construction are presented in Sec. \ref{boundTR} and follows the ideas first mentioned in \cite{Brizuela:2023vmb}. This procedure, when applied to, for example, Dirichlet, Neumann or Robin BC imposed on the scale factor, gives the family of covariant boundary actions that needs to be added in each case for the consistent variational problem. These are mentioned in Sec. \ref{bound_act}: Eqs. (\ref{eq:dbc_qbd_F}), (\ref{eq:Sbd_nbc_cov}) and (\ref{eq:Sbd_rbc_cov}) for the Dirichlet, Neumann and Robin BC cases, respectively. 

Such an analysis reveals that only certain choices for the boundary conditions are available for the fluctuation field $h_{ij}$, and the boundary condition cannot be arbitrarily imposed. Interestingly, in each case, the boundary condition $h_{ij} \br \rvert_{\rm bd}=0$ appears as the simplest allowed boundary condition motivated by the requirements of covariance, which in earlier studies was considered as a choice motivated by simplicity. Beside this, there are other available boundary choices which we do not consider in this paper and will form a part of our future works. 

After having finalized the boundary conditions that need to be imposed on the fields, we proceed to compute the one-loop effects coming from the path integral over the field variables $q(t)$ and $h_{ij}(t, {\bf x})$, going beyond the studies done in \cite{Barvinsky:1992dz} where the path integral is done only over the perturbations $h_{ij}(t, {\bf x})$ with fixed scale factor and lapse corresponding to Hartle-Hawking Universe. To sensibly define the path integral, appropriate gauge fixing and corresponding Faddeev-Popov ghosts are incorporated to prevent overcounting of the gauge orbits. Working in the background field formalism, the fields $q(t)$ and $h_{ij}$ are decomposed into background and fluctuation: $q(t) = \bar{q}(t) + Q(t)$ and $h_{ij} = \bar{h}_{ij} (t, {\bf x}) + H_{ij} (t, {\bf x})$ respectively. Requiring the background fields $\bar{q}(t)$ and $\bar{h}_{ij} (t, {\bf x})$ to satisfy equation of motion, we notice that in special case when $\bar{h}_{ij} (t=0,1, {\bf x}) =0$, then $\bar{h}_{ij} (t, {\bf x})=0$ for all times. Moreover, in this case it will not cause any backreaction on the evolution of $\bar{q}(t)$. In such a scenario, the full path integral decomposes into a path integral over $q(t)$ and path integral over $H_{ij}$, with an overall ordinary integration over lapse $N_c$. As discussed earlier, the path integral over $q(t)$ is exactly doable. The path integral over the fluctuation field $H_{ij}$, however, needs to be done with care. At one loop, it can be done exactly via mode decomposition and utilizing symmetry properties of the background $S^3$ geometry. The combined contribution from the path integral over $q(t)$ and $H_{ij}$ gives a quantum corrected lapse action ${\cal A}(N_c)$ mentioned in Eq. (\ref{eq:Nc_act_total}), which is new and has not been reported earlier in literature in the context of Lorentzian quantum cosmology.

We then proceed to systematically analyze the quantum corrected lapse action ${\cal A}(N_c)$, which is then utilized in the computation of the lapse $N_c$ integration. The lapse integration is performed via Picard-Lefschetz methodology in the complex $N_c$ plane. This is achieved by first determining the saddles of the lapse action ${\cal A}(N_c)$. For the case when $\bar{h}_{ij} (t=0,1, {\bf x}) =0$, at zeroth order in $\hbar$, the saddles are determined entirely from $S_{\rm grav}^{(\bar{q})}$. These are denoted by $N_{\sg}^{(q)}$, and they get $\hbar$ corrections at one loop. In the computation of transition amplitude up to one loop, it is seen that only the contribution from $N_{\sg}^{(q)}$ enters the computation. 

We then focus our attention on studying the one-loop corrected lapse action ${\cal A}(N_c)$ in more detail, and explore its behavior near the no-boundary saddles (recall that up to one loop, only the saddles of the background action $S^{\bar{(q)}}_{\rm grav}$ contribute). It is seen that the quantum corrected lapse action has ultraviolet divergences. One divergence arises as one approaches the no-boundary saddles but is otherwise well behaved at any other point. This is a log divergence, appearing when the initial size of the Universe approaches zero size, a feature of the ``no-boundary''  universe. Besides this, there are two more UV divergences: one arises due to an infinite summation over the modes of $H_{ij}$-field, while the other arises in a special case for the small final size of the Universe. Each of these UV divergences are systematically isolated and subtracted to obtain the one-loop corrected finite effective action for the lapse $N_c$, which is new and has not been reported earlier. 

UV-divergences are not alien to QFT studies of Einstein-Hilbert gravity, which is known to be perturbatively nonrenormalizable. However, in the context of path integral studies of Lorentzian quantum cosmology, this has not been reported previously. Such divergence, although expected, still causes worry as the lapse action is no longer finite. Our hope is that they may be more controlled in theories which are perturbatively renormalizable \cite{Stelle:1976gc, Salam:1978fd, Narain:2011gs, Narain:2016sgk} or in theories which are safely protected by the occurrence of nontrivial stable fixed point \cite{Buccio:2024hys}. In the current scenario, we overcome the divergences by adding suitable counterterms, thereby making the residual lapse action finite. The finite lapse action is then utilized to compute the transition amplitude via the usage of Picard-Lefschtez and the WKB methods. 

The finite transition amplitude is then analyzed. It is seen that the UV-finite transition amplitude of the no-boundary universe (where either Neumann or Robin is imposed at the initial boundary with Dirichlet at the final boundary), grows as the universe expands. This growth is attributed to the contribution of graviton loops leading to infrared divergent amplitude. In a sense, this is expected as we are studying the QFT in de Sitter spacetime, which is known to suffer from infrared divergent issues and secular growth of quantum corrections \cite{Akhmedov:2013vka, Akhmedov:2019cfd, Akhmedov:2024npw, Miao:2024shs}. Such infrared divergence could signal a breakdown of perturbation theory. Although there is no clear agreement on handling them, several proposals exist for dealing with them. These include resummation of the secularly growing terms \cite{Baumgart:2019clc, Honda:2023unh, Cespedes:2023aal, Huenupi:2024ztu}, regulating with an IR-cutoff \cite{Xue:2011hm, Huenupi:2024ksc}, or incorporating nonlocality \cite{Narain:2018rif}. In this paper, we refrain from dealing with them and they will be a part of our future studies. 

For completeness, we also study the case when perturbation $h_{ij}$ is nonvanishing at the final boundary. It is seen that background no-boundary saddles($N_\sigma^{(q)}$) get modified and are no longer the saddles of the full theory. After the corrections are incorporated, the ``no-boundary'' saddles of the theory without $h_{ij}$ field are no longer the saddles of the theory where $h_{ij}$-field corrections are incorporated. This implies that in these corrected geometries, the universe no longer starts with a zero size. We confirm this for both Neumann and Robin boundary conditions. Stability analysis of the uncorrected saddles $N_{\sg}^{(q)}$ shows that for the Neumann BC case, the saddles are well behaved, and the fluctuation is perturbatively stable, as has been observed in past literature \cite{DiTucci:2019bui, DiTucci:2019dji}. For the Robin BC case, no-boundary saddles are seen to be perturbatively stable, while for the non-no-boundary saddles, the perturbations do not remain Gaussian.

Being the first study to investigate the path integral of the fluctuation field along with the scale factor and lapse and the effects of various boundary conditions, offers a glimpse of the complications and issues that one encounters in the process of computing the transition amplitude involving quantum corrections. In this paper, we make a first attempt to address this. We largely work with a simplified problem with vanishing perturbations at the boundaries leading to vanishing of backreaction at one loop. While we make certain boundary choices in this paper leading to vanishing backreaction and subsequent simplification of the computation to get a glimpse of the nature of quantum corrections involved, the results obtained in this paper raise some interesting questions. An immediate issue to address will be the effect of other boundary choices, for which the backreaction need not vanish and cannot be ignored. This is something nontrivial to address which we could not incorporate in this paper and will return to it in future. Another important point that needs to be studied in detail is the nature of counterterms and their dependence on gauge-fixing conditions and boundary choices. Is it possible to obtain a covariant expression for the counterterms? At a deeper level, it is worth asking whether we are implementing the ``no-boundary'' requirements appropriately. This is something which has also been raised in recent articles \cite{Maldacena:2024uhs,Ivo:2024ill}. Finally, the issue of infrared divergences that appear needs to be further explored systematically, and we should look for possible ways to deal with them. We will return to these in the future.

%%%%%%%%%%%%%%%%%%%%%%%%%%%%%%%%%%%%%%%%%%%%%%%
\bigskip
\centerline{\bf Acknowledgements} 
%%%%%%%%%%%%%%%%%%%%%%%%%%%%%%%%%%%%%%%%%%%%%%

We are thankful to Romesh Kaul for various illuminating discussions 
during the course of this work. We would also 
like to thank Chethan Krishnan 
and Justin David for useful discussions at various stages of the work.

\appendix

%%%%%%%%%%%%%%%%%%%%%%%%%%%%%%%%%%%%%%%%%%%%%%%%
\section{Functional determinants via Gel'fand Yaglom method}
\label{app:gelfand}
%%%%%%%%%%%%%%%%%%%%%%%%%%%%%%%%%%%%%%%%%%%%%%%%

Functional determinants of differential operators play an important role in theoretical 
and mathematical physics, and in particular, in quantum field theories. 
However, they are not easy to compute in nontrivial cases. 
In situations where the functional determinant effectively reduces to 
a determinant of a one-dimensional system, an elegant method 
proposed by Gel'fand and Yaglom can be utilized to compute the 
determinant in a simple manner \cite{Kirsten:2004qv,Dunne:2007rt}. 
In this paper, we have used these methods either to verify the 
computation of functional determinant done alternatively or 
sometimes to explicitly use them to compute the determinant. 
Here, in this appendix, we demonstrate the methods 
when applied to functional determinants appearing in this paper. 

Consider the most general second-order Sturm-Liouville differential operator
\begin{equation}
\label{eq:STL_opp}
\mathcal{D}=\frac{d}{dt}\left(P(t)\frac{d}{dt}\right)+Q(t) \, ,
\end{equation}
where, $t\in [0,1]$, satisfying the following eigenvalue equations,
\begin{equation}
\label{eq:eigenEQ}
\mathcal{D}u_{(\lambda)}(t)=\lambda u_{(\lambda)}(t) \, .
\end{equation}
By defining $v_{(\lambda)}=P(t)\dot{u}_{(\lambda)}$ (``dot'' being $t$ derivative), 
the generalized linear boundary condition can be written as
\begin{equation}
\label{eq:mat_MN}
M\left(\begin{matrix}
u_{(\lambda)}(0)\\
v_{(\lambda)}(0)
\end{matrix}
\right)
+N\left(\begin{matrix}
u_{(\lambda)}(1)\\
v_{(\lambda)}(1)
\end{matrix}\right)
=\left( \begin{matrix}
0\\
0
\end{matrix}\right) \, .
\end{equation}
The metrics, $M$ and $N$ are specific to the given boundary condition.  According to the Gel'fand-Yaglom theorem, the determinant of the differential operator is given by
\begin{equation}
\label{eq:detMN_GY}
\det{\mathcal{D}}=\det\left[M+N\left(\begin{matrix}
u_1(1) &u_2(1)\\
\dot{u}_1(1) & \dot{u}_2(1)
\end{matrix}\right)\right] \, ,
\end{equation}
where, $u_1(t)$ and $u_2(t)$ are the two independent solutions to
\begin{equation}
\label{eq:IInd_sol}
\mathcal{D}u_i(t)=0 \, ,
\end{equation}
which is zero eigenvalue solutions of Eq. (\ref{eq:eigenEQ}), with initial conditions,
\begin{equation}
\label{eq:INC_cond_u1u2}
\begin{split}
&u_1(0)=1, \quad\quad  \dot{u}_1(0)=0\\
&u_2(0)=0, \quad\quad\dot{u}_2(0)=1/P(0) \, .
\end{split}
\end{equation}
%

%%%%%%%%%%%%%%%%%%%%%%%%%%%%%%%%%%%%%%%%%%%%%%%%
\subsection{Background}
\label{appsub:BG}
%%%%%%%%%%%%%%%%%%%%%%%%%%%%%%%%%%%%%%%%%%%%%%%%

In our studies of quantum cosmology, while dealing with the 
minisuperspace model whose action is mentioned in Eq. (\ref{eq:EHact_exp}),
we note that for theory without the fluctuation field $h_{ij}$,
the theory reduces to a one-dimensional quantum mechanical problem.
In such situations, one can set up an eigenvalue problem and determine the functional determinant
as discussed in this appendix. The differential operator of the 
corresponding Sturm-Liouville system is given by
\begin{equation}
\label{eq:diff_SLO_QC_bg}
\mathcal{D}_{q}=\frac{1}{N_c}\frac{d^2}{dt^2} \, .
\end{equation}
Solving $\mathcal{D}_{q}u_i(t)=0$ with the initial conditions mentioned, we have
\begin{equation}
\label{eq:u1u2_BGQC_mini}
u_1(t)=1,
\hspace{5mm}
u_2(t)=N_c t \, .
\end{equation}
\begin{itemize}
\item \emph{Dirichlet-Dirichlet :} $M$ and $N$ are given by,
\begin{equation}
M=\left(\begin{matrix}
    1&0\\
    0&0
\end{matrix}\right),\hspace{5mm}N= \left(\begin{matrix}
    0&0\\
    1&0
\end{matrix}\right) \, .
\end{equation}
So, the functional determinant becomes
\begin{equation}
    \det \mathcal{D}_{q}=u_2(1)=N_c \, .
\end{equation}
\item \emph{Neumann-Dirichlet :} $M$ and $N$ are given by
\begin{equation}
M=\left(\begin{matrix}
    0&0\\
    0&1
\end{matrix}\right),\hspace{5mm}N= \left(\begin{matrix}
    1&0\\
    0&0
\end{matrix}\right) \, .
\end{equation}
So, the functional determinant becomes
\begin{equation}
    \det \mathcal{D}_{q}=u_1(1)=1 .
\end{equation}
\item \emph{Robin-Dirichlet :} $M$ and $N$ are given by
\begin{equation}
M=\left(\begin{matrix}
    \beta&-\frac{3}{2}\\
    0&1
\end{matrix}\right),\hspace{5mm}N= \left(\begin{matrix}
    0&0\\
    1&0
\end{matrix}\right) \, ,
\end{equation}
So, the functional determinant becomes
\begin{equation}
    \det \mathcal{D}_{q}=\beta u_2(1)+\frac{3}{2}u_1(1)
    =\frac{3}{2}\left(1+\frac{2}{3}\beta N_c\right)\, .
\end{equation}
\end{itemize}

%%%%%%%%%%%%%%%%%%%%%%%%%%%%%%%%%%%%%%%%%%%%%%%%
\subsection{Fluctuations}
\label{appsub:FLuc}
%%%%%%%%%%%%%%%%%%%%%%%%%%%%%%%%%%%%%%%%%%%%%%%%

For the fluctuation, we have the following strum-Liouville differential operator 
\begin{equation}
    \mathcal{D}_{h}=\frac{d}{dt}\left(P(t)\frac{d}{dt}\right)+Q(t),
\end{equation}
where $P(t)$ and $Q(t)$ are given by,
\begin{equation}
\begin{split}
    P(t)=\frac{\bar{q}^2}{N_c},\hspace{5mm}
    Q(t)= -N_c\left(2(\Lambda \bar{q}-2 \kappa)-\frac{\dot{\bar{q}}^2+4\bar{q}\ddot{\bar{q}}}{2N_c^2}-(l(l+2)-2)\right).
\end{split}
\end{equation}
with 
\beq
\label{eq:qsol_gen1}
\bar{q}(t) = \frac{\Lam N_c^2}{3} t^2 + c_1 t + c_2 
= \frac{\Lam N_c^2}{3} (t-r_1)(t-r_2)\, ,
\eeq
For the \emph{Dirichlet-Dirichlet} boundary condition, with the same choice of $M$ and $N$ matrices, we have
\begin{equation}
    \det\mathcal{D}_{h}=u_2(1),
\end{equation}
Solving $\mathcal{D}_{h}u_2=0$ with the initial condition mentioned, we have
\begin{equation}
u_2(t)=\frac{N_c\left(\mathbb{P}_1^{\xi_l}(\tau_t)\mathbb{Q}_1^{\xi_l}(\tau_0)
-\mathbb{Q}_1^{\xi_l}(\tau_t)\mathbb{P}_1^{\xi_l}(\tau_0)\right)}{\sqrt{q(t)}\,\bar{q}(0)^{3/2}
\left(\dot{\mathbb{P}}_1^{\xi_l}(\tau_0))\mathbb{Q}_1^{\xi_l}(\tau_0)
-\dot{\mathbb{Q}}_1^{\xi_l}(\tau_0))\mathbb{P}_1^{\xi_l}(\tau_0)\right)},
\end{equation}
where, $\xi_l$ and $\tau_t$ are given in Eq. (\ref{eq:albt_form}). 
Utilizing the following Wronskian identity,
\begin{equation}
    \,\bar{q}(0)\left(\dot{\mathbb{P}}_1^{\xi_l}(\tau_0))\mathbb{Q}_1^{\xi_l}(\tau_0)
    -\dot{\mathbb{Q}}_1^{\xi_l}(\tau_0))\mathbb{P}_1^{\xi_l}(\tau_0)\right)=  N_c\, \mathbb{M}(\xi_{l},N_{c}) 
\end{equation}
we have
\begin{equation}
    \det\mathcal{D}_{h}=u_2(1)=\frac{\mathbb{P}_1^{\xi_l}(\tau_1)\mathbb{Q}_1^{\xi_l}(\tau_0)-\mathbb{Q}_1^{\xi_l}(\tau_1)\mathbb{P}_1^{\xi_l}(\tau_0)}{\mathbb{M}(\xi_{l},N_c)\sqrt{q_1\bar{q}(0)}}\, ,
\end{equation}
where
\beq
\label{eq:D_wron_def}
\mathbb{M}(\xi_{l},N_c) = e^{i(\xi_{l}+1)\pi}\frac{\Lam N_c}{6} (r_1 - r_2) \frac{\Gamma[2+\xi_{l}]}{\Gamma[2-\xi_{l}]}\,,
\eeq
At the saddle points where $\xi_l=l+1$ and becomes integer Jacobian $\mathbb{M}(\xi_{l},N_c)$ becomes ill defined due to the presence of $\Gamma(1-l)$ in the denominator. This arises because at $\xi_l=l+1$, $\mathbb{P}_1^{\xi_l}(\tau_t)$ identically vanishes for all $t$ due to the presence of $(-1)_{l+1}$ in the definition. However, one can get rid of these by giving a small imaginary perturbation and consider $\mathbb{P}_{1+i\lambda}^{\xi_l}(\tau_t)$ \cite{Barvinsky:1992dz}. In the end, one has to take $\lambda\rightarrow 0$ using the identity 
\begin{equation}\label{identity}
    \lim_{\lambda\rightarrow 0}\Gamma(-l+1+i\lambda)(-1-i\lambda)_{l+1}=(-1)^{l+1},\,\, l\in 2,3,4,...
\end{equation}
where, $(-1)_{l+1}$ is Pochhammer symbol.
We also note that whenever $\bar{q}(0)=0$, $\det\mathcal{D}_{per}$ 
becomes infinite (not well defined). This is expected since the Strum-Liouville 
operator has a well-defined regular eigenvalue problem only for 
$P(t)>0$ for all $t\in [0,1]$. However, in this case, $P(0)$ becomes zero.
%%%%%%%%%%%%%%%%%%%%%%%%%%%%%%%%%%%%%%%%%
\section{Infinite summation of Log-series}
\label{logsum}
%%%%%%%%%%%%%%%%%%%%%%%%%%%%%%%%%%%%%%%%%

In this appendix, we will sum the following infinite series that appeared in the quantum effective lapse action in eq.\ref{eq:Nc_act_total_nb} and extract the divergent piece out of it:
\begin{equation}
    \label{eq:Ser_sum}
    S(a) = \sum_{l=2}^\infty g_l \log (l+a) \, ,
\end{equation}
where $g_l = 2 (l-1)(l+3)$ and $Re[a]>0$. To proceed with the summation, we define a function 
\begin{equation}
    \label{eq:FF_def_l}
    F_l(a) = g_l \log (l+a) \, .
\end{equation}
It is understood that the infinite series started in Eq. (\ref{eq:Ser_sum}) is divergent and would need regularization to extract the piece causing the divergence. To sum the series, we make use of differential equation and complexified special functions to extract the divergence. We first note that derivatives of $F_l(a)$ satisfy the following relation:
\begin{eqnarray}
    \label{eq:fl_der_form1}
    \frac{F_l^{\prime\prime}(a)}{2(a-1)}
    - \frac{F_l^\prime(a)}{(a+1)(a-3)}
    &=& \frac{a^2-4 a+7}{ (a-3) (a^2-1)}
    \nonumber \\
    &&
    -\frac{(a-3)
   (a+1)}{(a-1) (a+l)^2}-\frac{2l}{(a-3) (a+1)} \, .
\end{eqnarray}
On this when we apply the summation over $l$, then the rhs is easy to sum using zeta functions. Each term on the rhs is summed over $l$ individually and later added together. We note that $S(a)$ satisfies the following differential equation: 
\begin{equation}
    \label{eq:diff_eq_Sa}
    \frac{S''(a)}{2(a-1)} - \frac{S'(a)}{(a+1)(a-3)}
    = \frac{(49-9 a) a-6 (a-3)^2 (a+1)^2 \psi
   ^{(1)}(a+2)-76}{6 (a-3) (a-1) (a+1)} \, ,
\end{equation}
where the function $\psi^{(1)}(a+2)$ is a Polygamma function. This differential equation is easily solved using Mathematica in terms of special functions and is given by
\begin{eqnarray}
    \label{eq:Sa_sol_gen}
    &&
    S(a) = \frac{1}{6} a (4 ((a-3) a-9) c_1
    +9 a-49)-2 (a-3) (a+1) \log\Gamma (a+2)
    \nonumber \\
    &&
    -4 \psi ^{(-3)}(a+2)+ 4
   (a-1) \psi ^{(-2)}(a+2)+c_2 \, ,
\end{eqnarray}
where $c_{1,2}$ are constants to be determined. It is seen that if we take $a=0$ in Eq. (\ref{eq:Ser_sum}) the series can be summed using the using a regularization:
\begin{equation}
    \label{eq:S0}
    S(0) = \sum_{l=2}^\infty g_l \log (l)
    = 4 \log (A)- 2\zeta '(-2)-\frac{1}{3}-3\log
   (2 \pi ) \, ,
\end{equation}
where $A$ is the Glaisher constant. Utilizing this in the expression of $S(a)$ obtained from the solution of the differential equation, we note
\begin{eqnarray}
    \label{eq:S0_diff_eq}
     &&
    S(0) = c_2- 4 \bigl\{\psi ^{(-3)}(2)+\psi ^{(-2)}(2) \bigr\} \, ,
    \nonumber \\
    &&
    c_2 = 12 \log (A)+\frac{\zeta (3)}{2 \pi
   ^2}-\frac{22}{3}+5\log (2 \pi ) \, .
\end{eqnarray}
Solving for $c_1$ needs a bit more work to compute it. To compute $c_1$, we start by first considering the series mentioned in Eq. (\ref{eq:Ser_sum}), and focus on its argument $F_l(a)$. It is seen that
\begin{equation}
\label{eq:Fl_der1}
    F_l^\prime(a) = \frac{g_l}{l+a }
    = 2(2 - a + l) + \frac{2(a+1)(a-3)}{l+a} \, .
\end{equation}
This will mean that $l$ summation is applied over the $F_l^\prime(a)$; it acts separately over the $2 - a + l$, and the term proportional to $1/(l+a)$. The former can be summed using the zeta function, while the latter exhibits a divergence. The coefficient $c_1$ can be identified by computing the $S'(0)$ summation explicitly, which is given by
\begin{equation}
\label{eq:S_prm_sum}
    S'(0)=\sum_{l=2}^\infty F'_l(0)=\sum_{l=2}^\infty 4 + 2l-\frac{6}{l}=-\frac{13}{6}-6\zeta(1) \, ,
\end{equation}
where $\zeta(1)$ is the divergent term in the above expression. To extract the divergence, we express $1 \sim e^{-\epsilon}$, where $\epsilon$ is a small quantity. The divergence is expected to show up in the limit $\epsilon\to0$. The expansion of zeta in powers of $\epsilon$ is given by
\begin{equation}
\label{eq:zeta_exp}
    \zeta(e^{-\epsilon})=-\frac{1}{\epsilon }+\left(\gamma -\frac{1}{2}\right)+\left(\gamma _1-\frac{1}{12}\right) \epsilon +\mathcal{O}(\epsilon^2) \, ,
\end{equation}
where $\gamma$ is the Euler-Mascheroni constant, and $\g_{1}$ is Stieltjjes-Gamma function. From the solution of the differential equation, we also have
\begin{equation}
    \label{eq:Spa_0}
    S^\prime(0) = -6 c_1-6 \gamma -\frac{13}{6} \, .
\end{equation}
The comparison of the two equations leads to an expression for $c_1$ given by
\begin{equation}
\label{eq:c1_exp}
    c_1 = \left(-\frac{1}{\epsilon } -\frac{1}{2} + (\g_1 - 1/12)\epsilon+\mathcal{O}(\epsilon^2)\right) \,.
\end{equation}
Using these expressions for $c_1$ and $c_2$ given in Eq. (\ref{eq:c1_exp}) and (\ref{eq:S0_diff_eq}) one can obtain the complete summation of the series mentioned in Eq. (\ref{eq:Ser_sum}), where the divergent and finite parts have been separated. The divergent term in the sum is given by
\beq
\label{eq:div_term}
S(a)_{div} = -\frac{2a(a^2-3a-9)}{3 \ep}\, .
\eeq
Moreover, this divergent term can also be extracted using Schwinger's proper-time approach, which relies on the integral representation of the $\log$ function. Specifically,
\bea\label{eq:log_int_repr}
\log{y} = \int_{0}^{\infty} \frac{ds}{s}\, (e^{-s} - e^{-sy}) \, .
\eea
Using the above relation, the $l$ sum over $F_{l}(a)$ would become
\bea
\label{eq:l_sum_schw}
S(a) = \sum_{l=2}^\infty g_l \log (l+a) = \sum_{l=2}^\infty g_l   \int_{0}^{\infty} ds\, \left( \frac{e^{-s}}{s}  - \frac{ e^{-s(l+a))}}{s}  \right)\, .
\eea
Performing the $l$ summation over the integrand gives
\bea
\label{eq:l_sum_schw}
S(a) = \int_{0}^{\infty} \frac{ds}{s} \left(\frac{8}{3} e^{-s} + \frac{6 e^{-s a} - 10 e^{(1-a)s}}{(e^{s} - 1)^3}\right) \, .
\eea
where we used zeta-function regularization in the first term $\sum_{l=2}^{\infty} g_l = 8/3$. The sum $S(a)$ is divergent at $s=0$. The divergent terms can be identified by expanding the integrand and performing the integral around $s=0$:
\bea
\label{eq:l_sum_schw_int}
S(a) =  \left(\frac{4}{3s^3} - \frac{2(a-1)}{s^2} + \frac{2(a^2-2a-4)}{s} + \frac{2a(a^2 - 3a -9)}{3} \log{s}  + \mathcal{O}(s) \right)\Bigg|_{0}^{\infty} \, .
\eea
Interestingly, the coefficient of $\log$-divergent term at $s=0$ matches exactly with the divergent term obtained in Eq. (\ref{eq:div_term}) using the zeta-function regularization approach. However, it should be specified that cutoff regularization does not respect gauge invariance, and in cases where the cutoff is implemented in a gauge-invariant manner by introducing higher-derivative terms, it compromises on unitarity. Gauge-invariant regularizations like dimensional regularization, zeta function or the one implemented here are preferred. 

%%%%%%%%%%%%%%%%%%%%%%%%%%%%%%%%%%%%%%%%%
\section{Quantum Corrected lapse action: exact expressions}
\label{app:qcLapse_ex}
%%%%%%%%%%%%%%%%%%%%%%%%%%%%%%%%%%%%%%%%%

In this appendix, we write down the exact expression for the quantum-corrected transition amplitude for different final sizes of the Universe ($q_f$). This consists of three cases:$q_f=3/\Lam$, $q_f<3/\Lam$ and $q_f>3/\Lam$. Quantum corrected lapse action for each is given below respectively.
%%%%%%%%%%%%%%%%%%%%%%%%%%%%%%%%%%%%%%%%%
\subsection{$q_f=3/\Lambda$} 
\label{app:subs_qf_eq_lam}
%%%%%%%%%%%%%%%%%%%%%%%%%%%%%%%%%%%%%%%%%
In the case of $q_f=3/\Lam$, the two no-boundary saddles merge, leading to a degenerate situation. In such a scenario, only one saddle contributes. In this case the real and imaginary part of the quantum corrected lapse action is given by (we write
$\mathcal{A}_{\rm fin}(N^{\rm nb}_\pm) = \mathcal{A}_{\rm fin}^{\rm real}(N^{\rm nb}_\pm) + i \mathcal{A}_{\rm fin}^{\rm imag}(N^{\rm nb}_\pm)$) ,
\begin{equation}
\label{eq:finite_re_img_act_nb_sad_RBC_3}
\begin{split}
&\mathcal{A}_{\rm fin}^{\rm real}(N^{\rm nb}_\pm)=0 \, ,\\
&\mathcal{A}_{\rm fin}^{\rm imag}(N^{\rm nb}_\pm)=-\frac{3V_3}{4\pi G \Lambda}
+\frac{\hbar}{2} \Bl[
\ln(1 - x)
-\frac{\zeta (3)}{2 \pi ^2} + \ln( \pi^3 A^{4})
\\
&
-\frac{31}{120}\ln{\Lambda}  - \frac{149}{60}\ln{2} + \frac{37}{20}\ln{3} + \frac{61}{6} 
\Br]\, .
\end{split}
\end{equation}
%
%%%%%%%%%%%%%%%%%%%%%%%%%%%%%%%%%%%%%%%%%
\subsection{$q_f< 3/\Lambda$}
\label{app:subs_qf<lam}
%%%%%%%%%%%%%%%%%%%%%%%%%%%%%%%%%%%%%%%%%

For the case of $q_f<3/\Lam$, only one saddle is relevant, and the quantum corrected lapse action computed at this saddle is given by the following:
\begin{equation}
\label{eq:finite_re_img_act_nb_sad_RBC_2}
\begin{split}
&\mathcal{A}_{\rm fin}^{\rm real}(N^{\rm nb}_+) =0 \, ,\\
&\mathcal{A}_{\rm fin}^{\rm imag}(N^{\rm nb}_+) = - \frac{V_3}{36 \pi G \Lam} \left[ 27 - \left(9-3q_f\Lam\right)^{3/2}\right] \\
& + \frac{\hbar}{2}\Bigg[ \ln\Bl\{1-x + x \sqrt{1- \frac{\Lam q_f}{3}}\Br\} 
+4\ln(\pi^3 A^{4})-\frac{\zeta(3)}{2\pi^2} - \frac{117}{40}\ln{q_f} - \frac{ 3 q_f \Lam + 20}{6} \\
&+ \frac{191}{60}\ln\Bl\{\frac{3 - \sqrt{9 - 3\Lam q_f}}{\Lam }\Br\} 
%\\& 
+  \frac{\sqrt{9 - 3q_f\Lam}}{54}(2q_f\Lam -27) - 4 \psi ^{(-3)} \bigl(3 + \sqrt{ 1-q_f\Lam/3} \bigr)  \\
& + \frac{4\sqrt{9-3q_f\Lam}}{3}  \psi ^{(-2)} \bigl(3 +\sqrt{ 1-q_f\Lam/3} \bigr) + \frac{18 + 2q_f \Lam}{3} \ln \biggl\{\Gamma\bigl(3 +  \sqrt{ 1- q_f\Lam/3} \bigr)\biggr\}\, 
\Bigg. \\
&\Bl.+\frac{1}{120} (782 \log (2)+191 \log (3))\Br]\,.
\end{split}
\end{equation}
In the limit $q_f \rightarrow 0$, one needs to add the counterterm mentioned in Eq. (\ref{eq:Nc_act_total_nb_ct}) to cancel the $\ln q_f$ divergence appearing in the above expression. By doing that, we get
\begin{equation}
\label{eq:finite_re_img_act_nb_sad_RBC_2_q_f_0}
\begin{split}
&\mathcal{A}_{\rm fin}^{\rm real}(N^{\rm nb}_+,q_f\rightarrow 0) =0 \, ,\\
&\mathcal{A}_{\rm fin}^{\rm imag}(N^{\rm nb}_+,q_f\rightarrow 0) = \mathcal{A}_{\rm fin}^{\rm imag}(N^{\rm nb}_+) - \frac{31\hbar}{240}\ln{q_f} .
\end{split}
\end{equation}
%
%%%%%%%%%%%%%%%%%%%%%%%%%%%%%%%%%%%%%%%%%
\subsection{$q_f>3/\Lambda$}
\label{app:subs_qf>lam}
%%%%%%%%%%%%%%%%%%%%%%%%%%%%%%%%%%%%%%%%%
For the case of $q_f>3/\Lam$ the quantum corrected lapse action at the no-boundary saddle is given by the following:
\begin{equation}
\label{eq:finite_re_img_act_nb_sad_RBC_1}
\begin{split}
&\mathcal{A}_{\rm fin}(N^{\rm nb}_\pm)=-\frac{V_3}{36\pi G \Lambda}\left[27i\pm (3q_f\Lambda-9)^{3/2}\right]\\
&+\frac{i\hbar}{2}
\Bl[
\ln\Bl(  1 - x \mp i  x\sqrt{\frac{ \Lambda  q_f}{3}-1}\Br) +4\ln( \pi^{3} A^{4} )
-\frac{\zeta (3)}{2 \pi ^2} 
-\frac{117}{40}\ln q_f - \frac{ 3 q_f \Lam + 20}{6} 
\\
&
+\frac{191}{60}\ln\Bl\{\frac{3 \pm i \sqrt{3\Lam q_f -9}}{\Lam }\Br\}
\mp i \frac{\sqrt{ 3q_f\Lam-9}}{54}(2q_f\Lam -27)
- 4 \psi ^{(-3)} \bigl(3 \mp i \sqrt{ q_f\Lam/3-1} \bigr)\\
&\mp i\frac{4\sqrt{3q_f\Lam-9}}{3}  \psi ^{(-2)} \bigl(3 \mp i \sqrt{ q_f\Lam/3-1} \bigr) + \frac{18 + 2q_f \Lam}{3} \ln \biggl\{\Gamma\bigl(3 \mp i \sqrt{ q_f\Lam/3-1} \bigr)\biggr\}\Br.\, \\
&\Bl.+\frac{1}{120} (782 \log (2)+191 \log (3))\Br].
\end{split}
\end{equation}

%%%%%%%%%%%%%%%%%%%%%%%%%%%%%%%%%%%%%%%%%

%%%%%%%%%%%%%%%%%%%%%%%%%%

\begin{thebibliography}{99}


%\cite{Gibbons:1978ac}
\bibitem{Gibbons:1978ac}
G.~W.~Gibbons, S.~W.~Hawking and M.~J.~Perry,
``Path Integrals and the Indefiniteness of the Gravitational Action,''
Nucl. Phys. B \textbf{138} (1978), 141-150
doi:10.1016/0550-3213(78)90161-X
%668 citations counted in INSPIRE as of 18 Oct 2024

%\cite{Kontsevich:2021dmb}
\bibitem{Kontsevich:2021dmb}
M.~Kontsevich and G.~Segal,
``Wick Rotation and the Positivity of Energy in Quantum Field Theory,''
Quart. J. Math. Oxford Ser. \textbf{72} (2021) no.1-2, 673-699
doi:10.1093/qmath/haab027
[arXiv:2105.10161 [hep-th]].
%112 citations counted in INSPIRE as of 25 Oct 2024

%\cite{Witten:2021nzp}
\bibitem{Witten:2021nzp}
E.~Witten,
``A Note On Complex Spacetime Metrics,''
[arXiv:2111.06514 [hep-th]].
%132 citations counted in INSPIRE as of 25 Oct 2024

%\cite{Lehners:2021mah}
\bibitem{Lehners:2021mah}
J.~L.~Lehners,
``Allowable complex metrics in minisuperspace quantum cosmology,''
Phys. Rev. D \textbf{105} (2022) no.2, 026022
doi:10.1103/PhysRevD.105.026022
[arXiv:2111.07816 [hep-th]].
%41 citations counted in INSPIRE as of 25 Sep 2024

%\cite{tHooft:1974toh}
\bibitem{tHooft:1974toh}
G.~'t Hooft and M.~J.~G.~Veltman,
``One loop divergencies in the theory of gravitation,''
Ann. Inst. H. Poincare A Phys. Theor. \textbf{20} (1974), 69-94
%78 citations counted in INSPIRE as of 17 Oct 2024

%\cite{Deser:1974nb}
\bibitem{Deser:1974nb}
S.~Deser, H.~S.~Tsao and P.~van Nieuwenhuizen,
``Nonrenormalizability of Einstein Yang-Mills Interactions at the One Loop Level,''
Phys. Lett. B \textbf{50} (1974), 491-493
doi:10.1016/0370-2693(74)90268-8
%79 citations counted in INSPIRE as of 08 Oct 2024

%\cite{Deser:1974cz}
\bibitem{Deser:1974cz}
S.~Deser and P.~van Nieuwenhuizen,
``One Loop Divergences of Quantized Einstein-Maxwell Fields,''
Phys. Rev. D \textbf{10} (1974), 401
doi:10.1103/PhysRevD.10.401
%570 citations counted in INSPIRE as of 22 Oct 2024

%\cite{Goroff:1985sz}
\bibitem{Goroff:1985sz}
M.~H.~Goroff and A.~Sagnotti,
``QUANTUM GRAVITY AT TWO LOOPS,''
Phys. Lett. B \textbf{160} (1985), 81-86
doi:10.1016/0370-2693(85)91470-4
%445 citations counted in INSPIRE as of 22 Oct 2024

%\cite{Goroff:1985th}
\bibitem{Goroff:1985th}
M.~H.~Goroff and A.~Sagnotti,
``The Ultraviolet Behavior of Einstein Gravity,''
Nucl. Phys. B \textbf{266} (1986), 709-736
doi:10.1016/0550-3213(86)90193-8
%870 citations counted in INSPIRE as of 08 Oct 2024

%\cite{vandeVen:1991gw}
\bibitem{vandeVen:1991gw}
A.~E.~M.~van de Ven,
``Two loop quantum gravity,''
Nucl. Phys. B \textbf{378} (1992), 309-366
doi:10.1016/0550-3213(92)90011-Y
%361 citations counted in INSPIRE as of 30 Sep 2024

%\cite{Stelle:1976gc}
\bibitem{Stelle:1976gc}
K.~S.~Stelle,
``Renormalization of Higher Derivative Quantum Gravity,''
Phys. Rev. D \textbf{16} (1977), 953-969
doi:10.1103/PhysRevD.16.953
%2622 citations counted in INSPIRE as of 25 Oct 2024

%\cite{Salam:1978fd}
\bibitem{Salam:1978fd}
A.~Salam and J.~A.~Strathdee,
``Remarks on High-energy Stability and Renormalizability of Gravity Theory,''
Phys. Rev. D \textbf{18} (1978), 4480
doi:10.1103/PhysRevD.18.4480
%156 citations counted in INSPIRE as of 25 Sep 2024

%\cite{Narain:2011gs}
\bibitem{Narain:2011gs}
G.~Narain and R.~Anishetty,
``Short Distance Freedom of Quantum Gravity,''
Phys. Lett. B \textbf{711} (2012), 128-131
doi:10.1016/j.physletb.2012.03.070
[arXiv:1109.3981 [hep-th]].
%37 citations counted in INSPIRE as of 25 Sep 2024

%\cite{Narain:2016sgk}
\bibitem{Narain:2016sgk}
G.~Narain,
``Exorcising Ghosts in Induced Gravity,''
Eur. Phys. J. C \textbf{77} (2017) no.10, 683
doi:10.1140/epjc/s10052-017-5249-z
[arXiv:1612.04930 [hep-th]].
%19 citations counted in INSPIRE as of 25 Sep 2024

%\cite{Buccio:2024hys}
\bibitem{Buccio:2024hys}
D.~Buccio, J.~F.~Donoghue, G.~Menezes and R.~Percacci,
``Physical Running of Couplings in Quadratic Gravity,''
Phys. Rev. Lett. \textbf{133} (2024) no.2, 021604
doi:10.1103/PhysRevLett.133.021604
[arXiv:2403.02397 [hep-th]].
%8 citations counted in INSPIRE as of 25 Sep 2024

%\cite{Solodukhin:2015ypa}
\bibitem{Solodukhin:2015ypa}
S.~N.~Solodukhin,
``Metric Redefinition and UV Divergences in Quantum Einstein Gravity,''
Phys. Lett. B \textbf{754} (2016), 157-161
doi:10.1016/j.physletb.2016.01.015
[arXiv:1509.04890 [hep-th]].
%5 citations counted in INSPIRE as of 25 Sep 2024

%\cite{Batalin:1977pb}
\bibitem{Batalin:1977pb}
I.~A.~Batalin and G.~A.~Vilkovisky,
``Relativistic S Matrix of Dynamical Systems with Boson and Fermion Constraints,''
Phys. Lett. B \textbf{69} (1977), 309-312
doi:10.1016/0370-2693(77)90553-6
%1094 citations counted in INSPIRE as of 25 Sep 2024

%\cite{Feldbrugge:2017kzv}
\bibitem{Feldbrugge:2017kzv}
J.~Feldbrugge, J.~L.~Lehners and N.~Turok,
``Lorentzian Quantum Cosmology,''
Phys. Rev. D \textbf{95} (2017) no.10, 103508
doi:10.1103/PhysRevD.95.103508
[arXiv:1703.02076 [hep-th]].
%202 citations counted in INSPIRE as of 11 Oct 2024

%\cite{Teitelboim:1981ua}
\bibitem{Teitelboim:1981ua}
C.~Teitelboim,
``Quantum Mechanics of the Gravitational Field,''
Phys. Rev. D \textbf{25} (1982), 3159
doi:10.1103/PhysRevD.25.3159
%365 citations counted in INSPIRE as of 25 Sep 2024

%\cite{Teitelboim:1983fk}
\bibitem{Teitelboim:1983fk}
C.~Teitelboim,
``The Proper Time Gauge in Quantum Theory of Gravitation,''
Phys. Rev. D \textbf{28} (1983), 297
doi:10.1103/PhysRevD.28.297
%147 citations counted in INSPIRE as of 25 Sep 2024

%\cite{Feldbrugge:2017mbc}
\bibitem{Feldbrugge:2017mbc}
J.~Feldbrugge, J.~L.~Lehners and N.~Turok,
``No rescue for the no boundary proposal: Pointers to the future of quantum cosmology,''
Phys. Rev. D \textbf{97}, no.2, 023509 (2018)
doi:10.1103/PhysRevD.97.023509
[arXiv:1708.05104 [hep-th]].
%114 citations counted in INSPIRE as of 04 Feb 2025

%\cite{Halliwell:1988wc}
\bibitem{Halliwell:1988wc}
J.~J.~Halliwell,
``Derivation of the Wheeler-De Witt Equation from a Path Integral for Minisuperspace Models,''
Phys. Rev. D \textbf{38} (1988), 2468
doi:10.1103/PhysRevD.38.2468
%314 citations counted in INSPIRE as of 25 Sep 2024

%\cite{Faddeev:1967fc}
\bibitem{Faddeev:1967fc}
L.~D.~Faddeev and V.~N.~Popov,
``Feynman Diagrams for the Yang-Mills Field,''
Phys. Lett. B \textbf{25} (1967), 29-30
doi:10.1016/0370-2693(67)90067-6
%2323 citations counted in INSPIRE as of 16 Oct 2024

%\cite{Ohta:2015zwa}
\bibitem{Ohta:2015zwa}
N.~Ohta and R.~Percacci,
``Ultraviolet Fixed Points in Conformal Gravity and General Quadratic Theories,''
Class. Quant. Grav. \textbf{33} (2016), 035001
doi:10.1088/0264-9381/33/3/035001
[arXiv:1506.05526 [hep-th]].
%45 citations counted in INSPIRE as of 11 Feb 2025

%\cite{Lin:2017ool}
\bibitem{Lin:2017ool}
H.~Lin and G.~Narain,
``AdS backgrounds and induced gravity,''
Mod. Phys. Lett. A \textbf{34} (2019) no.38, 2050057
doi:10.1142/S0217732320500571
[arXiv:1712.09995 [hep-th]].
%3 citations counted in INSPIRE as of 11 Feb 2025


%\cite{Teitelboim:1983fh}
\bibitem{Teitelboim:1983fh}
C.~Teitelboim,
``Causality Versus Gauge Invariance in Quantum Gravity and Supergravity,''
Phys. Rev. Lett. \textbf{50} (1983), 705
doi:10.1103/PhysRevLett.50.705
%100 citations counted in INSPIRE as of 25 Sep 2024

%\cite{Brizuela:2023vmb}
\bibitem{Brizuela:2023vmb}
D.~Brizuela and M.~de Cesare,
``Generalized boundary conditions in closed cosmologies,''
Phys. Rev. D \textbf{107} (2023) no.10, 104054
doi:10.1103/PhysRevD.107.104054
[arXiv:2303.04007 [gr-qc]].
%3 citations counted in INSPIRE as of 25 Sep 2024

%\cite{Chou:2024sgk}
\bibitem{Chou:2024sgk}
C.~Y.~Chou and J.~Nishimura,
``Monte Carlo studies of quantum cosmology by the generalized Lefschetz thimble method,''
[arXiv:2407.17724 [gr-qc]].
%1 citations counted in INSPIRE as of 25 Oct 2024

%\cite{Nishimura:2023dky}
\bibitem{Nishimura:2023dky}
J.~Nishimura, K.~Sakai and A.~Yosprakob,
``A new picture of quantum tunneling in the real-time path integral from Lefschetz thimble calculations,''
JHEP \textbf{09} (2023), 110
doi:10.1007/JHEP09(2023)110
[arXiv:2307.11199 [hep-th]].
%13 citations counted in INSPIRE as of 25 Oct 2024

%\cite{Lehners:2023yrj}
\bibitem{Lehners:2023yrj}
J.~L.~Lehners,
``Review of the no-boundary wave function,''
Phys. Rept. \textbf{1022} (2023), 1-82
doi:10.1016/j.physrep.2023.06.002
[arXiv:2303.08802 [hep-th]].
%32 citations counted in INSPIRE as of 17 Oct 2024

%\cite{DiTucci:2018fdg}
\bibitem{DiTucci:2018fdg}
A.~Di Tucci and J.~L.~Lehners,
``Unstable no-boundary fluctuations from sums over regular metrics,''
Phys. Rev. D \textbf{98} (2018) no.10, 103506
doi:10.1103/PhysRevD.98.103506
[arXiv:1806.07134 [gr-qc]].
%26 citations counted in INSPIRE as of 25 Sep 2024

%\cite{Lehners:2018eeo}
\bibitem{Lehners:2018eeo}
J.~L.~Lehners,
``No smooth beginning for spacetime,''
Int. J. Mod. Phys. D \textbf{28} (2018) no.02, 1930005
doi:10.1142/9789811258251\_0005
%6 citations counted in INSPIRE as of 14 Oct 2024

%\cite{Feldbrugge:2017fcc}
\bibitem{Feldbrugge:2017fcc}
J.~Feldbrugge, J.~L.~Lehners and N.~Turok,
``No smooth beginning for spacetime,''
Phys. Rev. Lett. \textbf{119} (2017) no.17, 171301
doi:10.1103/PhysRevLett.119.171301
[arXiv:1705.00192 [hep-th]].
%140 citations counted in INSPIRE as of 27 Sep 2024

%\cite{Barvinsky:1992dz}
\bibitem{Barvinsky:1992dz}
A.~O.~Barvinsky, A.~Y.~Kamenshchik and I.~P.~Karmazin,
``One loop quantum cosmology: Zeta function technique for the Hartle-Hawking wave function of the universe,''
Annals Phys. \textbf{219}, 201-242 (1992)
doi:10.1016/0003-4916(92)90347-O
%93 citations counted in INSPIRE as of 04 Feb 2025

%\cite{Krishnan:2016mcj}
\bibitem{Krishnan:2016mcj}
C.~Krishnan and A.~Raju,
``A Neumann Boundary Term for Gravity,''
Mod. Phys. Lett. A \textbf{32} (2017) no.14, 1750077
doi:10.1142/S0217732317500778
[arXiv:1605.01603 [hep-th]].
%58 citations counted in INSPIRE as of 25 Sep 2024

%\cite{Krishnan:2017bte}
\bibitem{Krishnan:2017bte}
C.~Krishnan, S.~Maheshwari and P.~N.~Bala Subramanian,
``Robin Gravity,''
J. Phys. Conf. Ser. \textbf{883} (2017) no.1, 012011
doi:10.1088/1742-6596/883/1/012011
[arXiv:1702.01429 [gr-qc]].
%30 citations counted in INSPIRE as of 25 Sep 2024

%\cite{DiTucci:2019bui}
\bibitem{DiTucci:2019bui}
A.~Di Tucci, J.~L.~Lehners and L.~Sberna,
``No-boundary prescriptions in Lorentzian quantum cosmology,''
Phys. Rev. D \textbf{100} (2019) no.12, 123543
doi:10.1103/PhysRevD.100.123543
[arXiv:1911.06701 [hep-th]].
%47 citations counted in INSPIRE as of 15 Oct 2024

%\cite{Narain:2021bff}
\bibitem{Narain:2021bff}
G.~Narain,
``On Gauss-bonnet gravity and boundary conditions in Lorentzian path-integral quantization,''
JHEP \textbf{05} (2021), 273
doi:10.1007/JHEP05(2021)273
[arXiv:2101.04644 [gr-qc]].
%11 citations counted in INSPIRE as of 22 Oct 2024

%\cite{Lehners:2021jmv}
\bibitem{Lehners:2021jmv}
J.~L.~Lehners,
``Wave function of simple universes analytically continued from negative to positive potentials,''
Phys. Rev. D \textbf{104} (2021) no.6, 063527
doi:10.1103/PhysRevD.104.063527
[arXiv:2105.12075 [hep-th]].
%20 citations counted in INSPIRE as of 14 Oct 2024

%\cite{DiTucci:2020weq}
\bibitem{DiTucci:2020weq}
A.~Di Tucci, M.~P.~Heller and J.~L.~Lehners,
``Lessons for quantum cosmology from anti\textendash{}de Sitter black holes,''
Phys. Rev. D \textbf{102} (2020) no.8, 086011
doi:10.1103/PhysRevD.102.086011
[arXiv:2007.04872 [hep-th]].
%25 citations counted in INSPIRE as of 14 Oct 2024

%\cite{Narain:2022msz}
\bibitem{Narain:2022msz}
G.~Narain,
``Surprises in Lorentzian path-integral of Gauss-Bonnet gravity,''
JHEP \textbf{04} (2022), 153
doi:10.1007/JHEP04(2022)153
[arXiv:2203.05475 [gr-qc]].
%8 citations counted in INSPIRE as of 22 Oct 2024

%\cite{Mazur:1989ch}
\bibitem{Mazur:1989ch}
P.~O.~Mazur and E.~Mottola,
``ABSENCE OF PHASE IN THE SUM OVER SPHERES,''
LA-UR-89-2118.
%4 citations counted in INSPIRE as of 23 Dec 2024


%\cite{DiTucci:2019dji}
\bibitem{DiTucci:2019dji}
A.~Di Tucci and J.~L.~Lehners,
``No-Boundary Proposal as a Path Integral with Robin Boundary Conditions,''
Phys. Rev. Lett. \textbf{122} (2019) no.20, 201302
doi:10.1103/PhysRevLett.122.201302
[arXiv:1903.06757 [hep-th]].
%61 citations counted in INSPIRE as of 25 Sep 2024

%\cite{Matsui:2022lfj}
\bibitem{Matsui:2022lfj}
H.~Matsui, S.~Mukohyama and A.~Naruko,
``No smooth spacetime in Lorentzian quantum cosmology and trans-Planckian physics,''
Phys. Rev. D \textbf{107} (2023) no.4, 043511
doi:10.1103/PhysRevD.107.043511
[arXiv:2211.05306 [gr-qc]].
%8 citations counted in INSPIRE as of 25 Oct 2024

%\cite{Matsui:2023hei}
\bibitem{Matsui:2023hei}
H.~Matsui and S.~Mukohyama,
``Hartle-Hawking no-boundary proposal and Ho\v{r}ava-Lifshitz gravity,''
Phys. Rev. D \textbf{109} (2024) no.2, 023504
doi:10.1103/PhysRevD.109.023504
[arXiv:2310.00210 [gr-qc]].
%5 citations counted in INSPIRE as of 25 Oct 2024

%\cite{Honda:2024aro}
\bibitem{Honda:2024aro}
M.~Honda, H.~Matsui, K.~Okabayashi and T.~Terada,
``Resurgence in Lorentzian quantum cosmology: No-boundary saddles and resummation of quantum gravity corrections around tunneling saddle points,''
Phys. Rev. D \textbf{110} (2024) no.8, 8
doi:10.1103/PhysRevD.110.083508
[arXiv:2402.09981 [gr-qc]].
%5 citations counted in INSPIRE as of 25 Oct 2024

%\cite{Matsui:2024bfn}
\bibitem{Matsui:2024bfn}
H.~Matsui,
``No smooth spacetime: Exploring primordial perturbations in Lorentzian quantum cosmology,''
Phys. Rev. D \textbf{110}, no.2, 023503 (2024)
doi:10.1103/PhysRevD.110.023503
[arXiv:2404.18609 [gr-qc]].
%3 citations counted in INSPIRE as of 25 Oct 2024

%\cite{Ailiga:2023wzl}
\bibitem{Ailiga:2023wzl}
M.~Ailiga, S.~Mallik and G.~Narain,
``Lorentzian Robin Universe,''
JHEP \textbf{01} (2024), 124
doi:10.1007/JHEP01(2024)124
[arXiv:2308.01310 [gr-qc]].
%7 citations counted in INSPIRE as of 22 Oct 2024

%\cite{Ailiga:2024mmt}
\bibitem{Ailiga:2024mmt}
M.~Ailiga, S.~Mallik and G.~Narain,
``Lorentzian path-integral of Robin Universe,''
[arXiv:2407.16692 [gr-qc]].
%1 citations counted in INSPIRE as of 22 Oct 2024


%\cite{Witten:2010cx}
\bibitem{Witten:2010cx}
E.~Witten,
``Analytic Continuation Of Chern-Simons Theory,''
AMS/IP Stud. Adv. Math. \textbf{50} (2011), 347-446
[arXiv:1001.2933 [hep-th]].
%497 citations counted in INSPIRE as of 18 Oct 2024

%\cite{Witten:2010zr}
\bibitem{Witten:2010zr}
E.~Witten,
``A New Look At The Path Integral Of Quantum Mechanics,''
[arXiv:1009.6032 [hep-th]].
%225 citations counted in INSPIRE as of 14 Oct 2024

%\cite{Basar:2013eka}
\bibitem{Basar:2013eka}
G.~Basar, G.~V.~Dunne and M.~Unsal,
``Resurgence theory, ghost-instantons, and analytic continuation of path integrals,''
JHEP \textbf{10} (2013), 041
doi:10.1007/JHEP10(2013)041
[arXiv:1308.1108 [hep-th]].
%133 citations counted in INSPIRE as of 22 Oct 2024

%\cite{Tanizaki:2014xba}
\bibitem{Tanizaki:2014xba}
Y.~Tanizaki and T.~Koike,
``Real-time Feynman path integral with Picard\textendash{}Lefschetz theory and its applications to quantum tunneling,''
Annals Phys. \textbf{351} (2014), 250-274
doi:10.1016/j.aop.2014.09.003
[arXiv:1406.2386 [math-ph]].
%106 citations counted in INSPIRE as of 22 Oct 2024

%\cite{DeWitt:1980jv}
\bibitem{DeWitt:1980jv}
B.~S.~DeWitt,
``A GAUGE INVARIANT EFFECTIVE ACTION,''
NSF-ITP-80-31.
%6 citations counted in INSPIRE as of 25 Sep 2024

%\cite{Abbott:1980hw}
\bibitem{Abbott:1980hw}
L.~F.~Abbott,
``The Background Field Method Beyond One Loop,''
Nucl. Phys. B \textbf{185} (1981), 189-203
doi:10.1016/0550-3213(81)90371-0
%1342 citations counted in INSPIRE as of 11 Oct 2024

%\cite{Gerlach:1978gy}
\bibitem{Gerlach:1978gy}
U.~H.~Gerlach and U.~K.~Sengupta,
``Homogeneous Collapsing Star: Tensor and Vector Harmonics for Matter and Field Asymmetries,''
Phys. Rev. D \textbf{18} (1978), 1773-1784
doi:10.1103/PhysRevD.18.1773
%73 citations counted in INSPIRE as of 26 Sep 2024

\bibitem{Sabir:1991PrJ}
M.~Sabir and S.~Rajagopalan,
``Path integral analysis of harmonic oscillators with time-dependent mass,''
Pramana \textbf{37} (1991) no.3, 253-260
doi:10.1007/BF02847479


%\cite{Khandekar:1986ib}
\bibitem{Khandekar:1986ib}
D.~C.~Khandekar and S.~V.~Lawande,
``Feynman Path Integrals: Some Exact Results and Applications,''
Phys. Rept. \textbf{137} (1986), 115-229
doi:10.1016/0370-1573(86)90029-3
%40 citations counted in INSPIRE as of 25 Sep 2024

%\cite{McKane:1995vp}
\bibitem{McKane:1995vp}
A.~J.~McKane and M.~B.~Tarlie,
``Regularization of functional determinants using boundary perturbations,''
J. Phys. A \textbf{28} (1995), 6931-6942
doi:10.1088/0305-4470/28/23/032
[arXiv:cond-mat/9509126 [cond-mat]].
%33 citations counted in INSPIRE as of 25 Sep 2024

\bibitem{Kleinert:1999Cher}
H.~Kleinert and A.~Chervyakov,
``Functional determinants from Wronski Green functions,''
Journal of Mathematical Physics \textbf{40} (1999) no.11, 6044–6051
doi:10.1063/1.533069
[arXiv:physics/9712048]


%\cite{Kleinert:1998rz}
\bibitem{Kleinert:1998rz}
H.~Kleinert and A.~Chervyakov,
``Simple explicit formulas for Gaussian path integrals with time dependent frequencies,''
Phys. Lett. A \textbf{245} (1998), 345-357
doi:10.1016/S0375-9601(98)00380-6
[arXiv:quant-ph/9803016 [quant-ph]].
%21 citations counted in INSPIRE as of 25 Sep 2024

%\cite{Lehners:2019ibe}
\bibitem{Lehners:2019ibe}
J.~L.~Lehners and K.~S.~Stelle,
``A Safe Beginning for the Universe?,''
Phys. Rev. D \textbf{100} (2019) no.8, 083540
doi:10.1103/PhysRevD.100.083540
[arXiv:1909.01169 [hep-th]].
%37 citations counted in INSPIRE as of 15 Oct 2024

%\cite{Lehners:2023fud}
\bibitem{Lehners:2023fud}
J.~L.~Lehners and K.~S.~Stelle,
``Higher-order gravity, finite action, and a safe beginning for the universe,''
Eur. Phys. J. Plus \textbf{139} (2024), 380
doi:10.1140/epjp/s13360-024-05125-y
[arXiv:2312.14048 [hep-th]].
%3 citations counted in INSPIRE as of 25 Sep 2024

%\cite{Akhmedov:2013vka}
\bibitem{Akhmedov:2013vka}
E.~T.~Akhmedov,
``Lecture notes on interacting quantum fields in de Sitter space,''
Int. J. Mod. Phys. D \textbf{23} (2014), 1430001
doi:10.1142/S0218271814300018
[arXiv:1309.2557 [hep-th]].
%144 citations counted in INSPIRE as of 24 Dec 2024

%\cite{Akhmedov:2019cfd}
\bibitem{Akhmedov:2019cfd}
E.~T.~Akhmedov, U.~Moschella and F.~K.~Popov,
``Characters of different secular effects in various patches of de Sitter space,''
Phys. Rev. D \textbf{99} (2019) no.8, 086009
doi:10.1103/PhysRevD.99.086009
[arXiv:1901.07293 [hep-th]].
%58 citations counted in INSPIRE as of 24 Dec 2024

%\cite{Akhmedov:2024npw}
\bibitem{Akhmedov:2024npw}
E.~T.~Akhmedov, V.~I.~Lapushkin and D.~I.~Sadekov,
``Light fields in various patches of de Sitter space-time,''
[arXiv:2411.11106 [hep-th]].
%0 citations counted in INSPIRE as of 24 Dec 2024

%\cite{Miao:2024shs}
\bibitem{Miao:2024shs}
S.~P.~Miao, N.~C.~Tsamis and R.~P.~Woodard,
``Leading Logarithm Quantum Gravity,''
[arXiv:2409.12003 [gr-qc]].
%2 citations counted in INSPIRE as of 24 Dec 2024

%\cite{Baumgart:2019clc}
\bibitem{Baumgart:2019clc}
M.~Baumgart and R.~Sundrum,
``De Sitter Diagrammar and the Resummation of Time,''
JHEP \textbf{07} (2020), 119
doi:10.1007/JHEP07(2020)119
[arXiv:1912.09502 [hep-th]].
%69 citations counted in INSPIRE as of 24 Dec 2024


%\cite{Honda:2023unh}
\bibitem{Honda:2023unh}
M.~Honda, R.~Jinno, L.~Pinol and K.~Tokeshi,
``Borel resummation of secular divergences in stochastic inflation,''
JHEP \textbf{08} (2023), 060
doi:10.1007/JHEP08(2023)060
[arXiv:2304.02592 [hep-th]].
%11 citations counted in INSPIRE as of 24 Dec 2024


%\cite{Cespedes:2023aal}
\bibitem{Cespedes:2023aal}
S.~C\'espedes, A.~C.~Davis and D.~G.~Wang,
``On the IR divergences in de Sitter space: loops, resummation and the semi-classical wavefunction,''
JHEP \textbf{04} (2024), 004
doi:10.1007/JHEP04(2024)004
[arXiv:2311.17990 [hep-th]].
%27 citations counted in INSPIRE as of 24 Dec 2024

%\cite{Starobinsky:1994bd}
\bibitem{Starobinsky:1994bd}
A.~A.~Starobinsky and J.~Yokoyama,
``Equilibrium state of a selfinteracting scalar field in the De Sitter background,''
Phys. Rev. D \textbf{50}, 6357-6368 (1994)
doi:10.1103/PhysRevD.50.6357
[arXiv:astro-ph/9407016 [astro-ph]].
%796 citations counted in INSPIRE as of 21 May 2025


%\cite{Huenupi:2024ztu}
\bibitem{Huenupi:2024ztu}
J.~Huenupi, E.~Hughes, G.~A.~Palma and S.~Sypsas,
``A note on loop resummation in de Sitter spacetime with the wavefunction of the universe approach,''
[arXiv:2412.01891 [hep-th]].
%0 citations counted in INSPIRE as of 24 Dec 2024


%\cite{Huenupi:2024ksc}
\bibitem{Huenupi:2024ksc}
J.~Huenupi, E.~Hughes, G.~A.~Palma and S.~Sypsas,
``Regularizing infrared divergences in de Sitter spacetime: loops, dimensional regularization and cutoffs,''
[arXiv:2406.07610 [hep-th]].
%2 citations counted in INSPIRE as of 24 Dec 2024


%\cite{Kamenshchik:2024ybm}
\bibitem{Kamenshchik:2024ybm}
A.~Kamenshchik and P.~Petriakova,
``IR finite correlation functions in de Sitter space, a smooth massless limit, and an autonomous equation,''
JHEP \textbf{04}, 127 (2025)
doi:10.1007/JHEP04(2025)127
[arXiv:2410.16226 [hep-th]].
%3 citations counted in INSPIRE as of 21 May 2025

%\cite{Kamenshchik:2025ses}
\bibitem{Kamenshchik:2025ses}
A.~Kamenshchik and P.~Petriakova,
``From the Fokker-Planck equation to perturbative QFT's results in de Sitter space,''
[arXiv:2504.20646 [hep-th]].
%0 citations counted in INSPIRE as of 21 May 2025


%\cite{Xue:2011hm}
\bibitem{Xue:2011hm}
W.~Xue, K.~Dasgupta and R.~Brandenberger,
``Cosmological UV/IR Divergences and de-Sitter Spacetime,''
Phys. Rev. D \textbf{83} (2011), 083520
doi:10.1103/PhysRevD.83.083520
[arXiv:1103.0285 [hep-th]].
%44 citations counted in INSPIRE as of 24 Dec 2024

%\cite{Narain:2018rif}
\bibitem{Narain:2018rif}
G.~Narain and N.~Kajuri,
``Non-local scalar field on deSitter and its infrared behaviour,''
Phys. Lett. B \textbf{791} (2019), 143-148
doi:10.1016/j.physletb.2019.02.030
[arXiv:1812.00947 [hep-th]].
%4 citations counted in INSPIRE as of 24 Dec 2024

%\cite{Maldacena:2024uhs}
\bibitem{Maldacena:2024uhs}
J.~Maldacena,
``Comments on the no boundary wavefunction and slow roll inflation,''
[arXiv:2403.10510 [hep-th]].
%22 citations counted in INSPIRE as of 26 Dec 2024

%\cite{Ivo:2024ill}
\bibitem{Ivo:2024ill}
V.~Ivo, Y.~Z.~Li and J.~Maldacena,
``The no boundary density matrix,''
[arXiv:2409.14218 [hep-th]].
%5 citations counted in INSPIRE as of 20 Feb 2025


%\cite{Kirsten:2004qv}
\bibitem{Kirsten:2004qv}
K.~Kirsten and A.~J.~McKane,
``Functional determinants for general Sturm-Liouville problems,''
J. Phys. A \textbf{37} (2004), 4649-4670
doi:10.1088/0305-4470/37/16/014
[arXiv:math-ph/0403050 [math-ph]].
%85 citations counted in INSPIRE as of 25 Sep 2024

%\cite{Dunne:2007rt}
\bibitem{Dunne:2007rt}
G.~V.~Dunne,
``Functional determinants in quantum field theory,''
J. Phys. A \textbf{41} (2008), 304006
doi:10.1088/1751-8113/41/30/304006
[arXiv:0711.1178 [hep-th]].
%108 citations counted in INSPIRE as of 21 Oct 2024



\end{thebibliography}
\end{document}